\documentclass[12pt]{article}
\pdfoutput=1
\usepackage[a4paper]{geometry}
\usepackage{jheppub, amsmath,amssymb,amsfonts,amsxtra, mathrsfs, makeidx,graphics,graphicx,amsthm,epsfig, ytableau,bm,longtable,float, color,tikz,mathtools,xfrac,footnote,rotating, lscape}
\usetikzlibrary{positioning}
\usetikzlibrary{chains}
\usetikzlibrary{arrows,fit,decorations.pathreplacing}
\tikzstyle{every picture}+=[remember picture]
\tikzstyle{na} = [baseline=-.5ex]

\usepackage{bbm}

\newcommand{\ie}{\textit{i.e.}}

\numberwithin{equation}{section}

\newcommand{\nn}{\nonumber}

\newcommand{\be}{\begin{equation}} \newcommand{\ee}{\end{equation}}
\newcommand{\bea}{\begin{equation} \begin{aligned}} \newcommand{\eea}{\end{aligned} \end{equation}}

\def\tilde{\widetilde}

\def\hat{\widehat}

\def\bar{\overline}

\def\rt2{\sqrt{2}}

\def\CH{{\cal H}}

\def\CN{{\cal N}}

\def\1{{\ds 1}}

\newcommand{\cN}{\mathcal{N}}

\newcommand{\cS}{\mathcal{S}}

\newcommand{\bZ}{\mathbb{Z}}

\newcommand{\e}{{\vec e}}

\def\repa{\raise4pt\hbox{$\square$}\mkern-14mu\raise-4pt\hbox{$\square$}}
\def\repab{\overline{\raise4pt\hbox{$\square$}\mkern-14mu\raise-4pt\hbox{$\square$}\mkern-1mu}}

\def\smileface{\ensuremath{\hbox{\large$\bigcirc$}\mkern-15mu\raise-1pt\hbox{\scriptsize$\smallsmile$}%
\mkern-10mu\raise4pt\hbox{..}\mkern4mu}}
\def\frownface{\ensuremath{\hbox{\large$\bigcirc$}\mkern-15mu\raise-1pt\hbox{\scriptsize$\smallfrown$}%
\mkern-10mu\raise4pt\hbox{..}\mkern4mu}}

\def\node#1#2{\overset{#1}{\underset{#2}{{\color{gray} \bullet}}}}
\def\Node#1#2{\overset{#1}{\underset{#2}{{ \bullet}}}}
\def\sqnode#1#2{\overset{#1}{\underset{#2}{{\color{gray} \blacksquare}}}}
\def\sqNode#1#2{\overset{#1}{\underset{#2}{{ \blacksquare}}}}
\def\ver#1#2{\overset{{\llap{$\scriptstyle#1$}\displaystyle{\color{gray} \blacksquare}{\rlap{$\scriptstyle#2$}}}}{\scriptstyle\vert}}
\def\Ver#1#2{\overset{{\llap{$\scriptstyle#1$}\displaystyle\blacksquare{\rlap{$\scriptstyle#2$}}}}{\scriptstyle\vert}}

\usetikzlibrary{arrows}
\tikzstyle{every picture}+=[remember picture]
\tikzstyle{na} = [baseline=-.5ex]

\newcommand{\ba}{\begin{array}}
\newcommand{\ea}{\end{array}}
\newcommand{\bi}{\begin{itemize}}
\newcommand{\ei}{\end{itemize}}
\def\vec#1{\bm{#1}}
\def\bea#1\eea{\allowdisplaybreaks \begin{align}#1\end{align}}
 \newcommand{\ben}{\begin{enumerate}}
\newcommand{\een}{\end{enumerate}}
\newcommand{\bean}{\begin{eqnarray*}}
\newcommand{\eean}{\end{eqnarray*}}
\newcommand{\eref}[1]{(\ref{#1})}

\newcommand{\tref}[1]{Table~\ref{#1}}
\newcommand{\fref}[1]{Figure~\ref{#1}}
\newcommand{\PE}{\mathop{\rm PE}}

\newcommand{\res}{\mathop{\rm Res}}
\newcommand{\tQ}{\widetilde{Q}}

\newcommand{\BC}{\mathbb{C}}

\newcommand{\BZ}{\mathbb{Z}}

\newcommand{\comment}[1]{}

\newcommand{\fflat}{\mathcal{F}^\flat}

\newcommand{\purple}{\color{purple}}
\newcommand{\blue}{\color{blue}}

\newcommand{\ns}{{\vec n}^{\vec \sigma}}

\title{$T^{\vec \sigma}_{\vec \rho}(G)$ Theories and Their Hilbert Series}
\author[a,b]{Stefano Cremonesi,}
\author[b]{Amihay Hanany,}
\author[c]{Noppadol Mekareeya,}
\author[d,e]{\\and Alberto Zaffaroni}

\affiliation[a]{Department of Mathematics, King's College London, \\
The Strand, London WC2R 2LS, United Kingdom}
\affiliation[b]{Theoretical Physics Group, Imperial College London, \\
Prince Consort Road, London, SW7 2AZ, UK}
\affiliation[c]{Theory Division, Physics Department, CERN, \\CH-1211, Geneva 23, Switzerland}
\affiliation[d]{Dipartimento di Fisica, Universit\`a di Milano-Bicocca,  \\ I-20126 Milano, Italy}
\affiliation[e]{INFN, sezione di Milano-Bicocca, I-20126 Milano, Italy}

\emailAdd{s.cremonesi@imperial.ac.uk}
\emailAdd{a.hanany@imperial.ac.uk}
\emailAdd{noppadol.mekareeya@cern.ch}
\emailAdd{alberto.zaffaroni@mib.infn.it}

\preprint{
{\small
\begin{flushright}
KCL-MTH-14-17\\
IMPERIAL-TP-14-AH-09\\
CERN-PH-TH-2014-135
\end{flushright}
}
}

\abstract{ We give an explicit formula for the Higgs and Coulomb branch Hilbert series for the class of $3d$ ${\cal N}=4$ superconformal gauge theories   $T^{\vec\sigma}_{\vec \rho}(G)$ corresponding to a set of D3 branes ending on NS5 and D5-branes, with or without O3 planes.  
Here $G$ is a classical group, $\vec \sigma$ is a partition of $G$ and   $\vec\rho$ a partition of the dual group $G^\vee$.  In deriving such a formula we make use of the recently discovered formula for the Hilbert series of the quantum Coulomb branch of ${\cal N}=4$ superconformal theories. The result can be expressed in terms of a generalization of a class of symmetric functions, the Hall-Littlewood polynomials, and can be interpreted in mathematical language in terms of localization. We mainly consider the case $G=SU(N)$ but some interesting results are also given for orthogonal and symplectic groups.
}

\begin{document}
\maketitle

\section{Introduction}

An efficient way of encoding the information on the chiral ring of a supersymmetric theory is given by the Hilbert series of the moduli space of supersymmetric vacua, which is the generating function for the  gauge invariant chiral operators. There has been recent progress in the analysis of the chiral ring and moduli space of an $\CN = 4$ superconformal gauge theory in $2+1$ dimensions.   We can now compute the Hilbert series for both the Higgs and Coulomb branch and use it to test dualities, most 
notably mirror symmetry  \cite{Intriligator:1996ex}. The Hilbert series for the Higgs branch, which is classical,  can be computed in a conventional way from the Lagrangian  using Molien-Weyl integrals. The Coulomb branch is more complicated, but in spite of the complex structure of the chiral ring and the quantum corrections, it is still possible to write the Hilbert series  by counting monopole operators dressed with scalar fields \cite{Cremonesi:2013lqa}.  We  refer to the Hilbert series for the Coulomb branch also as {\it monopole formula} \cite{Cremonesi:2013lqa}. The formula can be applied to any  $3d$ $\cN=4$ supersymmetric gauge theory that has a Lagrangian description and that is ``good'' or ``ugly'' in the sense of \cite{Gaiotto:2008ak}.\footnote{See also \cite{Razamat:2014pta}, where the monopole formula is recovered as a limit of the superconformal index.}

In this  paper we discuss the general properties of the Hilbert series for  the  three-dimensional superconformal field theories  known as $T^{\vec\sigma}_{\vec \rho}(G)$ \cite{Gaiotto:2008ak}. These are  linear quiver theories  associated with a partition $\vec \sigma$ of $G$ and a partition $\vec\rho$ of the GNO (or Langlands) dual group $G^\vee$ \cite{Goddard:1976qe}. They are defined in terms of  a general set of D3 branes ending on NS5 and D5-branes \cite{Hanany:1996ie}, possibly in the presence of O3 planes \cite{Feng:2000eq, Gaiotto:2008ak}. By construction, the mirror of $T^{\vec \sigma}_{\vec \rho}(G)$ is $T_{\vec \sigma}^{\vec \rho}(G^\vee)$.
These theories  serve as basic building blocks for constructing a large class of  more complicated theories.
 In \cite{Cremonesi:2014kwa,Cremonesi:2014vla} we already analyzed the special case
of the  Coulomb branch  of $T_{\vec \rho}(G)$, corresponding to $\vec\sigma=(1,\cdots,1)$, and we proposed a general formula for the Coulomb branch Hilbert series in terms of Hall-Littlewood polynomials.
In this paper we define generalized Hall-Littlewood functions that give a general expression for the Coulomb branch Hilbert series of $T^{\vec\sigma}_{\vec \rho}(G)$, or equivalently the Higgs branch Hilbert series of $T_{\vec\sigma}^{\vec \rho}(G^\vee)$, with background charges. The relevant formulae are \eref{mainHS} and \eref{mainHSG}.

The Hall-Littlewood polynomials are a class of symmetric function that have appeared in related context in both the mathematical and physical literature. In physics, they appeared as blocks
in the computation of a limit of the superconformal index  \cite{Romelsberger:2005eg,Kinney:2005ej} for class $\cS$ theories in four dimensions \cite{Gadde:2011uv}. The relation with our results for $T_{\vec \rho}(G)$ theories can be seen after compactification and mirror symmetry and was discussed in details in  \cite{Cremonesi:2014vla}.  
In mathematics, the Hall-Littlewood polynomials have appeared as characters of the cotangent bundle of flag varieties, which can be  computed by localization.%
\footnote{We thank Yuji Tachikawa and all the contributors to the  
MathOverflow discussion \cite{MathOverflow}. We kindly acknowledge the note \cite{Haiman:note} taken from a lecture given by Mark Haiman and linked therein.}
 The relation with our work comes from the fact that the moduli space of $T^{\vec \sigma}_{\vec \rho}(G)$ can be expressed in terms of nilpotent orbits of $G^\vee$ which have the  cotangent bundles of flag varieties as smooth resolutions.  We took inspiration from these mathematical results to derive our formula for the Hilbert series of  $T^{\vec \sigma}_{\vec \rho}(G)$. A similar approach has been successfully applied to the computation of the Hilbert series of instanton moduli spaces \cite{Nekrasov:2002qd} or the Hilbert series of non-compact Calabi-Yau's \cite{Martelli:2006yb}. 

In this paper we mainly focus on the case $G=SU(N)$ where all derivations are neat and we can make very general statements. The case of other classical groups, 
where there are complications with the algebraic description of nilpotent orbits and issues with discrete groups, is briefly discussed at the end of the paper. A regular and  interesting pattern seems to emerge also in other types, but we leave the general analysis for future work.

The paper is organized as follows. In section \ref{sec:quivers} we present  the quivers for the $T^{\vec \sigma}_{\vec \rho}(G)$ theory for a general classical group and generic  partitions $\vec\sigma$ and $\vec\rho$. We have been  able to find in the literature only particular examples and we state here the general result which follows from the constructions in \cite{Feng:2000eq,Gaiotto:2008ak}.  In section \ref{sec:gen} we
 discuss the brane construction of the $T^{\vec \sigma}_{\vec \rho}(SU(N))$ theories and we review the general expressions for the Coulomb branch Hilbert series (based on the monopole formula   \cite{Cremonesi:2013lqa}) and the Higgs branch Hilbert series (based on the Molien-Weyl formula). We allow for background 
magnetic charges for flavor symmetries in the Coulomb branch \cite{Kapustin:2011jm, Cremonesi:2014kwa} and baryonic charges in the Higgs branch \cite{Butti:2006au,Forcella:2007wk,Butti:2007jv}. The two set of parameters are exchanged by mirror symmetry and we provide a precise map in section \ref{mapping}. Section \ref{sec:Hilbert} contains the derivation of our main formula \eref{mainHS} for unitary groups. We first derive the baryonic Higgs branch Hilbert series of $T_{\vec \sigma}(SU(N))$ by a direct evaluation of the the Molien-Weyl integral. We complement the result with a second derivation based on a localization formula for the character
of the Higgs branch moduli space, which can be interpreted as a nilpotent orbit of $SU(N)$. The localization formula apply to the standard resolution of the
orbit as a cotangent bundle of a flag variety and can be  expressed  in terms of generalized Hall-Littlewood functions. In section \ref{sec:residue} we show that  the Higgs branch Hilbert series of $T^{\vec\rho}_{\vec \sigma}(SU(N))$ can be  obtained from that of the theory $T_{\vec \sigma}(SU(N))$ by taking residues with respect to the flavor fugacities. A mirror statement holds for the Coulomb branch: the  Coulomb branch Hilbert series of $T^{\vec \sigma}_{\vec \rho}(SU(N))$ can be obtained from that of $T^{\vec \sigma}(SU(N))$ by taking residues with respect to the fugacities for the topological symmetry. Section \ref{SUN} contains several explicit examples for unitary groups. Finally, section \ref{OSp} contains the generalization of our results to orthogonal and symplectic groups. After reviewing some general facts about partitions and orbit resolutions for orthogonal and symplectic groups,  we present  a generalised Hall-Littlewood formula \eref{mainHSG}  for a generic group $G$. We present a series of  results for $USp(4)$ and $SO(5)$ and discuss in details subtleties related to the choice of $SO/O$ gauge groups in the quiver. Other useful results, including explicit examples for other groups of low rank, are given in a series of appendices.

\section{Quiver diagrams for $T^{\vec \sigma}_{\vec \rho}(G)$ with $G$ a classical group}\label{sec:quivers}

The theories $T^{\vec \sigma}_{\vec \rho}(G)$  are a class of $3d$ ${\cal N}=4$ superconformal field theories arising as infrared limits of linear quivers with unitary or alternating orthogonal-symplectic gauge groups  \cite{Gaiotto:2008ak}.  $G$ is a classical group and   $\vec\sigma$  and  $\vec\rho$  are partitions of $G$   and  $G^\vee$, respectively, as defined below.  The  theories $T^{\vec \sigma}_{\vec \rho}(G)$ can be defined  in terms of configurations of D3 branes suspended between NS and D5-branes \cite{Hanany:1996ie}, possibly in the presence of an orientifold O3 plane \cite{Gaiotto:2008ak, Feng:2000eq}. $G$ is determined by the type of orientifold and the two partitions  $\vec\sigma$ and $\vec\rho$ specify how the D3 branes end on the D5-branes and the NS5-branes, respectively. An example for $G=SU(N)$ is depicted in figure \ref{fig:BraneTs322r2221a}. By construction, the mirror of $T^{\vec \sigma}_{\vec \rho}(G)$ is $T_{\vec \sigma}^{\vec \rho}(G^\vee)$.%
\footnote{More precisely, the Lie groups $G$ and $G^\vee$ should be replaced by their Lie algebras. Note also that the exotic case dubbed $G=USp'(2N)$ is self-dual: $G^\vee=G$.} 
 The quiver for  $T^{\vec \sigma}_{\vec \rho}(G)$ can be extracted from the brane construction using standard brane moves \cite{Hanany:1996ie} and paying attention to the presence of orientifolds  \cite{Feng:2000eq,Gaiotto:2008ak}.  
 
We could not find in the literature the quiver for the $T^{\vec \sigma}_{\vec \rho}(G)$ theory for a general classical group and generic  partitions $\vec\sigma$ and $\vec\rho$ and we present it here. We also discuss the $T^{\vec \sigma}_{\vec \rho}(USp'(2N))$ theories.  Note that the quivers for $T_{\vec \rho}(SO(N))$ have been explicitly written in \cite{Benini:2010uu}.   

Partitions of $G$ are defined as follows. A partition of $G=SU(N)$ is a non-increasing sequence of integer numbers (parts) corresponding to a partition of  $N$. Partitions for other classical groups are required to satisfy some constraints. A partition of $G = SO(N)$ is a  partition  of $N$ where any even part appears an even number of times. The partition  is called a $B$- or a $D$-partition if $N$ is odd or even, respectively. A partition of $G = USp(2N)$ is a partition  of $2N$ where any odd part  appears an even number of times. Such a partition is called a $C$-partition. With these definitions, partitions are in one-to-one correspondence with the nilpotent orbits of the group $G$ and also with the homomorphisms ${\rm Lie}(SU(2))\rightarrow {\rm Lie}(G)$ \cite{collingwood1993nilpotent}. The interpretation of these constraints in terms of D3 branes ending on D5-branes
in the presence of an O3 plane is given in  \cite{Gaiotto:2008ak, Feng:2000eq}.

\subsection{$T^{\vec \sigma}_{\vec \rho}(SU(N))$}

Let $\vec \rho= (\rho_1, \ldots, \rho_{\ell'})$ and $\vec \sigma= (\sigma_1, \ldots, \sigma_\ell)$ be two 
partitions of $N$: 
\bea\label{partitionsSU}
 \sigma_1 \geq \ldots \geq  \sigma_\ell > 0~, \qquad  \rho_1 \geq \ldots \geq  \rho_{\ell'} > 0~, \qquad \sum_{i=1}^\ell \sigma_i = \sum_{i=1}^{\ell'} \rho_i =N~.  
\eea
The quiver diagram for $T^{\vec \sigma}_{\vec \rho}(SU(N))$ is depicted in \eref{quivTsigrhoSUN}, where, according to standard notations,  round nodes denote gauge groups and square nodes flavor groups. The label $k$ at each node denotes a $U(k)$ group and $\ell'$ is the length of the partition $\vec \rho$.
\bea 
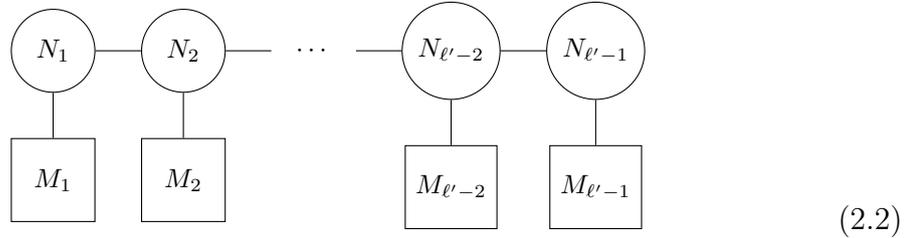
\label{quivTsigrhoSUN}
\begin{tikzpicture}[font=\footnotesize]
\begin{scope}[auto,%
  every node/.style={draw, minimum size=1.1cm}, node distance=0.6cm];
\node[circle] (UN1) at (0, 0) {$N_1$};
\node[circle, right=of UN1] (UN2) {$N_2$};
\node[draw=none, right=of UN2] (dots) {$\cdots$};
\node[circle, right=of dots] (UNlm1) {$N_{\ell'-2}$};
\node[circle, right=of UNlm1] (UNl) {$N_{\ell'-1}$};
\node[rectangle, below=of UN1] (UM1) {$M_1$};
\node[rectangle, below=of UN2] (UM2) {$M_2$};
\node[rectangle, below=of UNlm1] (UMlm1) {$M_{\ell'-2}$};
\node[rectangle, below=of UNl] (UMl) {$M_{\ell'-1}$};
\end{scope}
\draw (UN1) -- (UN2)
(UN2)--(dots)
(dots)--(UNlm1)
(UNlm1)--(UNl)
(UN1)--(UM1)
(UN2)--(UM2)
(UNlm1)--(UMlm1)
(UNl)--(UMl);
\end{tikzpicture}
\eea
The flavor symmetries $U(M_j)$, with $1\leq j \leq \ell'-1$, are determined from the transpose $\vec \sigma^T=(\hat \sigma_1, \ldots, \hat \sigma_{\hat \ell})$, with $\hat \sigma_1 \geq \ldots \geq  \hat \sigma_{\hat \ell} > 0$, of $\vec \sigma$ as follows:
\bea \label{flvgroups}
M_{j} &= 
\hat \sigma_{j} -\hat \sigma_{j+1} , \qquad \text{with}  \\
\hat \sigma_i &=0, \quad \text{for all}~ i \geq \hat \ell+1~.
\eea
Notice that $M_i=0$ for $i\geq \hat\ell +1$ so that  there are at most $\hat\ell$ flavor groups. 
The gauge symmetries $U(N_j)$, with $1\leq j \leq \ell'-1$, are given by
\bea
N_j &= \sum_{k=j+1}^{\ell'} \rho_k  - \sum_{i=j+1}^{\hat{\ell}} \hat \sigma_i~.
\eea
Notice that the theories $T^{\vec \sigma}_{\vec \rho}(SU(N))$ are defined only for $\vec\sigma^T \, < \, \vec\rho$. 
The quiver for $T^{\vec \sigma}_{\vec \rho}(SU(N))$ has first appeared in \cite{nakajimaquiver}.  Various properties of these theories have been studied recently using holography and three-sphere partition functions \cite{Nishioka:2011dq,Assel:2011xz,Assel:2012cp,Dey:2014tka,Assel:2014awa}.


\subsection{$T^{\vec \sigma}_{\vec \rho}(SO(2N))$}

These theories can be realised on the worldvolume of $N$ D3 branes parallel to an orientifold ${O3}^-$ plane and ending on systems of half D5 branes and of half NS5 branes. The partitions ${\vec \sigma}$ and ${\vec \rho}$ determine how the D3 branes end on the half D5 branes and on the half NS5 branes respectively. In this case both $\vec \sigma$ and $\vec \rho$ are $D$-partitions of $SO(2N)$, of lengths $\ell$ and $\ell'$.

The quiver diagram for $T^{\vec \sigma}_{\vec \rho}(SO(2N))$ consists of alternating $(S)O/USp$ groups depicted in \eref{quivTDsigrho}, where each grey nodes with a label $N$ denotes an $O(N)$ or $SO(N)$ group and each black node with a label $N$ denotes a $USp(N)$ group.
{\large
\bea \label{quivTDsigrho}
\Node{\ver{}{M_1}}{N_1}-\node{\Ver{}{M_2}}{N_2}-\cdots-\node{\Ver{}{M_{L-1}}}{N_{L-1}}-\Node{\ver{}{M_{L}}}{N_{L}}
\eea}
The length of the quiver \eref{quivTDsigrho} is $L= 2 \lfloor \ell'/2 \rfloor-1$, unless $N_L=M_L=0$, in which case we remove such nodes from the quiver and the length reduces to $L-1$.

The labels $M_j$, with $1\leq j \leq L$, for the flavor symmetries are determined by $\vec \sigma^T$ as in \eref{flvgroups}.   On the other hand, the labels $N_j$, with $1\leq j \leq L$, for the gauge symmetries are given by
\bea
N_j &= \left[ \sum_{k=j+1}^{\ell'} \rho_k  \right]_{+,-} - \left( \sum_{i=j+1}^{\hat{\ell}} \hat \sigma_i \right)~, \qquad \text{$+$ for $O/SO$ and $-$ for $USp$}~,
\eea
where $[n]_{+ (\text{resp.}\; -)}$ denotes the smallest (resp. largest) even integer $\geq n$ (resp. $\leq n$).

\subsection{$T^{\vec \sigma}_{\vec \rho}(SO(2N+1))$}

These theories can be realised on the worldvolume of $N$ D3 branes parallel to an orientifold $\widetilde{O3}^-$ plane and ending on systems of half D5 branes and of half NS5 branes. The partitions ${\vec \sigma}$ and ${\vec \rho}$ determine how the D3 branes end on the half D5 branes and on the half NS5 branes respectively: $\vec \sigma$ is a $B$-partition of $SO(2N+1)$, of length $\ell$,  and $\vec \rho$ is a $C$-partition of $USp(2N)$, of length $\ell'$.

The quiver diagram for $T^{\vec \sigma}_{\vec \rho}(SO(2N+1))$ consists of alternating $(S)O/USp$ groups depicted in \eref{quivTBsigrho}, where each grey nodes with a label $N$ denotes an $O(N)$ or $SO(N)$ group and each black node with a label $N$ denotes a $USp(N)$ group.
{\large
\bea \label{quivTBsigrho}
\Node{\ver{}{M_1}}{N_1}-\node{\Ver{}{M_2}}{N_2}-\cdots-\node{\Ver{}{M_{L-1}}}{N_{L-1}}-\Node{\ver{}{M_{L}}}{N_{L}}
\eea}
where the length of the quiver is given by $L= 2 \lfloor \ell'/2 \rfloor$.

The labels $M_j$, with $1\leq j \leq L$, for the flavor symmetries are determined by $\vec \sigma^T$ as in \eref{flvgroups}.   On the other hand, the labels $N_j$, with $1\leq j \leq L$, for the gauge symmetries are given by
\bea
N_j &= \left[1+ \sum_{k=j+1}^{\ell'} \rho_k  \right]_{\tilde{+},-} - \left( \sum_{i=j+1}^{\hat{\ell}} \hat \sigma_i \right)~, \qquad \text{$\tilde{+}$ for $O/SO$ and $-$ for $USp$}~,
\eea
where $[n]_{\tilde{+}}$ is the smallest odd integer $\geq n$.

Here and in the following, the distinction between $SO(N)$ and $O(N)$ groups is important. Theories with $SO(N)$ gauge groups  have typically more BPS gauge invariant operators compared with the same theory with gauge group $O(N)$ and  we have different theories according to the choice of $O/SO$ factors.
Examples are discussed in section \ref{OSp}.

\subsection{$T^{\vec \sigma}_{\vec \rho}(USp(2N))$}

These theories can be realised on the worldvolume of $N$ D3 branes parallel to an orientifold ${O3}^+$ plane and ending on systems of half D5 branes and of half NS5 branes. The partitions ${\vec \sigma}$ and ${\vec \rho}$ determine how the D3 branes end on the half D5 branes and on the half NS5 branes respectively: $\vec \sigma$ is a $C$-partition of $USp(2N)$, of length $\ell$, and $\vec \rho$ a $B$-partition of $SO(2N+1)$, of length $\ell'$.

The quiver diagram for $T^{\vec \sigma}_{\vec \rho}(USp(2N))$ consists of alternating $(S)O/USp$ groups depicted in \eref{quivTCsigrho}, where each grey nodes with a label $N$ denotes an $O(N)$ or $SO(N)$ group and each black node with a label $N$ denotes a $USp(N)$ group.
{\large
\bea \label{quivTCsigrho}
\node{\Ver{}{M_1}}{N_1}-\Node{\ver{}{M_2}}{N_2}-\cdots-\node{\Ver{}{M_{L}}}{N_{L}}
\eea}
where $L=2\lfloor \ell'/2 \rfloor$ and if $N_L$ and $M_L$ are both zero, we remove such nodes from the quiver, in which case the length of quiver \eref{quivTCsigrho} is $L-1$.

The labels $M_j$, with $1\leq j \leq L$, for the flavor symmetries are determined by $\vec \sigma^T$ as in \eref{flvgroups}.   On the other hand, the labels $N_j$, with $1\leq j \leq L$, for the gauge symmetries are given by
\bea
N_j &= \left[ \sum_{k=j+1}^{\ell'} \rho_k  \right]_{+,-} - \left( \sum_{i=j+1}^{\hat{\ell}} \hat \sigma_i \right)~, \qquad \text{$+$ for $O/SO$ and $-$ for $USp$}~.
\eea

\subsection{$T^{\vec \sigma}_{\vec \rho}(USp'(2N))$} \label{sec:TsigrhoUSpP}

These theories can be realised on the worldvolume of $N$ D3 branes parallel to an orientifold $\widetilde{O3}^+$ plane and ending on systems of half D5 branes and of half NS5 branes. The partitions ${\vec \sigma}$ and ${\vec \rho}$ determine how the D3 branes end on the half D5 branes and on the half NS5 branes respectively. In this case both $\vec \sigma$ and $\vec \rho$ are $C$-partitions of $USp(2N)$, of lengths $\ell$ and $\ell'$ respectively.

The quiver diagram for $T^{\vec \sigma}_{\vec \rho}(USp'(2N))$ consists of alternating $(S)O/USp$ groups depicted in \eref{quivTCsigrhoP}, where each grey nodes with a label $N$ denotes an $O(N)$ or $SO(N)$ group and each black node with a label $N$ denotes a $USp(N)$ group.
{\large
\bea \label{quivTCsigrhoP}
\node{\Ver{}{M_1}}{N_1}-\Node{\ver{}{M_2}}{N_2}-\cdots-\node{\Ver{}{M_{L}}}{N_{L}}
\eea}
We defined 
\bea
L= \begin{cases} \ell' -1  & \quad \text{$\ell'$ is even} \\
\ell'  & \quad \text{$\ell'$ is odd} \end{cases} 
\eea
and if both $N_L$ and $M_L$ are zero, the nodes are removed from the quiver and the length of the quiver \eref{quivTCsigrhoP} is $L-1$.

The labels $M_j$, with $1\leq j \leq L$, for the flavor symmetries are determined by $\vec \sigma^T$ as in \eref{flvgroups}.   On the other hand, the labels $N_j$, with $1\leq j \leq L$, for the gauge symmetries are given by
{\small
\bea
N_j = \begin{cases} 
\left[1+ \sum_{k=j+1}^{\ell'} \rho_k  \right]_{\tilde{+}} - \left( \sum_{i=j+1}^{\hat{\ell}} \hat \sigma_i \right) \quad \text{for the $O/SO$ node}~, & \text{if $\ell'$ is even}~, \\
 \left[\sum_{k=j+1}^{\ell'} \rho_k  \right]_{-} - \left( \sum_{i=j+1}^{\hat{\ell}} \hat \sigma_i \right) \quad \text{for the $USp$ node}~, & \text{if $\ell'$ is even}~, \\ 
\left[ \sum_{k=j+1}^{\ell'} \rho_k  \right]_{\tilde{+}} - \left( \sum_{i=j+1}^{\hat{\ell}} \hat \sigma_i \right) \quad \text{for the $O/SO$ node}~, & \text{if $\ell'$ is odd}~,\\ 
\left[ \sum_{k=j+1}^{\ell'} \rho_k  \right]_{+} - \left( \sum_{i=j+1}^{\hat{\ell}} \hat \sigma_i \right) \quad \text{for the $USp$ node}~, & \text{if $\ell'$ is odd}~.
\end{cases}
\eea}

\section{ The Hilbert series of  $T^{\vec \sigma}_{\vec \rho}(SU(N))$}\label{sec:gen}

In this section we state the general formulae for the Hilbert series of the Coulomb and Higgs branch of $T^{\vec \sigma}_{\vec \rho}(SU(N))$ theories. 
The Hilbert series for the Coulomb branch can be written using the monopole formula \cite{Cremonesi:2013lqa},
while the Hilbert series for the Higgs branch can be written as a Molien-Weyl integral. We introduce background magnetic fluxes for the flavor symmetries in the Coulomb branch \cite{Cremonesi:2014kwa} and baryonic charges in the Higgs branch \cite{Butti:2006au,Forcella:2007wk,Butti:2007jv}.
We explain how fugacities, fluxes and charges are related by mirror symmetry.  
We also provide a useful brane description of the theory.

\subsection{Brane configurations}

\begin{figure}[H]
\centering
\begin{tikzpicture} [scale=0.8, transform shape]
\draw (0,0)--(0,2.5) node[black,midway, xshift =0cm, yshift=1.5cm]{\footnotesize $x_1$};
\draw (1,0)--(1,2.5) node[black,midway, xshift =0cm, yshift=1.5cm]{\footnotesize $x_2$}; 
\draw (2,0)--(2,2.5) node[black,midway, xshift =-0.5cm, yshift=-1.7cm] {\footnotesize NS5} node[black,midway, xshift =0cm, yshift=1.5cm]{\footnotesize $x_3$};
\draw (3,0)--(3,2.5) node[black,midway, xshift =0cm, yshift=1.5cm]{\footnotesize $x_4$}; 
\draw [dashed,red] (-3,0)--(-3,2.5) node[black,midway, xshift =0cm, yshift=1.5cm] {\footnotesize $n_{1}$};
\draw [dashed,blue] (-4,0)--(-4,2.5) node[black,midway, xshift =0cm, yshift=-1.7cm] {\footnotesize D5} node[black,midway, xshift =0cm, yshift=1.5cm] {\footnotesize $n_{2}$};
\draw [dashed,purple] (-5,0)--(-5,2.5)  node[black,midway, xshift =0cm, yshift=1.5cm] {\footnotesize $n_{3}$};
\draw [red] (0,2)--(-3,2) node[black,midway, yshift=0.4cm] {\tiny D3}; \draw [red] (1,1.8)--(-3,1.8); \draw [red](2,1.6)--(-3,1.6);
\draw [blue] (0,1)--(-4,1); \draw [blue](1,0.8)--(-4,0.8);
\draw [purple](0,0.4)--(-5,0.4); \draw [purple](1,0.2)--(-5,0.2);
\draw (0,0.6)--(1,0.6); \draw (1,1.1)--(2,1.1); \draw (1,1.3)--(2,1.3); \draw (2,1.4)--(3,1.4); 
\end{tikzpicture}  
\caption{A brane construction for $T^{(3,2,2)}_{(2,2,2,1)}(SU(7))$. The partition $\vec \sigma = (3,2,2)$ gives the net number of D3 branes ending on each D5-brane from the interior to the exterior. The partition $\vec \rho = (2,2,2,1)$ gives the net number of D3 branes ending on each NS5-brane from the interior to the exterior. Here $x_i$ are the fugacities associated with each NS5 brane and $n_{j}$ are the background monopole charges associated with each D5-brane.}
\label{fig:BraneTs322r2221a}
\end{figure}
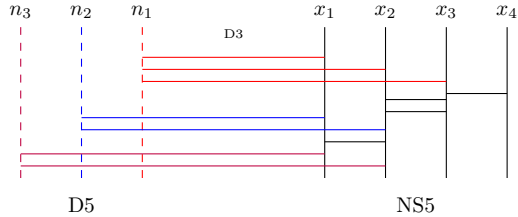

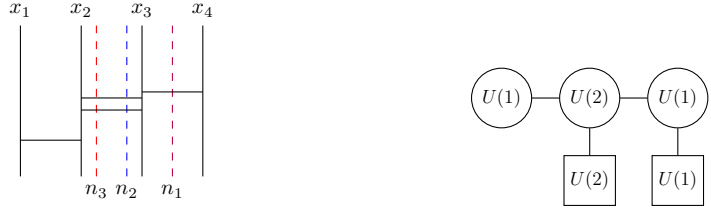
\begin{figure}[H]
\centering
\begin{tikzpicture} [scale=0.8, transform shape]
\draw (0,0)--(0,2.5) node[black,midway, xshift =0cm, yshift=1.5cm] {\footnotesize $x_{1}$};
\draw (1,0)--(1,2.5) node[black,midway, xshift =0cm, yshift=1.5cm] {\footnotesize $x_{2}$}; 
\draw (2,0)--(2,2.5) node[black,midway, xshift =0cm, yshift=1.5cm] {\footnotesize $x_{3}$};
\draw (3,0)--(3,2.5) node[black,midway, xshift =0cm, yshift=1.5cm] {\footnotesize $x_{4}$};
\draw [dashed,red] (1.25,0)--(1.25,2.5) node[black,midway, xshift =0cm, yshift=-1.5cm] {\footnotesize $n_{3}$};
\draw [dashed,blue] (1.75,0)--(1.75,2.5) node[black,midway, xshift =0cm, yshift=-1.5cm] {\footnotesize $n_{2}$};
\draw [dashed,purple] (2.5,0)--(2.5,2.5) node[black,midway, xshift =0cm, yshift=-1.5cm] {\footnotesize $n_{1}$};
\draw (0,0.6)--(1,0.6); \draw (1,1.1)--(2,1.1); \draw (1,1.3)--(2,1.3); \draw (2,1.4)--(3,1.4); 
\end{tikzpicture}  
\hspace{3cm}
\begin{tikzpicture}[scale=0.6, transform shape]
\begin{scope}[auto,%
  every node/.style={draw, minimum size=1.1cm}, node distance=0.6cm];
\node[circle] (UN1) at (0, 0) {$U(1)$};
\node[circle, right=of UN1] (UN2) {$U(2)$};
\node[circle, right=of UN2] (UN3) {$U(1)$};
\node[rectangle, below=of UN2] (UM2) {$U(2)$};
\node[rectangle, below=of UN3] (UM3) {$U(1)$};
\end{scope}
\draw (UN1) -- (UN2)
(UN2)--(UN3)
(UN2)--(UM2)
(UN3)--(UM3);
\end{tikzpicture}
\caption{Left: brane construction for $T^{(3,2,2)}_{(2,2,2,1)}(SU(7))$ after the D5-branes are moved inside the NS5-brane intervals. Right: the linear quiver is read off from the brane configuration.  
We adopt the convention that the $i$-th gauge group corresponds to the D3-brane interval between $x_i$ and $x_{i+1}$: hence $U(1)$, $U(2)$, $U(1)$ from left to right are regarded as the first, second and third gauge groups respectively, and similarly $U(2)$ and $U(1)$ are regarded as the second and third flavor groups respectively. }
\label{fig:BraneTs322r2221b}
\end{figure}

The theory $T^{\vec \sigma}_{\vec \rho}(SU(N))$ can be realized with $N$ D3 branes suspended between $\ell'$ NS5-branes and $\ell$ D5-branes, where $\ell'$ and $\ell$ are the length of the partitions $\vec\rho$ and $\vec\sigma$ respectively.
We order the branes in such a way that all the D5-branes are on one side of the NS5 branes (on the left in figure \ref{fig:BraneTs322r2221a}).  The parts of $\vec\rho= (\rho_1, \ldots, \rho_{\ell'})$ are the net number of D3 branes ending on the NS5 branes going from the interior to the exterior of the configuration and the parts $\vec \sigma= (\sigma_1, \ldots, \sigma_\ell)$ are the net number of D3 branes ending on the D5-branes again going from  the interior to the exterior. Since the 
partitions are ordered as in \eref{partitionsSU}, the smallest part of $\vec\rho$ and $\vec\sigma$ are associated with the most external NS5 and D5-branes, respectively, and they increase going into the interior.  The configuration must satisfy the s-rule requiring that no more than one D3 brane can end on the same pair of NS5 and D5-branes \cite{Hanany:1996ie}. The quiver can be read after splitting the D3 branes among the NS5 branes and moving the D5-branes inside the NS5 intervals as in the example in  \fref{fig:BraneTs322r2221a} and  \fref{fig:BraneTs322r2221b}. The result is the quiver  in \eref{quivTsigrhoSUN}.

Unless otherwise stated we always use the following convention in reading the quiver from the brane configuration. When talking about order we always refer
to the brane configuration where all the D5 are on one side of the NS5, as in figure \ref{fig:BraneTs322r2221a}. The gauge groups are numbered  by following the NS5 intervals from the interior to the exterior of the configuration, or, equivalently, in the direction which goes from the D5 to the NS5.
The first gauge group $U(N_1)$ in \eref{quivTsigrhoSUN} is the one living on the NS5 interval closer to the D5-branes. 

\begin{figure}[H]
\centering
\begin{tikzpicture} 
\draw[dashed,red ] (0,0)--(0,2.5) node[black,midway, xshift =0cm, yshift=1.5cm]{\footnotesize $x_1$};
\draw[dashed, blue] (0.5,0)--(0.5,2.5) node[black,midway, xshift =0cm, yshift=1.5cm]{\footnotesize $x_2$}; 
\draw[dashed,purple] (1,0)--(1,2.5) node[black,midway, xshift =-0.5cm, yshift=-1.7cm] {\footnotesize D5} node[black,midway, xshift =0cm, yshift=1.5cm]{\footnotesize $x_3$};
\draw[dashed, brown] (1.5,0)--(1.5,2.5) node[black,midway, xshift =0cm, yshift=1.5cm]{\footnotesize $x_4$}; 
\draw  (-3,0)--(-3,2.5) node[black,midway, xshift =0cm, yshift=1.5cm] {\footnotesize $n_{1}$};
\draw (-4,0)--(-4,2.5) node[black,midway, xshift =0cm, yshift=-1.7cm] {\footnotesize NS5} node[black,midway, xshift =0cm, yshift=1.5cm] {\footnotesize $n_{2}$};
\draw  (-5,0)--(-5,2.5)  node[black,midway, xshift =0cm, yshift=1.5cm] {\footnotesize $n_{3}$};
\draw [red] (0,1.6)--(-3,1.6) node[black,midway, yshift=0.6cm] {\tiny D3}; \draw [red] (0,1.8)--(-4,1.8);
\draw [blue](0.5,0.8)--(-3,0.8); \draw [blue] (0.5,1)--(-4,1); 
\draw [purple](1,0.5)--(-4,0.5); \draw [purple](1,0.3)--(-3,0.3);
\draw [brown] (1.5,0.1)--(-3,0.1);
\draw (-3,1.5)--(-4,1.5);\draw (-4,1.3)--(-5,1.3); \draw (-4,1.1)--(-5,1.1);
\end{tikzpicture}  
\\~\\
\begin{tikzpicture} 
\draw[dashed,red] (-4.75,0)--(-4.75,2.5) node[black,midway, xshift =0cm, yshift=-1.5cm]{\scriptsize $x_1$};
\draw[dashed, blue] (-4.5,0)--(-4.5,2.5) node[black,midway, xshift =0cm, yshift=-1.5cm]{\scriptsize $x_2$}; 
\draw[dashed,purple] (-4.2,0)--(-4.2,2.5) node[black,midway, xshift =0cm, yshift=-1.5cm]{\scriptsize $x_3$}; 
\draw[dashed, brown] (-3.5,0)--(-3.5,2.5) node[black,midway, xshift =0cm, yshift=-1.5cm]{\footnotesize $x_4$}; 
\draw  (-3,0)--(-3,2.5) node[black,midway, xshift =0cm, yshift=1.5cm] {\footnotesize $n_{1}$};
\draw (-4,0)--(-4,2.5) node[black,midway, xshift =0cm, yshift=1.5cm] {\footnotesize $n_{2}$};
\draw  (-5,0)--(-5,2.5)  node[black,midway, xshift =0cm, yshift=1.5cm] {\footnotesize $n_{3}$};
\draw (-3,1.5)--(-4,1.5);\draw (-4,1.3)--(-5,1.3); \draw (-4,1.1)--(-5,1.1);
\end{tikzpicture}  
\hspace{3cm}
\begin{tikzpicture}[scale=0.7, transform shape]
\begin{scope}[auto,%
  every node/.style={draw, minimum size=1.1cm}, node distance=0.6cm];
\node[circle] (UN1) at (0, 0) {$U(1)$};
\node[circle, right=of UN1] (UN2) {$U(2)$};
\node[rectangle, below=of UN1] (UM1) {$U(1)$};
\node[rectangle, below=of UN2] (UM2) {$U(3)$};
\end{scope}
\draw (UN1) -- (UN2)
(UN1)--(UM1)
(UN2)--(UM2);
\end{tikzpicture}
\caption{Top: brane construction for $T^{(2,2,2,1)}_{(3,2,2)}(SU(7))$, obtained by exchanging D5-branes and NS5-branes in \fref{fig:BraneTs322r2221a}. Bottom left: the D5-branes are moved inside the NS5-brane intervals. Bottom right: the quiver diagram read off from the bottom left brane configuration. 
 We adopt the convention that the $i$-th gauge group corresponds to the D3-brane interval between $n_i$ and $n_{i+1}$: hence $U(1)$ and $U(2)$ are regarded as the first and the second gauge groups respectively, and similarly $U(1)$ and $U(3)$ are regarded as the first and the second flavor groups respectively.}
\label{fig:BraneTs2221r322a}
\end{figure}

\subsection{The monopole formula for the Coulomb branch of $T^{\vec \sigma}_{\vec \rho}(SU(N))$}\label{monformula}

It is convenient to associate fugacities and fluxes   to the NS5 and D5-branes, respectively. We assign fugacities $\vec x= (x_1,\ldots, x_{\ell'}) $ to each NS5 brane and we order them from the interior to the exterior of the branes configuration as in \fref{fig:BraneTs322r2221a}. We also assign fluxes $\vec n =(n_1,\cdots n_\ell)$ to the D5-branes and we order them  again from the interior to the exterior of the branes configuration as in \fref{fig:BraneTs322r2221a}. 

The monopole formula \cite{Cremonesi:2013lqa} for quiver \eref{quivTsigrhoSUN} is given by\footnote{For convenience we rescale the fugacity $t \rightarrow t^2$ from our previous papers \cite{Cremonesi:2013lqa, Cremonesi:2014kwa, Cremonesi:2014vla}. $t$ is now a highest weight fugacity for the $SU(2)$ R-symmetry acting on the Coulomb branch.}
\bea\label{mon}
&H_{\text{mon}}[T^{\vec \sigma}_{\vec \rho}(SU(N))](t; \vec x; \vec {\tilde n_1}, \ldots, \vec {\tilde n_{\hat\ell}})  \nn \\
&= \Big( \prod_{j=1}^{L} {y_j}^{\sum_{i=1}^{M_j} \tilde n_{j,i}} \Big) \sum_{m_{1,1} \geq \ldots \geq m_{1,N_1} > -\infty} ~ \cdots~ \sum_{m_{L,1} \geq \ldots \geq m_{L,N_L} > -\infty}  \times \nn \\
& \qquad \Big( \prod_{i=1}^{L-1} t^{\sum_{k=1}^{N_i} \sum_{k'=1}^{N_{i+1}} |m_{i, k} - m_{i+1, k'}|} \Big)\Big( \prod_{i=1}^{L} t^{\sum_{k=1}^{N_i} \sum_{k'=1}^{M_{i}} |m_{i, k} - \tilde n_{i, k'}|} \Big) \times \nn \\
& \qquad \Big( \prod_{i=1}^{L} t^{-2\sum_{1\leq k < k' \leq N_i} |m_{i, k} - m_{i, k'}|} \Big) \Big( \prod_{i=1}^L P_{U(N_i)}(t; \vec  m_i) \Big) \Big( \prod_{j=1}^{L} z_j^{\sum_{k=1}^{N_j} m_{j,k}} \Big)~.
\eea
where $\vec m_j = (m_{j,1}, \ldots, m_{j, N_j})$ with $1 \leq j \leq L$ are dynamical magnetic charges associated with gauge group $U(N_j)$, $\vec {\tilde n_j} = (\tilde n_{j,1}, \ldots, \tilde n_{j, M_j})$ are background magnetic charges associated with the flavor group $U(M_j)$, and
\bea
L= \ell'-1~
\eea
is the number of gauge groups.
Here $z_j$ and $y_j$ are fugacities for the topological $U(1)$ symmetries associated with the gauge groups $U(N_j)$ and flavor groups $U(M_j)$ respectively. Since the flavor symmetry is actually $(\prod_{j=1}^L U(M_j))/U(1)$, these fugacities are not independent. Rather they satisfy the constraint
\bea \label{relzy}
\prod_{j=1}^L z_j^{N_j} y_j^{M_j}=1~,
\eea
which ensures that a shift of the magnetic charges corresponding to the removed $U(1)$ does not affect the monopole formula \eqref{mon}. We will refer to this as a shift symmetry in the following.

We now need to translate the topological fugacities $\vec z$ and $\vec y$ and the background magnetic fluxes $\vec{\tilde{n}}$ in terms of the previously defined variables $\vec x$ and $\vec n$.

The fugacities $\vec z$ and $\vec y$ are related to $\vec x$ as
\bea
z_j = x_{j+1} x_{j}^{-1}~, \quad y_j = x_1 \ldots x_j~,  \qquad 1\leq j \leq L~.
\eea
Then, \eref{relzy} translates to
\bea\label{constr}
\prod_{i=1}^{\ell'} x_i^{\rho_i} =1~.
\eea

Due to monopole operators, the topological symmetry $U(1)^{\ell'-1}$ is enhanced to 
\bea\label{symmetryC} 
S \left ( \prod_i U(\hat\rho_i-\hat\rho_{i+1})\right )\, ,
\eea
 where $\vec \rho^T=(\hat \rho_1, \ldots, \hat \rho_{\hat \ell'})$, with $\hat \rho_1 \geq \ldots \geq  \hat \rho_{\hat \ell'} > 0$ is the transpose partition of $\vec\rho$  \cite{Gaiotto:2008ak}. $\hat\rho_i-\hat\rho_{i+1}$
 is the number of parts of $\vec\rho$ equal to i.  As expected by mirror symmetry, \eref{symmetryC} is the flavor symmetry of the mirror theory $T_{\vec \sigma}^{\vec \rho}(SU(N))$. The $\vec x$ become 
 fugacities for the  non-abelian symmetry \eref{symmetryC}. We can split the $\vec x$ into $\hat\ell'$ pieces 
 \bea\label{pieces}
(\vec {\tilde x_1}, \ldots ,\vec {\tilde x_{\hat\ell'}})
 \eea
where, by definition, $ \vec {\tilde x_i}$  is the set of $x_k$ with $\rho_k=i$. The  $ \vec {\tilde x_i}$ are fugacities for the group $U(\hat\rho_i-\hat\rho_{i+1})$. The constraint \eref{constr} ensures that the overall $U(1)$  is removed in \eref{symmetryC}. Notice that the splitting \eref{pieces} reverses the order of the $x_i$. The $\vec {\tilde x_i}$ are constructed by collecting together all the fugacities  of the NS5 associated with the parts of $\vec\rho$ equal to $i$ and the index $i$ increases going in the direction which goes from the NS5 to the D5, from the exterior to the interior, while the $x_i$ are ordered in the opposite direction. 

For the flavour symmetry
\bea
S \left( \prod_i  U(M_i) \right) = S \left ( \prod_i U(\hat\sigma_i-\hat\sigma_{i+1})\right )\, ,
\eea
the background monopole fluxes $\vec {\tilde n_j}$  are related to the $\vec n =(n_1,\cdots, n_\ell)$ in a similar manner.  $\vec {\tilde n_j} $ is the set of fluxes $n_k$ with
$\sigma_k=j$.  The  $ \vec {\tilde n_i}$ are fugacities for the group $U(\hat \sigma_i-\hat \sigma_{i+1})$. Once we move the D5 inside the NS5 intervals, the fluxes in $\vec {\tilde n_j}$ are those associated with the D5-branes living in the $j$-th interval,
with the intervals ordered going from the D5 to the NS5 branes, according to our general convention.  Notice that, in this case also,   the splitting of the fluxes into the $\vec {\tilde n_i}$ pieces reverses the original order of the $n_i$.  

Let us discuss some examples. In \fref{fig:BraneTs322r2221a}, we have $\vec {\tilde{n}_{1}}=\emptyset, \vec {\tilde{n}_2}= (n_3, n_2), \vec {\tilde n_{3}}=(n_1)$, and $\vec {\tilde x_1}= (x_4),  \vec{\tilde x_2}= (x_3, x_2, x_1)$. Notice that the order of the $x_i$ and $n_i$  has been reversed.   The splitting of $n_i$, corresponding to the flavour symmetry, is manifest in \fref{fig:BraneTs322r2221b}. On the other hand, the splitting of the topological fugacities $x_i$ is not manifest in \fref{fig:BraneTs322r2221b}, but this becomes apparent in the mirror quiver depicted in Figure \ref{fig:BraneTs2221r322a}.

\subsection{The baryonic generating function for the Higgs branch of $T^{\vec \sigma}_{\vec \rho}(SU(N))$}\label{bar}

The baryonic Hilbert series for quiver \eref{quivTsigrhoSUN} is given by the Molien-Weyl integral \cite{Forcella:2007wk,Butti:2007jv} 
\bea\label{Molien-Weyl}
&g[T^{\vec \sigma}_{\vec \rho}(SU(N))](t; \vec w_1, \ldots , \vec w_{ \hat\ell}\, ; B_1, \ldots,  B_{\ell'-1})  = \int  \prod_{i=1}^L \left ( \frac{1}{N_i!}\prod_{1\leq k \leq N_i} \frac{1}{2\pi i}\frac{ds_{i,k}}{ s_{i,k}^{1+B_{i}}} \right )  \nn \\
& \prod_{i=1}^L \frac{\prod_{ k \neq p}^{N_i}  (1- s_{i,k}/s_{i,p}) \prod_{k\, , p}^{N_i} (1- t^2 s_{i,k}/s_{i,p})}{\prod_{ k=1}^{N_i} \prod_{p=1}^{N_{i+1}} \prod_{q=1}^{M_{i}} (1- t s_{i+1,p} / s_{i,k})(1- t s_{i,k} / s_{i+1,p})(1- t w_{i,q} / s_{i,k})(1- t s_{i,k} / w_{i,q})}
\eea
where $L= \ell'-1$ as before.  $\vec w_j = (w_{j,1}, \ldots, w_{j, M_j})$ with $1 \leq j \leq \hat\ell$ are fugacities for the flavor symmetry 
\bea\label{symmetryH}
S \left ( 
\prod_i U(M_i) \right) ~,
\eea
and the integration variables $s_{i,k}$ with $1 \leq k \leq N_i$ parameterise the Cartan of the gauge groups $U(N_i)$. The integration over the $U(1)$ center of each $U(N_i)$ factor selects the operators of baryonic charge $B_i$ for the leftover $SU(N_i)$ gauge groups. $1\leq i \leq L$ with the understanding that terms with occurrences of $s_{L+1,p}$ should not be included
in the product.  The numerator  in \eref{Molien-Weyl} contains the Haar measure and the contribution of the F-term relations while the denominator
receives contributions from the fundamental and bifundamental fields in the quiver.  

\subsection{Mapping of parameters under mirror symmetry}\label{mapping}

Under mirror symmetry $T^{\vec \sigma}_{\vec \rho}(SU(N))$ is exchanged with $T_{\vec \sigma}^{\vec \rho}(SU(N))$. The Coulomb branch of the former is identified with the Higgs branch of the latter and, at the level of Hilbert series, we have
\be\label{mirror}
\begin{split}
&H_{\text{mon}}[T^{\vec \sigma}_{\vec \rho}(SU(N))](t; \vec x; \vec {\tilde n_1},\cdots, \vec {\tilde n_{\hat\ell}} ) \\
&= \vec x^{\vec s(\vec n)} g[T_{\vec \sigma}^{\vec \rho}(SU(N))](t; \vec {\tilde x_1}, \ldots , \vec {\tilde x_{\hat\ell'}}; B_1, \ldots,  B_{\ell -1})\, , 
\end{split}
\ee
where the relation between fugacities and charges in the two sides of the equation can be determined by comparing global symmetries and following the brane configuration under S-duality. The result is the following.

The $\vec {\tilde x_i}$  are defined in terms of $\vec x$ as in \eref{pieces}. The  $\vec {\tilde x_i}$ with $i=1,\dots \hat\ell'$ are  fugacities for the global symmetry $S \left ( \prod_i U(\hat\rho_i-\hat\rho_{i+1})\right )$ which is the topological symmetry of $T^{\vec \sigma}_{\vec \rho}(SU(N))$ and the flavor symmetry of $T_{\vec \sigma}^{\vec \rho}(SU(N))$. The $\vec x$ are associated with the NS5-branes  in the Coulomb picture as in \fref{fig:BraneTs322r2221a} and with the D5-branes after S-duality, consistently with the identification made above.

The baryonic charges $B_i$, which can also be viewed as magnetic charges for the topological symmetry, are instead given by
\bea\label{bar}
B_i = n_{i}- n_{i+1}\, ,
\eea
where the $n_i$ and  $\vec {\tilde n_i}$ are related as discussed at the end of section \ref{monformula}.
Recall that the $n_i$ are associated with the D5-branes and ordered in the direction which goes from the NS5-branes to the D5. After an S-duality the $n_i$ are associated with the NS5-branes and ordered in the direction which goes from the D5-branes to the NS5 of the final configuration. The baryonic charge of the group living in the $i$-th NS5 interval is given by the difference between the fluxes on the two NS5 branes delimiting the interval. We follow the convention that  the gauge groups are ordered in the direction which goes from the D5-branes to the NS5 even after S-duality.

The prefactor $\vec x^{\vec s(\vec n)}$ is determined by the brane configuration of $T^{\vec \rho}_{\sigma}(SU(N))$ as follows.  First of all, write down the brane configuration of $T^{\vec \rho}_{\sigma}(SU(N))$ as obtained from mirror symmetry, labelling each NS5-brane by ${n_1}, n_2, \ldots, n_{\ell}$ from the interior to the exterior, and each D5-branes by $x_1, x_2, \ldots, x_{\ell'}$ from the interior to the exterior as in \fref{fig:BraneTs2221r322a}.  The relevant contributions to $\vec x^{\vec s(\vec n)}$ come from D3-branes that stretch between an NS5-brane and a D5-brane and not from those split between NS5-brane intervals. In particular, any D3-brane that stretches between a D5-brane labelled by $x_i$ and an NS5-brane labelled by 
$n_j$ contributes the monomial $x_i^{n_j-n_1}$ to the prefactor. Multiplying all such contributions, the prefactor $\vec x^{\vec s(\vec n)}$ is then given by \bea \label{prefactxs}
\vec x^{\vec s(\vec n)} = \prod_{i=1}^{\ell'} \prod_{j=1}^{\rho_i} x_i^{n_{j}-n_{1}}~.
\eea
The rationale for this prefactor comes from the residue computation presented in Appendix.
To illustrate the above procedure, we provide an example of $T^{(2,2,1,1)}_{(3,2,1)}(SU(6))$ in \eref{Ts2211r321pref}. 
 The dotted horizontal lines indicated in blue and red indicate the D3-brane segments that contribute non-trivially to the prefactor $\vec x^{\vec s(\vec n)}$.  In this example, $\vec x^{\vec s(\vec n)} = x_1^{n_2-n_1} x_2^{n_2-n_1}$. 


We have explicitly tested the relation \eref{mirror} in several different cases. 
Notice that there is an ambiguity in associating the $n_i$  corresponding to the same block $\vec {\tilde n_j}$ to the NS5 branes after S-duality. However, the Coulomb branch formula is manifestly invariant under permutations of fluxes belonging to the same flavor symmetry $U(M_i)$. One can check that also the Higgs branch formula is the same for set of fluxes $B_i$ obtained by permuting   the entries in the various blocks $\vec {\tilde n_j}$.

\section{The generalised Hall-Littlewood formula for $T^{\vec \sigma}_{\vec \rho}(SU(N))$}\label{sec:Hilbert}

In this section we provide a closed formula for the Hilbert series of the Coulomb branch of $T^{\vec \sigma}_{\vec \rho}(SU(N))$, or equivalently the Hilbert series of the Higgs branch of $T_{\vec \sigma}^{\vec \rho}(SU(N))$.   The Higgs branch part of the computation can be reinterpreted in the language of localization and generalizes a known connection between Hall-Littlewood polynomials and Hilbert series of cotangent bundles of flag varieties  \cite{haiman,haiman2}.
Subtleties and complications arising for other classical groups are discussed in section \ref{OSp}.

To state the formula we first need to repackage the magnetic fluxes in yet another form. We construct a string of $N$ integers by repeating $\sigma_{i}$ times each flux $n_i$  
\bea\label{flux}
\vec n^{\vec \sigma} = ( n_1^{\sigma_1} , n_2^{\sigma_{2}}, \cdots , n_\ell^{\sigma_\ell} ) ~,
\eea
where $n^a$ means $n$ repeated $a$ times.

\subsection{The  formula for $T^{\vec \sigma}_{\vec \rho}(SU(N))$}\label{HLformula}

The Coulomb branch formula  for $T^{\vec \sigma}_{\vec \rho}(SU(N))$ can be written as
\be \label{mainHS}
\begin{split}
H_{\text{mon}}[T^{\vec \sigma}_{\vec \rho}(SU(N))] (t; \vec x; \vec {\tilde n_i}) &= H[T^{\vec \sigma}_{\vec \rho}(SU(N))] (t; \vec x; \vec n^{\vec \sigma}) \\
&\equiv  t^{p_{\vec \sigma}(\ns)} (1-t^2)^{N-1} K_{\vec \rho}( \vec x;t) \widehat{Q}^{\ns}_{\vec \sigma} ( \vec a_{\vec \rho} (t, \vec x); t)\, ,
\end{split}
\ee
and it is valid when the fluxes are fully ordered $n_1\geq n_2\geq \cdots \geq n_\ell$. The notations are defined as follows.
\ben
\item $\widehat{Q}^{\ns}_{\vec \sigma}$ is a generalised Hall-Littlewood function for the group $SU(N)$, given by   
\be\label{Q}
\begin{split}
&\widehat{Q}^{\ns}_{\vec \sigma} (x_1, \ldots, x_r; t) \\
&=\frac{1}{\prod_i \sigma_i !} \sum_{w \in S_N} {\vec x}^{w(\ns)}
 {\purple \prod_{\vec \alpha \in \Delta_{\vec \sigma}} (1- {\vec x}^{-w(\vec \alpha)})(1-t^2 {\vec x}^{w(\vec \alpha)})} {\blue \prod_{\vec \gamma \in \Delta_+} \frac{1-t^2 {\vec x}^{-w(\vec \gamma)}}{1-{\vec x}^{-w(\vec \gamma)}}}~,
\end{split}
\ee
where 
\bi
\item $\Delta_+$ is the set of positive roots of $SU(N)$, which can be written in standard notation as ${\vec \alpha} =\e_i -\e_j$ (with $1 \leq i < j \leq N$).
\item $\Delta_{\vec \sigma}$ is the set of positive roots in the diagonal blocks associated with $\vec \sigma$: ${\vec \alpha} =\e_i -\e_j \in \Delta_{\vec\sigma}$ iff    $\sum_{j=1}^{k-1} \sigma_j < i < j \leq \sum_{j=1}^{k} \sigma_j$ for some $k=1,\dots,\ell$. 
\item the sum over $w$ is over  the Weyl group of $SU(N)$. 
\item $\ns$  determines a point in the weight lattice of $U(N)$. It is a dominant weight left invariant by the elements of the Weyl group $\prod_{i=1}^\ell S_{\sigma_i}$ that only permutes indices within the blocks  associated with $\vec\sigma$. 
\item The factor indicated in blue enters in the definition of the standard Hall-Littlewood polynomial. The factor indicated in purple is a modification appearing for non-trivial partitions $\vec \sigma\neq (1^N)$.
\ei
\item The power $p_{\vec \sigma}(\ns)$ is given by 
\bea
p_{\vec \sigma}(\ns) = \sum_{{\vec \alpha} \in \Delta_+(G)} \frac{1}{d_{\vec \sigma}({\vec \alpha})} {\vec \alpha}(\ns)~, \label{powerG}
\eea
where for the positive root ${\vec \alpha} =\e_i -\e_j$ (with $1 \leq i < j \leq N$), ${\vec \alpha}(\ns) =\ns_i -\ns_j$ and $d_{\vec \sigma}(\e_i - \e_j)$ depends only on the index $i$: it is equal to the size of block in the decomposition given by $\vec \sigma$ to which $\e_i$ belongs.
\item The partition $\vec\rho$ determines an embedding of $SU(2)$ in $SU(N)$ defined by the decomposition of the fundamental representation of $SU(N)$ in the sum of irreducible representations of $SU(2)$ of dimension $\rho_k$.
The argument $\vec a_{\vec \rho} (t, \vec x)$, which we shall henceforth abbreviate as $\vec a$, is determined by the following decomposition of the fundamental representation of $SU(N)$ to $G_{\vec \rho} \times {\vec \rho} (SU(2))$:
\bea \label{decompfund}
\chi^{SU(N)}_{{\bf fund}} (\vec a_{\vec \rho} ) = \sum_{k}  \chi^{G_{\rho_k}}_{{\bf fund}}( \vec {\tilde x_k}) \chi^{SU(2)}_{[\rho_k -1]} (t)~,
\eea
where $G_{\rho_k}=U(r_k)$ denotes a subgroup of $G_{\vec \rho}$ corresponding to the part $k$ of the partition $\vec \rho$ that appears $r_k$ times and the $\vec{\tilde x_k}$ are defined as in \eref{pieces}.  Formula \eref{decompfund} determines $\vec a$ as a function of $t$ and $\{ \vec {\tilde x_k} \}$ as required.  Of course, there are many possible choices for $\vec a$; choices related by outer automorphisms of $SU(N)$ are equivalent.
\item The prefactor $K_{\vec \rho}(\vec x; t)$ can be determined as follows.  The embedding specified by $\vec \rho$ induces the decomposition 
\begin{equation}
\chi^{SU(N)}_{\bf Adj} (\vec a) = \sum_{j \in \frac{1}{2}\bZ_{\ge 0}}  \chi^{G_{\vec \rho}}_{R_j}(\vec {\tilde x_j})  \chi^{SU(2)}_{[2j]}(t)~, \label{decompadj} 
\end{equation} 
where $\vec a$ on the left hand side is the same $\vec a$ as in \eref{decompfund}.  Each term in the previous formula gives rise to a plethystic exponential, giving 
\begin{equation}
K_{\vec \rho}(\vec x; t)=\PE \left[t^2 \sum_{j \in \frac{1}{2}\bZ_{\ge 0}}  t^{2j} \chi^{G_{\vec \rho}}_{R_j}(\vec {\tilde x_j} )\right].  \label{K}
\end{equation}
\een

\subsection{Derivation of  the  Hall-Littlewood formula for $T^{\vec \sigma}_{\vec \rho}(SU(N))$}

We first consider the Coulomb branch formula \eref{mon}  for  the theory  $T^{\vec \sigma}(SU(N))$, where omitted partitions are understood to be the trivial one $(1^N)$. By mirror symmetry we can equivalently compute the baryonic Higgs branch Hilbert series for $T_{\vec \sigma}(SU(N))$ using equation \eref{mirror}.

\subsubsection{$T_{\vec \sigma}(SU(N))$: computing residues for the gauge fugacities}\label{residues}

In the case of $T_{\vec \sigma}(SU(N))$  the quiver is 
\bea
[N]- (\sum_{k=2}^\ell \sigma_k)  - \cdots - (\sigma_\ell +\sigma_{\ell-1}) -  (\sigma_\ell) 
\eea
where  $(n)$ and $[n]$ indicate a $U(n)$ gauge and flavor group  respectively. 
By defining 
$N_0 =N$ and $s_{0,k}=x_k$, we can rewrite the Molien-Weyl formula as follows
\be\label{Molien-Weyl2}
\begin{split}
&g[T_{\vec \sigma}(SU(N))](t; \vec x; B_1, \ldots,  B_{\ell-1})  = \\
&  \int  \prod_{i=1}^{\ell -1} \left [\left (\frac{1}{N_i!}\prod_{k=1}^{N_i} \frac{1}{2\pi i }\frac{ds_{i,k}}{ s_{i,k}^{1+B_{i}}} \right ) \frac{ \prod_{ k \neq p}^{N_i}  (1- s_{i,k}/s_{i,p}) \prod_{k \, , p}^{N_i} (1- t^2 s_{i,k}/s_{i,p})}{ \prod_{k=1}^{N_i} \prod_{p=1}^{N_{i-1}} (1- t s_{i-1,p} / s_{i,k})(1- t s_{i,k} / s_{i-1,p})} \right ]
\end{split}
\ee
where  $N_i= \sum_{k=i+1}^\ell \sigma_k$ with $i=0,\cdots, \ell -1$.

We need to identify the poles that contributes to the integral \eref{Molien-Weyl2}. We choose to evaluate the integral adding (minus) the contributions from all the poles outside the unit circle. For positive baryonic charges $B_i\geq 0$ there are no poles at $s_{i,k}=\infty$. Assuming $|t|<1$, the poles for the fugacities  of the gauge group $U(N_i)$ are of the form   
\bea\label{pole0}
s_{i,k} =  s_{i-1,p_k}/t\, , \qquad k=1,\cdots , N_i
\eea
for a choice of $N_i$ fugacities $ s_{i-1,p_k}$  of the gauge group $U(N_{i-1})$. Most of these poles give the same contribution to the integral due to the permutation symmetry of the fugacities $s_{i,k}$ for each $i$ and this contribution is compensated by the factors $N_i!$.  Let us consider  the  contribution of the pole
\bea\label{pole}
s_{i,k}= s_{i-1,k}/t\, \, , \qquad k=1,\cdots , N_i
\eea
for the gauge group $U(N_i)$.  The residue of the $i$-th term in the product  in \eref{Molien-Weyl2} is 
\bea
& \underset{s_{i,k}\rightarrow s_{i-1,k}/t}{{\rm Res}}  \frac{  \prod_{k=1}^{N_i} s_{i,k}^{-1-B_i} \prod_{ k \neq p}^{N_i}  (1- s_{i,k}/s_{i,p}) \prod_{k \, , p}^{N_i} (1- t^2 s_{i,k}/s_{i,p})}{ \prod_{k=1}^{N_i} \prod_{p=1}^{N_{i-1}} (  1- t s_{i-1,p}/s_{i,k})(1- t s_{i,k} / s_{i-1,p})} = \nn \\
& (-1)^{N_i} t^{ B_i N_i  } \prod_{k=1}^{N_i} s_{i-1,k}^{-B_i} \prod_{k\leq N_{i}}\prod_{N_i< p\leq N_{i-1}}  (1- t^2 s_{i-1,p}/s_{i-1,k})^{-1}(1- s_{i-1,k}/s_{i-1,p})^{-1}\, .
\eea
Combining the contributions of all the groups and observing that, by iteration, $s_{i,k}= x_k/t^i$ and $s_{i,k}/s_{i,p}=x_k/x_p$ we obtain the contribution
\bea\label{uff}
\frac{t^{\sum_i  i B_i N_i} (x_1\cdots x_{\sigma_\ell})^{-B_1\cdots - B_{\ell-1}} (x_{\sigma_\ell +1}\cdots x_{\sigma_\ell+\sigma_{\ell -1}})^{-B_1\cdots- B_{\ell-2}}\cdots}{\prod_{(k,p)\in P} (1-  t^2 x_p/x_k) (1-  x_k/x_p)}
\eea 
where $P$ runs over all the entries $(k,p)$ of the upper triangular part of an $N\times N$ matrix with diagonal blocks of sizes $(\sigma_\ell,\cdots, \sigma_1 )$ removed. Here $\sigma_\ell$ corresponds to the block on the top of the matrix. 
All other poles in \eref{pole0} give contributions that are obtained by permuting the $x_i$. Permutations  that exchange only indices belonging to same blocks can be reabsorbed by a  permutation of the $s_{i,k}$ and do not lead to new contributions.

We can rewrite the result in a more compact form in terms of roots. Using the conventions where the positive roots of $SU(N)$ ($e_i -e_j$  with $i<j$)  corresponds to the entries $(i, j)$ of the hermitian  matrix in the Lie algebra ${\rm Lie}(SU(N))$, we find 
\bea \label{HL1}
t^{p_\sigma(\ns)}  \sum_{w \in W_{SU(N)}/W_{L(\vec\sigma)}} {\vec x}^{w(\ns)}
 \prod_{\vec \alpha \in \Delta_+ \setminus \Delta_{\vec\sigma}} \frac{1}{(1-{\vec x}^{-w(\vec \alpha)})(1-t^2 \, {\vec x}^{w(\vec \alpha)})}
\eea
where $\Delta_+$ is the set of positive roots of $SU(N)$, $\Delta_{\vec\sigma}$ is the set of positive roots in the diagonal blocks of size $\sigma_i$ and $W_{L(\vec\sigma)}$ is the subgroup of the Weyl group of $SU(N)$ which just permutes roots inside the various blocks.%
\footnote{Notice that, compared with \eref{uff}, we have reversed the order of the blocks. Here $\sigma_1$ denotes the diagonal block on the left top of the matrix, $\sigma_2$ the adjacent diagonal block and so on. The contribution \eref{uff} corresponds to the permutation $(1,2,\cdots,N) \rightarrow (N,\cdots,2,1)$ in the sum \eref{HL1}. We used  \eref{bar} and the fact that  $\prod_j x_j=1$.}  $p_\sigma(\ns)$ is defined in \eref{powerG} and $\ns$  in  \eref{flux}. 

It is convenient to write the expression \eref{HL1} as follows 
\be \label{HL2}
\begin{split}
 & t^{p_\sigma(\ns)} \prod_{\vec \alpha \in \Delta_+ } \frac{1}{(1-t^2 {\vec x}^{\vec \alpha})(1-t^2 {\vec x}^{-\vec \alpha})}  \\
 &\sum_{w \in W_{SU(N)}/W_{L(\vec\sigma)}} {\vec x}^{w(\ns)}
 \prod_{\vec \alpha \in \Delta_{\vec \sigma}} (1- {\vec x}^{-w(\vec \alpha)})(1-t^2 {\vec x}^{w(\vec \alpha)})\prod_{\vec \gamma \in \Delta_+} \frac{1-t^2 {\vec x}^{-w(\vec \gamma)}}{1-{\vec x}^{-w(\vec \gamma)}} \\
& =  t^{p_\sigma(\ns)} (1-t^2)^{N-1} K_{(1,\cdots,1)}(\vec x ,t) \widehat{Q}^{\ns}_{\vec \sigma} (\vec x; t)
\end{split}
\ee
where $\widehat{Q}^{\ns}_{\vec \sigma} (\vec x; t)$ and $K_{\vec\rho}$ have been defined in \eref{Q} and \eref{K} respectively. We can extend the sum to the entire Weyl group since the fluxes are equal inside the blocks.
We have thus recovered the expression \eref{mainHS}. 

The computation is valid for $B_i\geq 0$ which, using \eref{bar} and \eref{flux},  correspond to fully ordered fluxes $n_1\geq n_2\geq \cdots \geq n_\ell$. For other values of $B_i$, extra poles at infinity might affect the result and give a more complicated expression.

\subsubsection{The localisation formula}\label{loc}
We can reinterpret the previous computation in terms of localisation. 
A similar approach has been successfully 
used for the computation of the Hilbert series of non-compact Calabi-Yaus \cite{Martelli:2006yb} and the Hilbert series of instanton moduli spaces \cite{Nekrasov:2002qd}.  We use localisation in the following form. Suppose that  we have a line bundle ${\cal L}$ over a smooth manifold $X$ with a holomorphic action of a  torus $\mu: T \rightarrow X$ with isolated fixed points.
The Lefschetz fixed point formula states that \cite{atiyah1967lefschetz, atiyah1968lefschetz}
\bea  \label{lefschetz}
\sum_i (-1)^i {\rm Tr}\{ \mu | H^{(0,i)}(X, {\cal L})\} = \sum_P \frac{q^{m^L_P}}{\prod_j (1- q^{m^j_P})}
\eea
where $P$ are the fixed points of the torus action, $m^j_P,\,\, j=1,\cdots, {\rm dim} X$ are the weights of the linearization of the torus action $\mu$ on the tangent space of $X$ at the point $P$
and $m^L_P$ is the weight of the action of $\mu$ on the fiber of the line bundle at $P$. $q$ denotes a set of fugacities for the action of $T$. Whenever the higher cohomology groups $H^{(0,i)}(X, {\cal L})\, , i\ge 1$ vanish the left hand side of \eref{lefschetz} computes the Hilbert series counting holomorphic sections of the line bundle ${\cal L}$.

In order to use formula \eref{lefschetz} we need to find an algebraic description  of the Higgs branch of $T_{\vec \sigma}(SU(N))$, a smooth resolution of it, and the conditions under which the higher
cohomology groups vanish.

 It is  known that, as an algebraic variety, the Higgs branch of   $T_{\vec \sigma}(SU(N))$ is the closure of the nilpotent orbit of Jordan type $\vec\sigma^T$  \cite{kraftprocesi,nakajimaquiver,kobak}
\bea
\label{higgsTsigma}
{\rm Higgs}(T_{\vec \sigma}(SU(N))) = \bar O_{\vec \sigma^T}\, .
\eea
Recall that a partition $\vec \lambda=(\lambda_1 \cdots, \lambda_l)$ of $N$ 
\bea \lambda_1 \ge \cdots \geq \lambda_l \,, \qquad \qquad \sum_{i=1}^l \lambda_i = N \eea
naturally identifies  a nilpotent matrix $N_{\vec\lambda}$ in ${\rm Lie}(SU(N))$  with Jordan blocks of dimension $\lambda_i$. The nilpotent orbit $O_{\vec\lambda}$  of type $\vec\lambda$ is, by definition,  the orbit of $N_{\vec\lambda}$ under the adjoint action of $SU(N)$. Notice that the transpose of $\vec\sigma$ enters in \eref{higgsTsigma}.

It is also well known that the singular variety $\bar O_{\vec \sigma^T}$ has a smooth resolution, called the Springer resolution, 
\bea  
\label{springer}
\mu\,  :\,  T^*(SU(N)/P) \rightarrow \bar O_{\vec \sigma^T}\, ,
\eea
in terms of the cotangent bundle of a flag variety.  $P$ here is a parabolic subgroup of $SU(N)$ consisting of the upper triangular block matrices with blocks of dimensions $\sigma_i$. The non-zero entries in $P$ are  those belonging to diagonal blocks of dimensions $\sigma_i\times \sigma_i$ in addition to all the entries above the diagonal.  The homogeneous space $SU(N)/P$  parametrizes all the possible flags of type $\vec\sigma$ in $\mathbb{C}^N$, i.e. the set of vector subspaces $V_0=\{0\}\subset V_1 \cdots \subset V_N=\mathbb{C}^N$ with relative dimension ${\rm dim} (V_{i+1}/V_i)= \sigma_{\ell -i}$. 

We can also give a different characterization of $T^*(SU(N)/P)$ which is sometime useful. 
The elements in $P$ belonging to the diagonal blocks form a subgroup $S(\prod_i U(\sigma_i))$ of $P$, called the Levi subgroup and denoted by $L(\vec \sigma)$. Accordingly, the Lie algebra of $P$ decomposes as
\bea  {\rm Lie}(P) = {\rm Lie} (L(\vec\sigma)) \oplus n(P)\, , \eea
where the nil-radical $n(P)$ consists of nilpotent matrices. The cotangent bundle $T^*(SU(N)/P)$ can be written as $SU(N) \times_{P} n(P)$, which is the quotient
of $SU(N) \times n(P)$ by the equivalence relation
\bea 
(g,n) \sim ( g^\prime ,n^\prime) \Leftrightarrow g^\prime = g\,  p\, , n^\prime = p^{-1}\,  n \, p\, ,  \qquad p\in P \, .
\eea  
The resolution  in \eref{springer} is just given by  $\mu\, :  (g,n) \rightarrow g n g^{-1}$.

We can now use the localisation formula \eref{lefschetz}. We can apply the formula to $X=T^*(SU(N)/P)$ since it has the same holomorphic functions as $\bar{O}_{\vec\sigma}$.
The torus action is induced by the Cartan subgroup of $SU(N)$ and by the scaling symmetry, with associated fugacities $\vec x$ and $t$. The Cartan subgroup of $SU(N)$ acts
in the obvious way on the cosets in $SU(N)/P$ and its action is naturally extended to the cotangent bundle. The scaling symmetry acts on the cotangent fiber.   
This torus action has  isolated fixed points. A coset $gP$ in  $SU(N)/P$ is fixed by the action of the Cartan torus $T\subset SU(N)$ if and only if  $T \subset g P g^{-1}$ and this selects $g\in W_{SU(N)}/W_{L(\vec\sigma)}$ where   $W_{SU(N)}$ is the Weyl group of $SU(N)$ and $W_{L(\vec\sigma)}$ the Weyl group of the Levi subgroup of $P$.
The fiber at the fixed points must be zero because of the scaling symmetry. In order to use \eref{lefschetz} we need to linearize the torus action around the fixed points.
The tangent space to $T^*(SU(N)/P)$ at a fixed point can be written as 
\bea {\rm Lie} (SU(N))/ {\rm Lie} (P) \oplus {\rm Lie} (SU(N))^*/ {\rm Lie} (P)^* \, ,\eea
where the first factor is the tangent space to the flag manifolfd $SU(N)/P$ and the second to the cotangent fiber. The torus action on an element of the root space $\alpha$ in ${\rm Lie} (SU(N))/ {\rm Lie} (P)$ is $\vec x^{\vec\alpha}$ while on the corresponding element in  the dual space ${\rm Lie} (SU(N))^*/ {\rm Lie} (P)^*$ is $t^2 \vec x^{-\vec\alpha}$. We also consider a line bundle ${\cal L}$  associated with the fluxes
$\vec n$, which give the weight of the representation of the Cartan subgroup of $SU(N)$ on the fiber.

The right hand side of  \eref{lefschetz} then reads
\bea \label{HL11}
 \sum_{w \in W_{SU(N)}/W_{L(\vec\sigma)}} {\vec x}^{w(\ns)}
 \prod_{\vec \alpha \in \Delta_+ \setminus \Delta_{\vec\sigma}} \frac{1}{(1-{\vec x}^{-w(\vec \alpha)})(1-t^2 \, {\vec x}^{w(\vec \alpha)})}
\eea
where $\Delta_+$ is the set of positive roots of $SU(N)$, while  $\Delta_{\vec\sigma}$ is the set of positive roots in $L(\vec\sigma)$. The product in \eref{HL11} covers  all the roots in   ${\rm Lie} (SU(N))/ {\rm Lie} (P)$ which correspond to the entries
in the lower triangular part of the matrix with the exclusion of those living in the diagonal blocks.  When $\vec\sigma=(1,\cdots ,1)$ the sum in \eref{HL11} runs over all the positive roots and the expression in \eref{HL11} becomes a (dual) Hall-Littlewood polynomial \cite{haiman}.

The expression \eref{HL11} is the baryonic Hilbert series of the Higgs branch of $T_{\vec \sigma}(SU(N))$ provided the higher cohomology group of the line bundle ${\cal L}$ vanish. Sufficient conditions for the vanishing have been discussed in \cite{broer} (see Proposition 3.7)
and require that $\ns$ is a dominant weight and it is fixed by the action of $W_{L(\vec\sigma)}$. This requires that all the entries in $\ns$ are ordered and equal in the blocks corresponding to the partition $\vec\sigma$
\bea \label{higher}
\ns = ( n_1^{\sigma_1}, n_2^{\sigma_2}, \cdots , n_l^{\sigma_l} ) \, , \quad n_1\ge n_2 \ge \cdots \ge n_l \, ,
\eea  
where $n^a$ means $n$ repeated $a$ times.  

The rest of the computation is the same as  in section \ref{residues}. We can  manipulate  expression \eref{HL11}  and obtain again the  final formula \eref{HL2}.  The sum is extended to the entire Weyl group assuming the condition \eref{higher} on the fluxes.
In this approach the overall prefactor $t^{p_\sigma(\ns)}$ is found by an explicit comparison with the monopole formula.

\subsubsection{Computing residues in the flavor fugacities}\label{sec:residue}
We state the following general observation:
\begin{quote}
The Higgs branch Hilbert series of $T^{\vec\rho}_{\vec \sigma}(SU(N))$ can be obtained from that of the theory $T^{(1^N)}_{\vec \sigma}(SU(N))$ by taking residues with respect to the flavor fugacities. 
\end{quote}
The Higgs branch Hilbert series  \eref{Molien-Weyl2} for $T^{\vec \rho}_{\vec \sigma}(SU(N))$ has poles corresponding to a particular limit of the fugacities. The residue at this pole is the Higgs branch Hilbert series  for $T^{\vec \rho^\prime}_{\vec \sigma}(SU(N))$, where $\vec\rho^\prime$ is obtained from $\vec\rho$ by moving the last box in the partition $\vec\rho$ to a previous column.  For example, we can go from the trivial partition $\vec\rho =(1^N)$ to $\vec\rho^\prime =(2,1^{N-1})$ as follows:
\be \label{H1hNm1}
\begin{split}
& \res_{z \rightarrow 1}  g[T^{(1^N)}_{\vec \sigma}(SU(N))] (t; \vec w_1; \vec B) \Bigg |_{\substack{w_{1,1}= t z x_1 \\ w_{1,q}=x_q \,\, q=2,\cdots, N-1\\ w_{1,N}=  (tz)^{-1} x_1} } =  \\
& \frac{1}{2} x_1^{-B_1} {\rm PE}  \left[t^2+t \sum_{q=1}^{N-1} (x_1 x_{q}^{-1}+ x_1^{-1} x_{q})\right] g[T^{(2,1^{N-1})}_{\vec \sigma}(SU(N))] (t; \vec {\tilde w_1},\vec {\tilde w_{2}}; \vec B)
 ~,
\end{split}
\ee
where $\vec{\tilde w_1} = (x_{N-1},\cdots, x_{2})$, $\vec {\tilde w_2}= (x_1)$ and the first line receives the contribution from the residue
\bea
s_{1, N_1} = t w_{1,N} = x_1 z^{-1}~.
\eea
We give a proof and a  generalization of this formula to partitions $\vec \rho$ and $\vec \rho^\prime$ of $SU(N)$ which are related by moving a single box in Appendix \ref{app:Higgsresidue}.
Any partition $\vec\rho$ can be obtained from the trivial partition $(1^N)$ by an iteration of the previous move. Therefore by repeated residue computations we may extract the Higgs branch Hilbert series of any $T^{\vec \rho}_{\vec \sigma}(SU(N))$ theory from that of $T_{\vec \sigma}^{(1^N)}(SU(N))$. 

We can do a completely analogous computation in the Coulomb branch. The mirror of the previous observation is: 
\begin{quote}
  The  Coulomb branch Hilbert series of $T^{\vec \sigma}_{\vec \rho}(SU(N))$ can be obtained from that of $T^{\vec \sigma}_{(1^N)}(SU(N))$ by taking residues with respect to the topological fugacities. 
  \end{quote}
 As discussed in section 6 of \cite{Cremonesi:2014kwa}, the monopole formula for $T^{\vec \sigma}_{\vec \rho}(SU(N))$ has poles 
corresponding to a particular limit of the fugacities. The residue at this pole gives the monopole formula for $T^{\vec \sigma}_{\vec \rho^\prime}(SU(N))$, where $\vec\rho^\prime$ is obtained from $\vec\rho$ by moving the last box in the partition $\vec\rho$ to a previous column. This was proven in 
\cite{Cremonesi:2014kwa} for the case $\vec\sigma=(1^N)$ but it can straightforwardly extended to the case of a general $\vec\sigma$.
For example, we can go from the trivial partition $\vec\rho =(1^N)$ to $\vec\rho^\prime =(2,1^{N-1})$ by computing
\be \label{resHiggs}
\begin{split}
& \res_{z\to 1} \,  H_{\text{mon}}[T^{\vec \sigma}_{(1^N)}(SU(N))] (t; x_1\, t \, z, x_2,\cdots, x_{N-1}, x_1 t^{-1} z^{-1}; \tilde{\vec n}) = \\
&\frac{1}{2} {\rm PE} \left[t^2+t \sum_{q=1}^{N-1} \left(\frac{x_1}{x_{q}}+ \frac{x_{q}}{x_1}\right)\right] H_{\text{mon}}[T^{\vec \sigma}_{(2,1^{N-1})}(SU(N))] (t; x_1, x_2, \cdots, x_{N-1}; \tilde{\vec n})
\end{split}
\ee
By repeated residue computations we may extract the Coulomb branch Hilbert series of any $T^{\vec \sigma}_{\vec \rho}(SU(N))$ theory from that of $T^{\vec \sigma}_{(1^N)}(SU(N))$. 

By carefully mapping the fugacities under mirror symmetry, we see that the two previous observations are consistent with \eref{mirror}. 
Notice that in taking residues with respect to the flavor symmetries we obtain a prefactor with powers of $x_i$ in the Higgs branch computation but not in the Coulomb branch one. This is consistent with and explains the prefactor \eref{prefactxs} in \eref{mirror}.

The observations can be now used  to conclude our proof of  \eref{mainHS}.  The Higgs branch Hilbert series of  $T^{\vec \rho}_{\vec \sigma}(SU(N))$ can be obtained from \eref{HL2} by taking residues with respect to the flavor fugacities. Notice that the poles in formula \eref{HL2} come only from the  factor  $K_{(1^N)}(\vec x,t)$.  The partition $\vec\rho$ can be obtained from $(1^N)$ by a set of moves like those in \eref{movebox}.   It is not difficult to see that this set of moves has the effect of replacing   $\vec x$ with $\vec a_{\vec \rho} (t, \vec x)$ given in \eref{decompfund}.  The multiplicative factors in \eref{H1hNm1} and \eref{resgener} cancel some terms in the denominator of $K_{(1^N)}(\vec x,t)$ and transform it into $K_{\vec\rho}(\vec x,t)$. They also introduce a prefactor  which coincides with \eref{prefactxs}.  In this way we obtain the general expression for the  Higgs branch Hilbert series $T_{\vec \sigma}^{\vec \rho}(SU(N))$. Removing  the prefactor according to  \eref{mirror},  we obtain precisely the general expression for the Coulomb branch Hilbert series of the mirror $T^{\vec \sigma}_{\vec \rho}(SU(N))$ given in \eref{mainHS}.

%
%
%
%
%
%

Geometrically, the structure of the factor $K_{\vec\rho}(\vec x,t)$ is related to the fact that, as an algebraic variety, the Coulomb branch of $T^{\vec \sigma}_{\vec \rho}(SU(N))$, equivalently the Higgs branch of $T_{\vec \sigma}^{\vec \rho}(SU(N))$, is the intersection of the nilpotent orbit
of type $\vec\sigma^T$ with the Slodowy slice of type $\vec\rho$  \cite{nakajimaquiver, Gaiotto:2008ak},
\bea
\bar{O}_{\vec \sigma^T} \cap S_{\vec \rho}\, .
\eea
The Slodowy slice is defined as follows. The  partition $\vec\rho$  identifies a homomorphism $\vec \rho: {\rm Lie}(SU(2))\rightarrow {\rm Lie}(SU(N))$ by saying that the image of $J_+= J_1+i J_2$, where $J_i$ are the standard generators of $SU(2)$, is a nilpotent matrix of Jordan type $\vec\rho$. A theorem by Jacobson and Morozov guarantees that the map between partitions and homomorphisms is one-to-one \cite{collingwood1993nilpotent}. The Lie algebra of $SU(N)$ decomposes
under the homomorphism $\vec\rho$ into a set of irreducible  $ G_{\vec\rho} \times SU(2)$ representations $[R_j;2j]$   as in \eref{decompadj}. Let $t_j$ be the $SU(2)$ lowest weight in each representation $[R_j;2j]$. The Slodowy  slice associated with the partition $\vec\rho$ is the  subspace of ${\rm Lie}(SU(N))$ consisting of the elements of the form
\bea
\vec\rho (J_1+i J_2) + \sum_j c_j t_j \, . \eea
The various terms entering \eref{K} schematically correspond to such description of the slice.

\section{Applications of the Hall-Littlewood formula for $T^{\vec \sigma}_{\vec \rho}(SU(N))$} \label{SUN}
In this section we demonstrate the previous results for unitary groups of small rank. 
\subsection{$T^{\vec \sigma}(SU(4))$}
In the following we focus on cases in which the partition $\vec \rho$ is trivial, namely $\vec \rho=(1,1,1, 1)$.
{\small
\begin{table}[H]
\begin{center}
\begin{tabular}{|l|l|l|l|l| l|}
\hline
Partition $\vec \sigma$ & Quiver for $T^{\vec \sigma}(SU(4))$ & $\Delta_{\vec \sigma}$ & Levi subgroup  \\
\hline
$(1,1,1,1)$ & $(3,4)-(2,0)-(1,0)$	    &    	$\emptyset$ &  $S(U(1)^4)$	 \\				
$(2,1,1)$	& $(2,2)-(2,1)-(1,0)$	     &    $\{ {\vec e}_1-{\vec e}_2 \}$&  $S(U(2) \times U(1)^2)$	\\				  
$(2,2)$	& $(1,0)-(2,2)-(1,0)$	&	$\{ {\vec e}_1-{\vec e}_2, \; {\vec e}_3-{\vec e}_4\}$&  $S(U(2) \times U(2))$	\\		
$(3,1)$	& $(1,1)-(1,0)-(1,1)$	&	$\{ {\vec e}_1-{\vec e}_2, \; {\vec e}_1-{\vec e}_3, \; {\vec e}_2-{\vec e}_3\}$ & $S(U(3) \times U(1))$ \\
\hline
\end{tabular}
\caption{Parameters for the partitions of $SU(4)$, whose positive roots are ${\vec e}_i - {\vec e}_j$ with $1\leq i < j \leq 4$.  The shorthand notation $(k_1, N_1)-(k_2, N_2) - \ldots -(k_\ell, N_\ell)$ denotes the quiver with the gauge group $U(k_1)\times \cdots \times U(k_\ell)$ with $\ell-1$ bifundamental hypermultiplets and $N_i$ fundamental flavors charged under the gauge group $U(k_i)$ for all $1 \leq i \leq \ell$.}
\label{tab:paraSU4}
\end{center}
\end{table}


Let us map the results obtained from the right-hand side of \eref{mainHS} to those obtained from the monopole formula.

\subsubsection{${\vec \sigma} =(3,1)$}  
The brane configurations corresponding to $T^{(3,1)}(SU(4))$ are
\bea
\begin{tikzpicture}
\draw (0,0)--(0,2.5) node[black,midway, xshift =0cm, yshift=1.5cm]{\footnotesize $x_1$};
\draw (1,0)--(1,2.5) node[black,midway, xshift =0cm, yshift=1.5cm]{\footnotesize $x_2$}; 
\draw (2,0)--(2,2.5) node[black,midway, xshift =-0.5cm, yshift=-1.7cm] {\footnotesize NS5} node[black,midway, xshift =0cm, yshift=1.5cm]{\footnotesize $x_3$};
\draw (3,0)--(3,2.5) node[black,midway, xshift =0cm, yshift=1.5cm]{\footnotesize $x_4$}; 
\draw [dashed,red] (-1,0)--(-1,2.5) node[black,midway, xshift =0cm, yshift=1.5cm] {\footnotesize $n_{1}$};
\draw [dashed,blue] (-2,0)--(-2,2.5) node[black,midway, xshift =0cm, yshift=-1.7cm] {\footnotesize D5} node[black,midway, xshift =0cm, yshift=1.5cm] {\footnotesize $n_{2}$};
\draw [red] (0,2)--(-1,2) node[black,midway, yshift=0.4cm] {\tiny D3}; \draw [red] (1,1.8)--(-1,1.8); \draw [red](2,1.6)--(-1,1.6);
\draw [blue] (0,1)--(-2,1);
\draw (0,0.6)--(1,0.6); \draw (1,1.1)--(2,1.1); \draw (2,1.4)--(3,1.4); 
\draw[thick,->] (4,1) -- (5,1);
\end{tikzpicture}  
\qquad 
\begin{tikzpicture}
\draw (0,0)--(0,2.5) node[black,midway, xshift =0cm, yshift=1.5cm]{\footnotesize $x_1$};
\draw (1,0)--(1,2.5) node[black,midway, xshift =0cm, yshift=1.5cm]{\footnotesize $x_2$}; 
\draw (2,0)--(2,2.5) node[black,midway, xshift =-0.5cm, yshift=-1.7cm] {} node[black,midway, xshift =0cm, yshift=1.5cm]{\footnotesize $x_3$};
\draw (3,0)--(3,2.5) node[black,midway, xshift =0cm, yshift=1.5cm]{\footnotesize $x_4$}; 
\draw [dashed,red] (2.5,0)--(2.5,2.5) node[black,midway, xshift =0cm, yshift=1.5cm] {\footnotesize $n_{1}$};
\draw [dashed,blue] (0.5,0)--(0.5,2.5) node[black,midway, xshift =0cm, yshift=-1.7cm] {} node[black,midway, xshift =0cm, yshift=1.5cm] {\footnotesize $n_{2}$};
\draw (0,0.6)--(1,0.6); \draw (1,1.1)--(2,1.1); \draw (2,1.4)--(3,1.4); 
\end{tikzpicture}  
\eea

The monopole formula for the Coulomb branch of $T^{(3,1)}(SU(4))$ reads
\be
\begin{split}
H_{\rm{mon}} [T^{(3,1)}(SU(4))] (t; \vec x; n_2, n_1) &= y_1^{n_2} y_3^{n_1} \sum_{m_1 = -\infty}^\infty ~\sum_{m_2 = -\infty}^\infty~ \sum_{m_3 = -\infty}^\infty  z_1^{m_1} z_2^{m_2} z_3^{m_3} \times \\
& \qquad t^{|m_1-m_2|+|m_2-m_3|+|m_1-n_2|+|m_3-n_1|} \times  \\
& \qquad P_{U(1)} (t; m_1) P_{U(1)} (t; m_2) P_{U(1)} (t; m_3) ~,
\end{split}
\ee
where $z_1, z_2, z_3$ and $y_1, y_3$ are fugacities for the topological charges, with
\bea \label{consz31}
z_1 = x_2x_1^{-1}~, \qquad z_2 = x_3 x_2^{-1}~, \qquad z_3 = x_4 x_3^{-1}~, \qquad y_1 = x_1~, \qquad y_3 =x_1 x_2 x_3~.
\eea
Let us emphasize that $n_1, n_2$ are the background fluxes for the {\it third} and the {\it first} $U(1)$ flavor symmetries in the quiver reading from left to right, respectively. 

Due to the shift symmetry, the fugacities $\vec z$ and $\vec y$ satisfy
\bea
z_1 z_2 z_3 y_1 y_3 =1~
\eea
and so
\bea \label{consx31}
x_1 x_2 x_3 x_4 =1~.
\eea
We find that
\bea
& H_{\rm{mon}}[T^{(3,1)}(SU(4))] (t; \vec x; n_2, n_1) \nn \\
& = \begin{cases} 
H[T^{(3,1)}(SU(4))](t; x_1, \dots, x_4;  n_1,n_1,n_1,n_2) ~, & \qquad n_2\geq n_1-3 \\
H[T^{(3,1)}(SU(4))](t; x_1^{-1}, \dots, x_4^{-1};  n_2,n_2,n_2,n_1) ~, & \qquad n_1\geq n_2-3 ~.
\end{cases}
\eea
where $H[T^{(3,1)}(SU(4))]$ is given by \eref{mainHS} with relevant data given by \tref{tab:paraSU4}. 

We know that $H_{{\text{mon}}}$ and $H$ coincide for fully ordered fluxes. We see that in certain specific cases this constraint can be relaxed.
In general, whenever there are two background fluxes present in the theory, it is always possible to find an ordering of such fluxes in the generalised Hall-Littlewood formula to match the result obtained from the monopole formula.  The reason is the symmetry of the monopole formula under permutation of the fluxes belonging to the same flavor group and under change of sign of all the background fluxes together with a reflection $x\rightarrow x^{-1}$ of the fugacities. 
We present another example in the next subsection.  Note that when there are three or more background fluxes, this is not always possible; we comment on this below \eref{matchTr321s2211}.

Let us compare this result to the baryonic generating function of the mirror $T_{(3,1)}(SU(4)): (1)-[4]$. 
\bea
g[T_{(3,1)}(SU(4))] (t; (x_4, \ldots,x_1); B) = \oint_{|b|=1} \frac{1}{ 2\pi i b^{B+1}} (1-t^2) \PE \left[t b \sum_{i=1}^4 x_i^{-1} + t b^{-1} \sum_{i=1}^4 x_i  \right]~.
\eea
After the constraints \eref{consz31} and \eref{consx31} are imposed, we find that
\be
\begin{split}
&g[T_{(3,1)}(SU(4))] (t; (x_4, x_3, x_2, x_1); n_1-n_2) \\
&= H_{{\rm mon}}[T^{(3,1)}(SU(4))] (t; x_1, x_2, x_3, x_4; n_1,n_2) ~.
\end{split}
\ee

\subsubsection{${\vec \sigma} =(2^2)$}  
The brane configurations corresponding to $T^{(2,2)}(SU(4))$ are
\bea
\begin{tikzpicture}
\draw (0,0)--(0,2.5) node[black,midway, xshift =0cm, yshift=1.5cm]{\footnotesize $x_1$};
\draw (1,0)--(1,2.5) node[black,midway, xshift =0cm, yshift=1.5cm]{\footnotesize $x_2$}; 
\draw (2,0)--(2,2.5) node[black,midway, xshift =-0.5cm, yshift=-1.7cm] {\footnotesize NS5} node[black,midway, xshift =0cm, yshift=1.5cm]{\footnotesize $x_3$};
\draw (3,0)--(3,2.5) node[black,midway, xshift =0cm, yshift=1.5cm]{\footnotesize $x_4$}; 
\draw [dashed,red] (-1,0)--(-1,2.5) node[black,midway, xshift =0cm, yshift=1.5cm] {\footnotesize $n_{1}$};
\draw [dashed,blue] (-2,0)--(-2,2.5) node[black,midway, xshift =0cm, yshift=-1.7cm] {\footnotesize D5} node[black,midway, xshift =0cm, yshift=1.5cm] {\footnotesize $n_{2}$};
\draw [red] (0,2)--(-1,2) node[black,midway, yshift=0.4cm] {\tiny D3}; \draw [red] (1,1.8)--(-1,1.8); 
\draw [blue] (0,1)--(-2,1); \draw [blue] (1,0.8)--(-2,0.8); 
\draw (0,0.6)--(1,0.6); \draw (1,1.1)--(2,1.1); \draw (1,1.3)--(2,1.3); \draw (2,1.4)--(3,1.4); 
\draw[thick,->] (4,1) -- (5,1);
\end{tikzpicture}  
\qquad 
\begin{tikzpicture}
\draw (0,0)--(0,2.5) node[black,midway, xshift =0cm, yshift=1.5cm]{\footnotesize $x_1$};
\draw (1,0)--(1,2.5) node[black,midway, xshift =0cm, yshift=1.5cm]{\footnotesize $x_2$}; 
\draw (2,0)--(2,2.5) node[black,midway, xshift =-0.5cm, yshift=-1.7cm] {} node[black,midway, xshift =0cm, yshift=1.5cm]{\footnotesize $x_3$};
\draw (3,0)--(3,2.5) node[black,midway, xshift =0cm, yshift=1.5cm]{\footnotesize $x_4$}; 
\draw [dashed,red] (1.25,0)--(1.25,2.5) node[black,midway, xshift =0cm, yshift=-1.5cm] {\footnotesize $n_1$};
\draw [dashed,blue] (1.75,0)--(1.75,2.5) node[black,midway, xshift =0cm, yshift=-1.5cm] {\footnotesize $n_2$} node[black,midway, xshift =0cm, yshift=1.5cm] {};
\draw (0,0.6)--(1,0.6); \draw (1,1.1)--(2,1.1); \draw (1,1.3)--(2,1.3); \draw (2,1.4)--(3,1.4); 
\end{tikzpicture}  
\eea
The monopole formula for the Coulomb branch of $T^{(2,2)}(SU(4))$ reads
\bea
H_{{\rm mon}} [T^{(2,2)}(SU(4))] (t; \vec x; n_1, n_2) &= y_2^{n_1+n_2}  \sum_{u_1=-\infty}^\infty ~ \sum_{m_1 \geq m_2 > -\infty} ~ \sum_{u_3=-\infty}^\infty z_1^{u_1} z_2^{m_1+m_2} z_3^{u_3} \nn \\
&  t^{\sum_{i=1}^2(|u_1-m_i| + |u_3-m_i|) +\sum_{i=1}^2 \sum_{j=1}^2 |m_i-n_j|-2|m_1-m_2|}  \times \nn \\
& P_{U(1)}(t; u_1) P_{U(2)}(t;\vec m) P_{U(1)}(t; u_3) ~.
\eea
Due to the shift symmetry, the fugacities $\vec z$ and $\vec y$ satisfy
\bea
z_1 z_2^2 z_3 y_2^2 =1~. \label{consz22}
\eea
Setting
\bea \label{consx22}
z_1 = x_2x_1^{-1}~, \qquad z_2 = x_3 x_2^{-1}~, \qquad z_3 = x_4 x_3^{-1}~, \qquad y_2 = x_1 x_2~,
\eea
the above constraint translates to
\bea \label{consxsig22}
x_1 x_2 x_3 x_4=1~.
\eea

We find that
\bea
& H_{{\rm mon}}[T^{(2,2)}(SU(4))]  (t;\vec x; n_2, n_1) \nn \\
&=  \begin{cases}  H[T^{(2,2)}](SU(4))]  (t; x_1, \ldots, x_4; n_1,n_1,n_2,n_2)~, &\quad  n_1-n_2 \geq -1 \\
H [T^{(2,2)}(SU(4))]  (t; x_1, \ldots, x_4; n_2,n_2,n_1,n_1)~, &\quad  n_2-n_1 \geq -1 \end{cases} ,
\eea
where $H[T^{(2,2)}(SU(4))]$ is given by \eref{mainHS} with relevant data given by Tables \ref{tab:paraSU4}.  

Let us compare this to the baryonic generating function of $T_{(2,2)}(SU(4)): (2)-[4]$.
\bea
g[T_{(2,2)}(SU(4))] (t; (x_4, \ldots x_1); B) &= \oint_{|b_1|=1}  \frac{{\rm d}b_1}{2 \pi i b_1^{1+B}}  \oint_{|b_2|=1} \frac{{\rm d}b_2}{2 \pi i b_2^{1+B}}  (1-b_1 b_2^{-1})(1-b_2 b_1^{-1})  \nn \\
&  \PE \Bigg[t  (b_1+b_2) \sum_{i=1}^4 x_i^{-1} + t(b_1^{-1}+b_2^{-1}) \sum_{i=1}^4 x_i   \nn \\
& \qquad - t^2 (b_1+b_2)(b_1^{-1}+b_2^{-1})\Bigg]~.
\eea
Then after the constraints \eref{consz22}, \eref{consx22} and \eref{consxsig22} are imposed, we find that
\bea
& g[T_{(2,2)}(SU(4))] (t; (x_4, \ldots, x_1); n_1-n_2) \nn \\
&= H_{{\rm mon}}[T^{(2,2)}(SU(4))] (t; x_1, \ldots, x_4; n_2,n_1)~, \qquad x_1 x_2 x_3 x_4=1~.
\eea

\subsubsection{${\vec \sigma} =(2,1^2)$} 
The brane configurations corresponding to $T^{(2,1,1)}(SU(4))$ are
\bea
\begin{tikzpicture} 
\draw (0,0)--(0,2.5) node[black,midway, xshift =0cm, yshift=1.5cm]{\footnotesize $x_1$};
\draw (1,0)--(1,2.5) node[black,midway, xshift =0cm, yshift=1.5cm]{\footnotesize $x_2$}; 
\draw (2,0)--(2,2.5) node[black,midway, xshift =-0.5cm, yshift=-1.7cm] {\footnotesize NS5} node[black,midway, xshift =0cm, yshift=1.5cm]{\footnotesize $x_3$};
\draw (3,0)--(3,2.5) node[black,midway, xshift =0cm, yshift=1.5cm]{\footnotesize $x_4$}; 
\draw [dashed,red] (-1,0)--(-1,2.5) node[black,midway, xshift =0cm, yshift=1.5cm] {\footnotesize $n_{1}$};
\draw [dashed,blue] (-2,0)--(-2,2.5) node[black,midway, xshift =0cm, yshift=-1.7cm] {\footnotesize D5} node[black,midway, xshift =0cm, yshift=1.5cm] {\footnotesize $n_{2}$};
\draw [dashed,purple] (-3,0)--(-3,2.5)  node[black,midway, xshift =0cm, yshift=1.5cm] {\footnotesize $n_{3}$};
\draw [red] (0,2)--(-1,2) node[black,midway, yshift=0.4cm] {\tiny D3}; \draw [red] (1,1.8)--(-1,1.8); 
\draw [blue] (0,1)--(-2,1);
\draw [purple](0,0.2)--(-3,0.2);
\draw (0,0.4)--(1,0.4);\draw (0,0.6)--(1,0.6); \draw (1,0.9)--(2,0.9); \draw (1,1.1)--(2,1.1); \draw (2,1.4)--(3,1.4); 
\draw[thick,->] (4,1) -- (5,1);
\end{tikzpicture}  
\qquad
\begin{tikzpicture}
\draw (0,0)--(0,2.5) node[black,midway, xshift =0cm, yshift=1.5cm]{\footnotesize $x_1$};
\draw (1,0)--(1,2.5) node[black,midway, xshift =0cm, yshift=1.5cm]{\footnotesize $x_2$}; 
\draw (2,0)--(2,2.5) node[black,midway, xshift =-0.5cm, yshift=-1.7cm] {} node[black,midway, xshift =0cm, yshift=1.5cm]{\footnotesize $x_3$};
\draw (3,0)--(3,2.5) node[black,midway, xshift =0cm, yshift=1.5cm]{\footnotesize $x_4$}; 
\draw [dashed,red] (1.5,0)--(1.5,2.5) node[black,midway, xshift =0cm, yshift=-1.5cm] {\footnotesize $n_{1}$};
\draw [dashed,blue] (0.75,0)--(0.75,2.5) node[black,midway, xshift =0cm, yshift=-1.7cm] {} node[black,midway, xshift =0cm, yshift=-1.5cm] {\footnotesize $n_{2}$};
\draw [dashed,purple] (0.25,0)--(0.25,2.5) node[black,midway, xshift =0cm, yshift=-1.7cm] {} node[black,midway, xshift =0cm, yshift=-1.5cm] {\footnotesize $n_{3}$};
\draw (0,0.4)--(1,0.4);\draw (0,0.6)--(1,0.6); \draw (1,0.9)--(2,0.9); \draw (1,1.1)--(2,1.1); \draw (2,1.4)--(3,1.4); 
\end{tikzpicture} 
\eea
The monopole formula reads
\be \label{mon211}
\begin{split}
& H_{\text{mon}}[T^{(2,1^2)}(SU(4))] (t; \vec x; n_1, n_2, n_3) \\
&=  y_1^{n_2+n_3} y_2^{n_1}  \sum_{u_1 \geq u_2 > -\infty}~ \sum_{m_1 \geq m_2 > -\infty}~ \sum_{v=-\infty}^\infty ~ z_1^{u_1+u_2} z_2^{m_1+m_2} z_3^{v} \times \\
& \qquad t^{\sum_{i=1}^2 \sum_{j=1}^2 |u_i-m_j|+\sum_{j=1}^2 |v-m_j|+\sum_{i=1}^2 \sum_{k=2}^3 |u_i-n_k|+ \sum_{j=1}^2 |m_j-n_1|} \times \\
& \qquad t^{-2|m_1-m_2| -2|u_1-u_2|} P_{U(2)} (t; \vec u) P_{U(2)} (t; \vec m) P_{U(1)} (t; v)   ~,
\end{split}
\ee
where $z_1, z_2, z_3$ are topological fugacities for the $U(2)$, $U(2)$ and $U(1)$ gauge groups from left to right; $y_1, y_2$ are topological fugacities for the $U(2)$ and $U(1)$ flavor groups from left to right; $n_1$ is the background charge for the $U(1)$ flavor symmetry and $n_2, n_3$ are those for the $U(2)$ flavor symmetry.
The relations between $\vec z$, $\vec y$ and $\vec x$ are
\bea
z_1 = x_2x_1^{-1}~, \qquad z_2 = x_3 x_2^{-1}~, \qquad z_3 = x_4 x_3^{-1}~, \qquad  y_1= x_1~, \qquad y_2= x_1 x_2~.
\eea
As before, these fugacities satisfy
\bea
z_1^2 z_2^2 z_3 y_1^2 y_2 =1 \quad \Leftrightarrow \quad  x_1 x_2 x_3x_4=1~.
\eea

The brane configuration of the mirror theory $T_{(2,1,1)}(SU(4)): [4]-(2)-(1)$ is
\bea
\begin{tikzpicture} 
\draw [dashed,red] (0,0)--(0,2.5) node[black,midway, xshift =0cm, yshift=1.5cm]{\footnotesize $x_1$};
\draw [dashed,blue] (1,0)--(1,2.5) node[black,midway, xshift =0cm, yshift=1.5cm]{\footnotesize $x_2$}; 
\draw [dashed,purple] (2,0)--(2,2.5) node[black,midway, xshift =-0.5cm, yshift=-1.7cm] {\footnotesize D5} node[black,midway, xshift =0cm, yshift=1.5cm]{\footnotesize $x_3$};
\draw [dashed, brown] (3,0)--(3,2.5) node[black,midway, xshift =0cm, yshift=1.5cm]{\footnotesize $x_4$}; 
\draw  (-1,0)--(-1,2.5) node[black,midway, xshift =0cm, yshift=1.5cm] {\footnotesize $n_{1}$};
\draw  (-2,0)--(-2,2.5) node[black,midway, xshift =0cm, yshift=-1.7cm] {\footnotesize NS5} node[black,midway, xshift =0cm, yshift=1.5cm] {\footnotesize $n_{2}$};
\draw  (-3,0)--(-3,2.5)  node[black,midway, xshift =0cm, yshift=1.5cm] {\footnotesize $n_{3}$};
\draw [red] (0,2)--(-1,2) node[black,midway, yshift=0.4cm] {\tiny D3}; \draw [blue] (1,1.8)--(-1,1.8); \draw [purple] (2,1.6)--(-1,1.6); \draw [brown] (3,1.4)--(-1,1.4); 
\draw (-2,0.4)--(-1,0.4);\draw (-2,0.6)--(-1,0.6); \draw (-3,0.9)--(-2,0.9); 
\draw[thick,->] (4,1) -- (5,1);
\end{tikzpicture} 
\qquad 
\begin{tikzpicture} 
\draw [dashed,red] (-1.2,0)--(-1.2,2.5) node[black,midway, xshift =0cm, yshift=-1.5cm]{};
\draw [dashed,blue] (-1.4,0)--(-1.4,2.5) node[black,midway, xshift =0cm, yshift=-1.5cm]{}; 
\draw [dashed,purple] (-1.6,0)--(-1.6,2.5) node[black,midway, xshift =-0.5cm, yshift=-1.7cm] {} node[black,midway, xshift =0cm, yshift=-1.5cm]{};
\draw [dashed, brown] (-1.8,0)--(-1.8,2.5) node[black,midway, xshift =0cm, yshift=-1.5cm]{}; 
\draw  (-1,0)--(-1,2.5) node[black,midway, xshift =0cm, yshift=1.5cm] {\footnotesize $n_{1}$};
\draw  (-2,0)--(-2,2.5) node[black,midway, xshift =0cm, yshift=-1.7cm] {} node[black,midway, xshift =0cm, yshift=1.5cm] {\footnotesize $n_{2}$};
\draw  (-3,0)--(-3,2.5)  node[black,midway, xshift =0cm, yshift=1.5cm] {\footnotesize $n_{3}$};
\draw (-2,0.4)--(-1,0.4);\draw (-2,0.6)--(-1,0.6); \draw (-3,0.9)--(-2,0.9); 
\end{tikzpicture} 
\eea
Reading from right to left, the first gauge group is $U(2)$ and the second gauge group is $U(1)$.  The baryonic generating function reads 
\be \label{bar211}
\begin{split}
& g[T_{(2,1,1)}(SU(4))] (t; (x_4, \ldots, x_1); B_1, B_2) \\
& = \oint_{|b_1|=1} \frac{{\rm d} b_1}{2\pi ib_1} \oint_{|b_2|=2} \frac{{\rm d} b_2}{2\pi i b_2} \oint_{|b_3|=2} \frac{{\rm d} b_3}{2\pi i b_3} (b_1 b_2)^{-B_1} b_3^{-B_2} \\
& \quad \PE \Big[ (b_1^{-1}+b_2^{-1}) (x_1 +\ldots +x_4) t + (b_1+b_2) (x_1^{-1} +\ldots +x_4^{-1}) t  \\
&  \quad + (b_1+b_2) b_3^{-1} t+(b_1^{-1}+b_2^{-1})b_3  t- t^2 -t^2(b_1+b_2)(b_1^{-1}+b_2^{-2})  \Big] ~,
\end{split}
\ee
where $B_1$ is the baryonic charge associated with the $U(2)$ group and $B_2$ is that associated with the $U(1)$ group.

Formulae \eref{mon211} and \eref{bar211} can be matched as follows:
\be
\begin{split}
&g[T_{(2,1^2)}(SU(4))] (t; (x_4, \ldots, x_1); n_1-n_2, n_2-n_3)  \\
& = H_{\text{mon}}[T^{(2,1^2)}(SU(4))] (t; \vec x; n_1, n_2, n_3)~, \qquad x_1x_2 x_3 x_4=1~.
\end{split}
\ee

The monopole formula \eref{mon211} can also be related to the generalised Hall-Littlewood formula \eref{mainHS}.  When the monopole fluxes in the former are ordered, we find that
\be
\begin{split}
&H_{\text{mon}}[T^{(2,1^2)}(SU(4))] (t; x_1, \ldots, x_4; n_3, n_2, n_1) \\
&= H[T^{(2,1^2)}(SU(4))] (t; x_1, \ldots, x_4; n_1, n_1, n_2, n_3)~, \qquad n_1 \geq n_2 \geq n_3~,
\end{split}
\ee
where $H[T^{(2,1^2)}(SU(4))] (t; \vec x; \vec n)$ is given by \eref{mainHS} with $p_{(2,1^2)}(\vec n) = 2n_1-2n_3$.

\subsection{Examples of $T^{\vec \sigma}_{\vec \rho} (SU(N))$}
Below we present some examples for $T^{\vec \sigma}_{\vec \rho} (SU(N))$ theories.

\subsubsection{$T^{(3,2,1)}_{(2,2,1,1)} (SU(6))$}
The brane configuration and quiver diagram for $T^{(3,2,1)}_{(2,2,1,1)} (SU(6))$ are depicted in Figures \ref{fig:BraneTs321r2211a} and \ref{fig:BraneTs321r2211b}.  

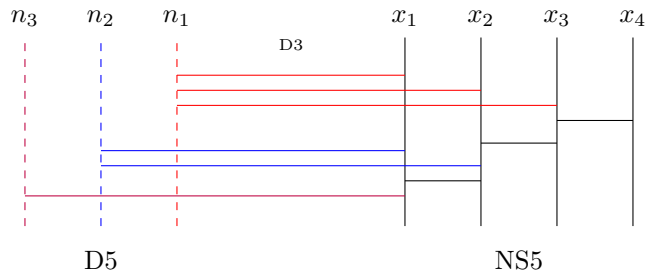
\begin{figure}[H]
\centering
\begin{tikzpicture} 
\draw (0,0)--(0,2.5) node[black,midway, xshift =0cm, yshift=1.5cm]{\footnotesize $x_1$};
\draw (1,0)--(1,2.5) node[black,midway, xshift =0cm, yshift=1.5cm]{\footnotesize $x_2$}; 
\draw (2,0)--(2,2.5) node[black,midway, xshift =-0.5cm, yshift=-1.7cm] {\footnotesize NS5} node[black,midway, xshift =0cm, yshift=1.5cm]{\footnotesize $x_3$};
\draw (3,0)--(3,2.5) node[black,midway, xshift =0cm, yshift=1.5cm]{\footnotesize $x_4$}; 
\draw [dashed,red] (-3,0)--(-3,2.5) node[black,midway, xshift =0cm, yshift=1.5cm] {\footnotesize $n_{1}$};
\draw [dashed,blue] (-4,0)--(-4,2.5) node[black,midway, xshift =0cm, yshift=-1.7cm] {\footnotesize D5} node[black,midway, xshift =0cm, yshift=1.5cm] {\footnotesize $n_{2}$};
\draw [dashed,purple] (-5,0)--(-5,2.5)  node[black,midway, xshift =0cm, yshift=1.5cm] {\footnotesize $n_{3}$};
\draw [red] (0,2)--(-3,2) node[black,midway, yshift=0.4cm] {\tiny D3}; \draw [red] (1,1.8)--(-3,1.8); \draw [red](2,1.6)--(-3,1.6);
\draw [blue] (0,1)--(-4,1); \draw [blue](1,0.8)--(-4,0.8);
\draw [purple](0,0.4)--(-5,0.4);
\draw (0,0.6)--(1,0.6); \draw (1,1.1)--(2,1.1); \draw (2,1.4)--(3,1.4); 
\end{tikzpicture}  
\caption{Brane construction for $T^{(3,2,1)}_{(2,2,1,1)}(SU(6))$.  The corresponding quiver diagram is depicted in \fref{fig:BraneTs321r2211b}.}
\label{fig:BraneTs321r2211a}
\end{figure}

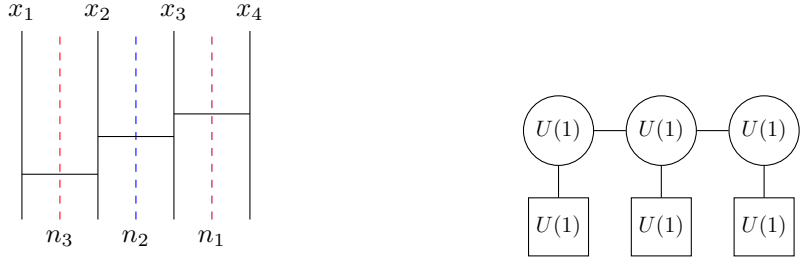
\begin{figure}[H]
\centering
\begin{tikzpicture} 
\draw (0,0)--(0,2.5) node[black,midway, xshift =0cm, yshift=1.5cm] {\footnotesize $x_{1}$};
\draw (1,0)--(1,2.5) node[black,midway, xshift =0cm, yshift=1.5cm] {\footnotesize $x_{2}$}; 
\draw (2,0)--(2,2.5) node[black,midway, xshift =0cm, yshift=1.5cm] {\footnotesize $x_{3}$};
\draw (3,0)--(3,2.5) node[black,midway, xshift =0cm, yshift=1.5cm] {\footnotesize $x_{4}$};
\draw [dashed,red] (0.5,0)--(0.5,2.5) node[black,midway, xshift =0cm, yshift=-1.5cm] {\footnotesize $n_{3}$};
\draw [dashed,blue] (1.5,0)--(1.5,2.5) node[black,midway, xshift =0cm, yshift=-1.5cm] {\footnotesize $n_{2}$};
\draw [dashed,purple] (2.5,0)--(2.5,2.5) node[black,midway, xshift =0cm, yshift=-1.5cm] {\footnotesize $n_{1}$};
\draw (0,0.6)--(1,0.6); \draw (1,1.1)--(2,1.1); \draw (2,1.4)--(3,1.4); 
\end{tikzpicture}  
\hspace{3cm}
\begin{tikzpicture}[scale=0.7, transform shape]
\begin{scope}[auto,%
  every node/.style={draw, minimum size=1.1cm}, node distance=0.6cm];
\node[circle] (UN1) at (0, 0) {$U(1)$};
\node[circle, right=of UN1] (UN2) {$U(1)$};
\node[circle, right=of UN2] (UN3) {$U(1)$};
\node[rectangle, below=of UN1] (UM1) {$U(1)$};
\node[rectangle, below=of UN2] (UM2) {$U(1)$};
\node[rectangle, below=of UN3] (UM3) {$U(1)$};
\end{scope}
\draw (UN1) -- (UN2)
(UN2)--(UN3)
(UN1)--(UM1)
(UN2)--(UM2)
(UN3)--(UM3);
\end{tikzpicture}
\caption{Left: brane construction for $T^{(3,2,1)}_{(2,2,1,1)}(SU(6))$ after the D5-branes are moved into the NS-brane intervals. Right: the linear quiver read off from the brane configuration.  We adopt the convention that the $i$-th gauge group corresponds to the D3-brane interval between $x_i$ and $x_{i+1}$.}
\label{fig:BraneTs321r2211b}
\end{figure}

\begin{figure}[H]
\centering
\begin{tikzpicture} 
\draw[dashed,red ] (0,0)--(0,2.5) node[black,midway, xshift =0cm, yshift=1.5cm]{\footnotesize $x_1$};
\draw[dashed, blue] (0.5,0)--(0.5,2.5) node[black,midway, xshift =0cm, yshift=1.5cm]{\footnotesize $x_2$}; 
\draw[dashed,purple] (1,0)--(1,2.5) node[black,midway, xshift =-0.5cm, yshift=-1.7cm] {\footnotesize D5} node[black,midway, xshift =0cm, yshift=1.5cm]{\footnotesize $x_3$};
\draw[dashed, brown] (1.5,0)--(1.5,2.5) node[black,midway, xshift =0cm, yshift=1.5cm]{\footnotesize $x_4$}; 
\draw  (-3,0)--(-3,2.5) node[black,midway, xshift =0cm, yshift=1.5cm] {\footnotesize $n_{1}$};
\draw (-4,0)--(-4,2.5) node[black,midway, xshift =0cm, yshift=-1.7cm] {\footnotesize NS5} node[black,midway, xshift =0cm, yshift=1.5cm] {\footnotesize $n_{2}$};
\draw  (-5,0)--(-5,2.5)  node[black,midway, xshift =0cm, yshift=1.5cm] {\footnotesize $n_{3}$};
\draw [red] (0,1.6)--(-3,1.6) node[black,midway, yshift=0.6cm] {\tiny D3}; \draw [red] (0,1.8)--(-4,1.8);
\draw [blue](0.5,0.8)--(-3,0.8); \draw [blue] (0.5,1)--(-4,1); 
\draw [purple](1,0.4)--(-3,0.4);
\draw [brown] (1.5,0.1)--(-3,0.1);
\draw (-3,1.4)--(-4,1.4); \draw (-4,1.1)--(-5,1.1);
\end{tikzpicture}  
\\~\\
\begin{tikzpicture} 
\draw[dashed,red] (-4.75,0)--(-4.75,2.5) node[black,midway, xshift =0cm, yshift=-1.5cm]{\footnotesize $x_1$};
\draw[dashed, blue] (-4.25,0)--(-4.25,2.5) node[black,midway, xshift =0cm, yshift=-1.5cm]{\footnotesize $x_2$}; 
\draw[dashed,purple] (-3.75,0)--(-3.75,2.5) node[black,midway, xshift =0cm, yshift=-1.5cm]{\footnotesize $x_3$}; 
\draw[dashed, brown] (-3.25,0)--(-3.25,2.5) node[black,midway, xshift =0cm, yshift=-1.5cm]{\footnotesize $x_4$}; 
\draw  (-3,0)--(-3,2.5) node[black,midway, xshift =0cm, yshift=1.5cm] {\footnotesize $n_{1}$};
\draw (-4,0)--(-4,2.5) node[black,midway, xshift =0cm, yshift=1.5cm] {\footnotesize $n_{2}$};
\draw  (-5,0)--(-5,2.5)  node[black,midway, xshift =0cm, yshift=1.5cm] {\footnotesize $n_{3}$};
\draw (-3,1.4)--(-4,1.4); \draw (-4,1.1)--(-5,1.1);
\end{tikzpicture}  
\hspace{3cm}
\begin{tikzpicture}[scale=0.7, transform shape]
\begin{scope}[auto,%
  every node/.style={draw, minimum size=1.1cm}, node distance=0.6cm];
\node[circle] (UN1) at (0, 0) {$U(1)$};
\node[circle, right=of UN1] (UN2) {$U(1)$};
\node[rectangle, below=of UN1] (UM1) {$U(2)$};
\node[rectangle, below=of UN2] (UM2) {$U(2)$};
\end{scope}
\draw (UN1) -- (UN2)
(UN1)--(UM1)
(UN2)--(UM2);
\end{tikzpicture}
\caption{Top: brane construction for $T^{(2,2,1,1)}_{(3,2,1)}(SU(6))$. This diagram is obtained by exchanging D5-branes and NS5 branes in \fref{fig:BraneTs321r2211a}. Bottom left: the D5-branes are moved into the NS5-brane intervals. Bottom right: the quiver diagram read off from the bottom left brane configuration.  We adopt the convention that the $i$-th gauge group corresponds to the D3-brane interval between $n_i$ and $n_{i+1}$.}
\label{fig:BraneTs2221r321a}
\end{figure}

\paragraph{The monopole formula.} The monopole formula for this theory reads
\be
\begin{split}
& H_{\text{mon}}(t; \vec x; n_3, n_2, n_1) = y_1^{n_3} y_2^{n_2} y_3^{n_1} (1-t^2)^{-3} \times \\
&\sum_{u_1=-\infty}^\infty \; \sum_{u_2=-\infty}^\infty \; \sum_{u_3=-\infty}^\infty z_1^{u_1} z_2^{u_2} z_3^{u_3}~ t^{|u_1-u_2|+|u_2-u_3|+|u_1-n_3|+|u_2-n_2|+|u_3-n_1|} ~,
\end{split}
\ee
where we set
\be
\begin{split}
& z_1= x_2 x_1^{-1}~, \quad z_2 = x_3 x_1^{-1}~, \quad z_3 = x_4 x_1^{-1}~,\\
& y_1 = x_1, \quad y_2 =x_1 x_2~, \quad y_3 =x_1x_2x_3~,
\end{split}
\ee
and by shift symmetry we impose the following conditions:
\bea
z_1 z_2 z_3 y_1 y_2 y_3=1 \qquad \Leftrightarrow \qquad x_1^2 x_2^2 x_3 x_4=1~.
\eea

\paragraph{The baryonic generating function.}  From \fref{fig:BraneTs2221r321a}, the baryonic generating function of $T^{(2,2,1,1)}_{(3,2,1)}(SU(6))$ reads
\bea
&g[T^{(2,2,1,1)}_{(3,2,1)}(SU(6))](t, (x_4,x_3),(x_2,x_1); B_1, B_2) = \oint_{|b_1|=1} \frac{{\rm d} b_1}{2 \pi i b_1^{1+B_1}}  \oint_{|b_2|=1} \frac{{\rm d} b_2}{2 \pi i b_2^{1+B_2}} \times \nn \\
& \qquad  \PE \Big[b_1^{-1} (x_3+x_4) +b_1 (x_3^{-1}+x_4^{-1}) + b_2^{-1} (x_1+x_2) +b_2(x_1^{-1}+x_2^{-1})  \nn \\\
& \qquad  +b_1 b_2^{-1}+ b_2 b_1^{-1} - 2t^2 \Big]~.
\eea
This can be equated to the monopole formula as follows:
\be
\begin{split}
& x_1^{n_2-n_1} x_2^{n_2-n_1} g[T^{(2,2,1,1)}_{(3,2,1)}(SU(6))](t, (x_4,x_3),(x_2,x_1); n_1-n_2, n_2-n_3) \\
&= H_{\text{mon}}(t; \vec x; n_3, n_2, n_1) ~, \quad x_1^2 x_2^2 x_3 x_4=1~.
\end{split}
\ee
According to \eref{prefactxs}, the prefactor $x_1^{n_2-n_1} x_2^{n_2-n_1}$ in the first line is due to the dotted D3-branes in the interval $n_2-n_1$ in the following diagram:
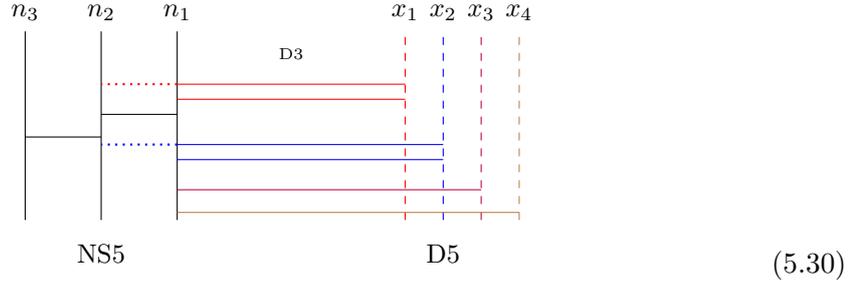
\bea \label{Ts2211r321pref}
\begin{tikzpicture} 
\draw[dashed,red ] (0,0)--(0,2.5) node[black,midway, xshift =0cm, yshift=1.5cm]{\footnotesize $x_1$};
\draw[dashed, blue] (0.5,0)--(0.5,2.5) node[black,midway, xshift =0cm, yshift=1.5cm]{\footnotesize $x_2$}; 
\draw[dashed,purple] (1,0)--(1,2.5) node[black,midway, xshift =-0.5cm, yshift=-1.7cm] {\footnotesize D5} node[black,midway, xshift =0cm, yshift=1.5cm]{\footnotesize $x_3$};
\draw[dashed, brown] (1.5,0)--(1.5,2.5) node[black,midway, xshift =0cm, yshift=1.5cm]{\footnotesize $x_4$}; 
\draw  (-3,0)--(-3,2.5) node[black,midway, xshift =0cm, yshift=1.5cm] {\footnotesize $n_{1}$};
\draw (-4,0)--(-4,2.5) node[black,midway, xshift =0cm, yshift=-1.7cm] {\footnotesize NS5} node[black,midway, xshift =0cm, yshift=1.5cm] {\footnotesize $n_{2}$};
\draw  (-5,0)--(-5,2.5)  node[black,midway, xshift =0cm, yshift=1.5cm] {\footnotesize $n_{3}$};
\draw [thick, dotted, red] (-3,1.8)--(-4,1.8); \draw [red] (0,1.6)--(-3,1.6) node[black,midway, yshift=0.6cm] {\tiny D3}; \draw [red] (0,1.8)--(-3,1.8);
\draw [blue](0.5,0.8)--(-3,0.8); \draw [blue] (0.5,1)--(-3,1); \draw [thick, dotted, blue] (-3,1)--(-4,1); 
\draw [purple](1,0.4)--(-3,0.4);
\draw [brown] (1.5,0.1)--(-3,0.1);
\draw (-3,1.4)--(-4,1.4); \draw (-4,1.1)--(-5,1.1);
\end{tikzpicture}  
\eea
The dotted red line contributes $x_1^{n_2-n_1}$ to the prefactor and the dotted blue line contributes $x_2^{n_2-n_1}$.

\paragraph{The generalised Hall-Littlewood formula.} From \eref{mainHS}, the generalised Hall-Littlewood formula for the Coulomb branch Hilbert series of $T^{(3,2,1)}_{(2,2,1,1)} (SU(6))$ reads 
\bea
& H[T^{(3,2,1)}_{(2,2,1,1)} (SU(6))](t; x_1, \ldots, x_4; n_1, n_2, n_3) \\
&= t^{3n_1-n_2-2n_3}  (1-t^2)^5 K_{(2,2,1,1)} (x_1, x_2, x_3; t) \widehat{Q}^{(n_1^3, n_2^2, n_3)}_{(2,2,1,1)} ( t x_1, t^{-1} x_1, tx_2, t^{-1} x_2 , x_3, x_4; t)~, \nn
\eea
where
\be
\begin{split}
&K_{(2,2,1,1)} (x_1, \ldots, x_4; t) = \PE \Big[ t^2 (3+ x_1 x_2^{-1}+x_2 x_1^{-1}+ x_3 x_4^{-1}+x_4 x_3^{-1}) \\
& \qquad + t^4 (2+ x_1 x_2^{-1} + x_2 x_1^{-1}) +t^3 \sum_{i=1,2}\sum_{j=3,4} (x_i x_j^{-1}+x_j x_i^{-1})\Big]~, \\
& \Delta_{(3,2,1)} = \{ \e_1-\e_2, \e_1-\e_3, \e_2-\e_3, \e_4-\e_5 \}~.
\end{split}
\ee
The HL formula agrees with the monopole formula when the fluxes are ordered:
\be \label{matchTr321s2211}
\begin{split}
&H[T^{(3,2,1)}_{(2,2,1,1)} (SU(6))](t; x_1, \ldots, x_4; n_1,n_1,n_1, n_2,n_2, n_3)  \\
&= H_{\text{mon}}[T^{(3,2,1)}_{(2,2,1,1)} (SU(6))](t; x_1, \ldots, x_4; n_3, n_2, n_1)~, \quad n_1\geq n_2 \geq n_3 \in \BZ~.
\end{split}
\ee
Note that for some values of fluxes $(n_1, n_2, n_3)$ that are not ordered, such as $(n_1, n_2, n_3)= (0,1,0)$, the two formulae cannot be matched with each other. This is due to novanishing contributions to the index \eqref{lefschetz} from higher cohomology groups.

\subsubsection{$T^{(2,2)}_{(2,1,1)} (SU(4)): (1)-[2]$}
The quiver diagram for $T^{(2,2)}_{(2,1,1)} (SU(4))$ is the same as that of $T(SU(2))$ theory, namely $(1)-[2]$.  Although this example violates the condition ${\vec \sigma}^T < \vec \rho$, we demonstrate below that the monopole formula and the generalised Hall-Littlewood formula can still be applied. 
\bea
\begin{tikzpicture}
\draw (0,0)--(0,2.5) node[black,midway, xshift =0cm, yshift=1.5cm]{\footnotesize $x_1$};
\draw (1,0)--(1,2.5) node[black,midway, xshift =0cm, yshift=1.5cm]{\footnotesize $x_2$}; 
\draw (2,0)--(2,2.5) node[black,midway, xshift =-0.5cm, yshift=-1.7cm] {\footnotesize NS5} node[black,midway, xshift =0cm, yshift=1.5cm]{\footnotesize $x_3$};
\draw [dashed,red] (-1,0)--(-1,2.5) node[black,midway, xshift =0cm, yshift=1.5cm] {\footnotesize $n_{1}$};
\draw [dashed,blue] (-2,0)--(-2,2.5) node[black,midway, xshift =0cm, yshift=-1.7cm] {\footnotesize D5} node[black,midway, xshift =0cm, yshift=1.5cm] {\footnotesize $n_{2}$};
\draw [red] (0,2)--(-1,2) node[black,midway, yshift=0.4cm] {\tiny D3}; \draw [red] (1,1.8)--(-1,1.8); 
\draw [blue] (0,1)--(-2,1); \draw [blue] (1,0.8)--(-2,0.8); 
 \draw (1,1.1)--(2,1.1);
\draw[thick,->] (3,1) -- (4,1);
\end{tikzpicture}
\qquad 
\begin{tikzpicture}
\draw (0,0)--(0,2.5) node[black,midway, xshift =0cm, yshift=1.5cm]{\footnotesize $x_1$};
\draw (1,0)--(1,2.5) node[black,midway, xshift =0cm, yshift=1.5cm]{\footnotesize $x_2$}; 
\draw (2,0)--(2,2.5) node[black,midway, xshift =-0.5cm, yshift=-1.7cm] {} node[black,midway, xshift =0cm, yshift=1.5cm]{\footnotesize $x_3$};
\draw [dashed,red] (1.25,0)--(1.25,2.5) node[black,midway, xshift =0cm, yshift=-1.7cm] {\footnotesize $n_{1}$};
\draw [dashed,blue] (1.75,0)--(1.75,2.5) node[black,midway, xshift =0cm, yshift=-1.7cm] {\footnotesize $n_{2}$} node[black,midway, xshift =0cm, yshift=1.5cm] {};
 \draw (1,1.1)--(2,1.1);
\end{tikzpicture}
\eea

From \eref{mainHS} the Coulomb branch Hilbert series of $T^{(2,2)}_{(2,1,1)}(SU(4))$ reads 
\be \label{HT22211}
\begin{split}
& H[T^{(2,2)}_{(2,1,1)}(SU(4))](t; x_1, x_2, x_3; n_1, n_2)  \\
&=t^{2n_1 - 2n_2}  (1-t^2)^3 K_{(2,1,1)} (x_1, x_2, x_3; t) \widehat{Q}^{(n_1^2, n_2^2)}_{(2,2)} ( t x_1, t^{-1} x_1, x_2, x_3; t)~,
\end{split}
\ee
where
\bea
K_{(2,1,1)} (x_1, x_2, x_3; t) &= \PE \left[ t^4+t^3 \left(\frac{x_1}{x_2}+\frac{x_1}{x_3}+\frac{x_2}{x_1}+\frac{x_3}{x_1}\right)+t^2 \left(\frac{x_2}{x_3}+\frac{x_3}{x_2}+2\right) \right] ~.
\eea

The Coulomb branch Hilbert series can also be computed directly from the monopole formula
\bea \label{HTSU2}
H_{\text{mon}}[T^{(2,2)}_{(2,1,1)}(SU(4))](t; z_1, z_2; y_2; n_2, n_1) = y_2^{n_1+n_2}  \sum_{u=-\infty}^\infty t^{\sum_{i=1,2}|n_i-u|} z_2^u (1-t^{2})^{-1}~,
\eea
where we take
\bea
y_2= x_1 x_2~, \qquad z_2 = x_3 x_2^{-1}~.
\eea
By the shift symmetry, 
\bea
y_2^2 z_2 =1 \qquad \Rightarrow \qquad x_1^2 x_2 x_3=1~.
\eea
Formulae \eref{HT22211} and \eref{HTSU2} can be matched as follows:
\be
\begin{split}
&H[T^{(2,2)}_{(2,1,1)}(SU(4))](t; x_1, x_2, x_3; n_1, n_1, n_2,n_2)   \\
&= H_{\text{mon}}[T^{(2,2)}_{(2,1,1)}(SU(4))](t; x_1,x_2,x_3; n_2, n_1)~.
\end{split}
\ee

Let us compare this result to the baryonic generating function of $T^{(2,1,1)}_{(2,2)}(SU(4))$.
\bea
\begin{tikzpicture} 
\draw [dashed,red] (0,0)--(0,2.5) node[black,midway, xshift =0cm, yshift=1.5cm]{\footnotesize $x_1$} node[black,midway, xshift =0cm, yshift=-1.7cm] {\footnotesize D5} ;
\draw [dashed,blue] (1,0)--(1,2.5) node[black,midway, xshift =0cm, yshift=1.5cm]{\footnotesize $x_2$}; 
\draw [dashed,purple] (2,0)--(2,2.5) node[black,midway, xshift =0cm, yshift=1.5cm]{\footnotesize $x_3$}; 
\draw  (-1,0)--(-1,2.5) node[black,midway, xshift =0cm, yshift=1.5cm] {\footnotesize $n_{1}$};
\draw  (-2,0)--(-2,2.5) node[black,midway, xshift =0cm, yshift=-1.7cm] {\footnotesize NS5} node[black,midway, xshift =0cm, yshift=1.5cm] {\footnotesize $n_{2}$};
\draw [red] (0,2)--(-1,2) node[black,midway, yshift=0.4cm] {\tiny D3}; 
\draw [red] (0,1.8)--(-2,1.8);
\draw [blue] (1,1.5)--(-1,1.5);
\draw [purple] (2,1.2)--(-1,1.2);
\draw (-2,0.6)--(-1,0.6);
\draw[thick,->] (3,1) -- (4,1);
\end{tikzpicture} 
\qquad 
\begin{tikzpicture} 
\draw [dashed,purple] (-1.25,0)--(-1.25,2.5) node[black,midway, xshift =0cm, yshift=-1.5cm]{\footnotesize $x_3$};
\draw [dashed,blue] (-1.75,0)--(-1.75,2.5) node[black,midway, xshift =0cm, yshift=-1.5cm]{\footnotesize $x_2$}; 
\draw [dashed,red] (-3,0)--(-3,2.5) node[black,midway, xshift =0cm, yshift=-1.5cm]{\footnotesize $x_1$}; 
\draw  (-1,0)--(-1,2.5) node[black,midway, xshift =0cm, yshift=1.5cm] {\footnotesize $n_{1}$};
\draw  (-2,0)--(-2,2.5) node[black,midway, xshift =0cm, yshift=-1.7cm] {} node[black,midway, xshift =0cm, yshift=1.5cm] {\footnotesize $n_{2}$};
\draw (-2,0.6)--(-1,0.6); 
\end{tikzpicture} 
\eea
\bea
&g[T^{(2,1,1)}_{(2,2)}(SU(4))] (t; (x_3, x_2); B) \nn \\
&= \oint_{|b|=1} \frac{{\rm d}b}{2\pi i b} b^{-B} \PE \left[ b^{-1} (x_2+x_3) t+ b (x_2^{-1}+x_3^{-1}) t - t^2 \right]~.
\eea
This can be equated to the monopole formula as follows:
\be
\begin{split}
& H_{\text{mon}}[T^{(2,2)}_{(2,1,1)}(SU(4))](t; x_1, x_2, x_3; n_2, n_1) \\
&= x_1^{n_2-n_1} g[T^{(2,1,1)}_{(2,2)}(SU(4))] (t; (x_3, x_2); n_1-n_2)~,
\end{split}
\ee
where the prefactor $x_1^{n_2-n_1}$ is due to the D3-brane indicated by the dotted red horizontal line in the diagram below.
\bea
\begin{tikzpicture} 
\draw [dashed,red] (0,0)--(0,2.5) node[black,midway, xshift =0cm, yshift=1.5cm]{\footnotesize $x_1$} node[black,midway, xshift =0cm, yshift=-1.7cm] {\footnotesize D5} ;
\draw [dashed,blue] (1,0)--(1,2.5) node[black,midway, xshift =0cm, yshift=1.5cm]{\footnotesize $x_2$}; 
\draw [dashed,purple] (2,0)--(2,2.5) node[black,midway, xshift =0cm, yshift=1.5cm]{\footnotesize $x_3$}; 
\draw  (-1,0)--(-1,2.5) node[black,midway, xshift =0cm, yshift=1.5cm] {\footnotesize $n_{1}$};
\draw  (-2,0)--(-2,2.5) node[black,midway, xshift =0cm, yshift=-1.7cm] {\footnotesize NS5} node[black,midway, xshift =0cm, yshift=1.5cm] {\footnotesize $n_{2}$};
\draw [red] (0,2)--(-1,2) node[black,midway, yshift=0.4cm] {\tiny D3}; 
\draw [red] (0,1.8)--(-1,1.8);
\draw [thick, dotted, red] (-2,1.8)--(-1,1.8);
\draw [blue] (1,1.5)--(-1,1.5);
\draw [purple] (2,1.2)--(-1,1.2);
\draw (-2,0.6)--(-1,0.6);
\end{tikzpicture} 
\eea

\subsubsection{$T^{(2,1,1)}_{(2,1,1)} (SU(4)): [1]-(1)-(1)-[2]$}
The brane configurations are given by
\bea
\begin{tikzpicture} 
\draw (0,0)--(0,2.5) node[black,midway, xshift =0cm, yshift=1.5cm]{\footnotesize $x_1$};
\draw (1,0)--(1,2.5) node[black,midway, xshift =0cm, yshift=1.5cm]{\footnotesize $x_2$}; 
\draw (2,0)--(2,2.5) node[black,midway, xshift =-0.5cm, yshift=-1.7cm] {\footnotesize NS5} node[black,midway, xshift =0cm, yshift=1.5cm]{\footnotesize $x_3$};
\draw [dashed,red] (-1,0)--(-1,2.5) node[black,midway, xshift =0cm, yshift=1.5cm] {\footnotesize $n_{1}$};
\draw [dashed,blue] (-2,0)--(-2,2.5) node[black,midway, xshift =0cm, yshift=-1.7cm] {\footnotesize D5} node[black,midway, xshift =0cm, yshift=1.5cm] {\footnotesize $n_{2}$};
\draw [dashed,purple] (-3,0)--(-3,2.5)  node[black,midway, xshift =0cm, yshift=1.5cm] {\footnotesize $n_{3}$};
\draw [red] (0,2)--(-1,2) node[black,midway, yshift=0.4cm] {\tiny D3}; \draw [red] (1,1.8)--(-1,1.8); 
\draw [blue] (0,1)--(-2,1);
\draw [purple](0,0.2)--(-3,0.2);
\draw (0,0.4)--(1,0.4); \draw (1,0.9)--(2,0.9);
\draw[thick,->] (3,1) -- (4,1);
\end{tikzpicture}  
\qquad
\begin{tikzpicture}
\draw (0,0)--(0,2.5) node[black,midway, xshift =0cm, yshift=1.5cm]{\footnotesize $x_1$};
\draw (1,0)--(1,2.5) node[black,midway, xshift =0cm, yshift=1.5cm]{\footnotesize $x_2$}; 
\draw (2,0)--(2,2.5) node[black,midway, xshift =-0.5cm, yshift=-1.7cm] {} node[black,midway, xshift =0cm, yshift=1.5cm]{\footnotesize $x_3$};
\draw [dashed,red] (1.5,0)--(1.5,2.5) node[black,midway, xshift =0cm, yshift=-1.5cm] {\footnotesize $n_{1}$};
\draw [dashed,blue] (0.75,0)--(0.75,2.5) node[black,midway, xshift =0cm, yshift=-1.7cm] {} node[black,midway, xshift =0cm, yshift=-1.5cm] {\footnotesize $n_{2}$};
\draw [dashed,purple] (0.25,0)--(0.25,2.5) node[black,midway, xshift =0cm, yshift=-1.7cm] {} node[black,midway, xshift =0cm, yshift=-1.5cm] {\footnotesize $n_{3}$};
\draw (0,0.4)--(1,0.4); \draw (1,0.9)--(2,0.9); 
\end{tikzpicture} 
\eea
From \eref{mainHS}, the Hilbert series of the Coulomb branch of $T^{(2,1,1)}_{(2,1,1)} (SU(4))$ reads 
\be \label{HT211211}
\begin{split}
&H[T^{(2,1,1)}_{(2,1,1)}(SU(4))](t; x_1, x_2, x_3; n_1, n_1, n_2, n_3)  \\
&= t^{2(n_1-n_3)} (1-t^2)^3 K_{(2,1,1)}(t; x_1, x_2, x_3) \widehat{Q}^{(n_1, n_1, n_2, n_3)}_{(2,1,1)} (t x_1, t^{-1} x_1, x_2, x_3; t)~,
\end{split}
\ee
where 
\be
\begin{split}
& K_{(2,1,1)}(t; x_1, x_2, x_3)   \\
&= \PE \Big[t^2 (2+ x_2 x_3^{-1} +x_3 x_2^{-1}) + t^3(x_1 x_2^{-1} +x_2 x_1^{-1}+ x_1 x_3^{-1}+x_3 x_1^{-1}) +t ^4  \Big]~.
\end{split}
\ee

\paragraph{The monopole formula.} The Coulomb branch Hilbert series of $T^{(2,1,1)}_{(2,1,1)} (SU(4))$ can be computed directly for any values of the magnetic fluxes from the monopole formula
\be \label{gT211211}
\begin{split}
&H_{\text{mon}}\left[T^{(2,1,1)}_{(2,1,1)}(SU(4)) \right](t; \vec x; n_3, n_2, n_1) \\
&= y_1^{n_2+n_3} y_2^{n_1} \sum_{u=-\infty}^\infty~ \sum_{v=-\infty}^\infty t^{|u-v|+|u-n_1|+|v-n_2|+|v-n_3|}  z_1^v  z_2^u (1-t^2)^{-2}~,
\end{split}
\ee
where $n_1$ is the background flux for the $U(1)$ flavor symmetry and $n_2, n_3$ are those for the $U(2)$ flavor symmetry.  The fugacities are subject to the constraint
\bea \label{relT211211} z_1 y_1^2 z_2 y_2=1~. \eea
Setting
\bea
z_1=x_2 x_1^{-1}~, \qquad  z_2=x_3 x_2^{-1}~, \qquad y_1 = x_1~, \qquad  y_2 = x_1 x_2~,
\eea
the constraint \eref{relT211211} becomes
\bea 
x_1^2 x_2 x_3=1~.
\eea
When the background fluxes are ordered, we find as expected that
\be
\begin{split}
& H_{\text{mon}}[T^{(2,1^2)}_{(2,1^2)}(SU(4))](t; x_1, x_2, x_3; n_3, n_2, n_1) \\
&= H[T^{(2,1,1)}_{(2,1,1)}(SU(4))](t; x_1, x_2, x_3; n_1, n_1, n_2, n_3), \qquad n_1\geq n_2 \geq n_3 \in \BZ~.
\end{split}
\ee

\paragraph{The baryonic generating function.} The baryonic generating function of the mirror $T^{(2,1,1)}_{(2,1,1)} (SU(4))$ is
\bea
& g[T^{(2,1,1)}_{(2,1,1)} (SU(4))](t; (x_3, x_2), x_1;  B_1, B_2) = \oint_{|b_1|=1}  \frac{{\rm d}b_1}{2 \pi i b_1^{1+B_1}} \oint_{|b_2|=1}  \frac{{\rm d}b_2}{2 \pi i b_2^{1+B_2}} \times \\ 
& \qquad \PE \Big[ \left \{ b_2 x_1^{-1} + b_2^{-1} x_1t + b_1 b_2^{-1} + b_2 b_1^{-1} + b_1^{-1} (x_2+x_3)+b_1(x_2^{-1}+x_3^{-1}) \right \} t- 2t^2 \Big]~. \nn
\eea
This can be equated with the monopole formula as follows:
\be
\begin{split}
&H_{\text{mon}}\left[T^{(2,1,1)}_{(2,1,1)}(SU(4)) \right](t; x_1, x_2, x_3; n_3, n_2, n_1)  \\
&= x_1^{n_2-n_1} g[T^{(2,1,1)}_{(2,1,1)} (SU(4))](t;( x_3, x_2), x_1;  n_1-n_2, n_2-n_3)~, \qquad x_1^2 x_2 x_3=1~.
\end{split}
\ee
The prefactor $x_1^{n_2-n_1}$ is due to the D3-brane indicated by the dotted red horizontal line in the diagram below.
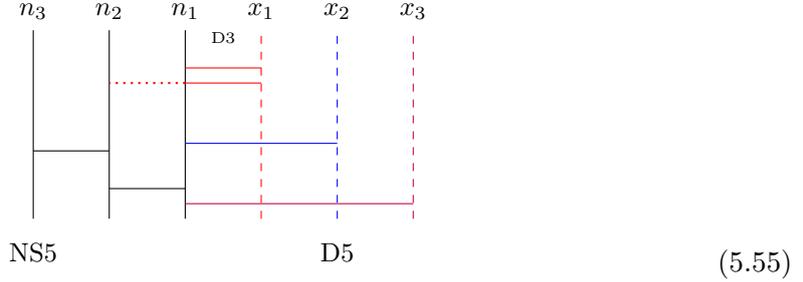
\bea
\begin{tikzpicture} 
\draw (0,0)--(0,2.5) node[black,midway, xshift =0cm, yshift=1.5cm]{\footnotesize $n_1$};
\draw (-1,0)--(-1,2.5) node[black,midway, xshift =0cm, yshift=1.5cm]{\footnotesize $n_2$}; 
\draw (-2,0)--(-2,2.5) node[black,midway, xshift =0cm, yshift=-1.7cm] {\footnotesize NS5} node[black,midway, xshift =0cm, yshift=1.5cm]{\footnotesize $n_3$};
\draw [dashed,red] (1,0)--(1,2.5) node[black,midway, xshift =0cm, yshift=1.5cm] {\footnotesize $x_{1}$};
\draw [dashed,blue] (2,0)--(2,2.5) node[black,midway, xshift =0cm, yshift=-1.7cm] {\footnotesize D5} node[black,midway, xshift =0cm, yshift=1.5cm] {\footnotesize $x_{2}$};
\draw [dashed,purple] (3,0)--(3,2.5)  node[black,midway, xshift =0cm, yshift=1.5cm] {\footnotesize $x_{3}$};
\draw [red] (0,2)--(1,2) node[black,midway, yshift=0.4cm] {\tiny D3}; \draw [red] (1,1.8)--(0,1.8); 
\draw [thick, dotted, red] (0,1.8)--(-1,1.8); 
\draw [blue] (0,1)--(2,1);
\draw [purple](0,0.2)--(3,0.2);
\draw (0,0.4)--(-1,0.4); \draw (-1,0.9)--(-2,0.9);
\end{tikzpicture}  
\eea

\section{Orthogonal and symplectic groups}\label{OSp}

In this section we discuss the generalisation of our results to other classical groups $SO(N)$ and $USp(2N)$. We consider the case of $USp'(2N)$ in Appendix \ref{sec:TsigrhoUSpP_Higgs}.  Life is much harder in other types.  
Complications with orthogonal and symplectic nilpotent orbits and issues with discrete groups make it difficult to state general results.
We present few examples for the case of  orthogonal and symplectic groups with low  rank, mostly for the Coulomb branch of  $T^{\vec\sigma}(G)$  and the  Higgs branch of $T_{\vec\sigma}(G)$ at zero external fluxes, leaving the general analysis for future work.  We  also provide a generalised Hall-Littlewood formula  and discuss its condition of validity.  A regular and  interesting pattern seems to emerge, which would be interesting to study in more detail.

A non-exhaustive list of differences with the unitary case is the following. 
\begin{itemize}
\item Many of the $T^{\vec \sigma}_{\vec \rho}(G)$ theories with orthogonal and symplectic groups are {\it bad} in the language of  \cite{Gaiotto:2008ak}, meaning that the dimension of some monopole operator computed using the ultraviolet  R-symmetry violates the unitary bound. The monopole formula written in terms of the Lagrangian data is ill-defined and divergent. Complications arise also in the Higgs branch where typically there is no complete Higgsing.  Such theories are supposed to flow to an interacting superconformal point in the IR but, unfortunately, we have no general description of it.  

\item Recall that, for $T^{\vec \sigma}_{\vec \rho}(G)$,    $\vec\sigma$  and  $\vec\rho$  are partitions of $G$   and  $G^\vee$, respectively. Partitions of $G$, as defined in section \ref{sec:quivers}, are in one-to-one
correspondence with the nilpotent orbits of the group $G$ and also with the homomorphisms ${\rm Lie}(SU(2))\rightarrow {\rm Lie}(G)$ \cite{collingwood1993nilpotent}. The Coulomb branch of $T^{\vec \sigma}_{\vec \rho}(G)$, equivalently the Higgs branch of $T_{\vec \sigma}^{\vec \rho}(G^\vee)$, as an algebraic variety, can be still written as an intersection of a  nilpotent orbit  with a Slodowy slice, but  this time  of the group $G^\vee$ \cite{Gaiotto:2008ak}
\bea\label{orbit}
\bar{O}_{\vec \sigma^\vee} \cap S_{\vec \rho}\, .
\eea
$\vec\rho$  is indeed a partition of $G^\vee$ and determines the Slodowy slice through the homomorphism $\vec\rho : {\rm Lie}(SU(2))\rightarrow {\rm Lie}(G^\vee)$. To determine the orbit itself we need a map $\vee : \sigma \rightarrow \sigma^\vee$ from partitions of $G$ to partitions of $G^\vee$. Such a map is well known in the mathematical literature \cite{spaltenstein,barbash} and we discuss it below. It has also explicitly appeared in the physical literature in the context of the $(2,0)$ theory compactified on Riemann surfaces with punctures \cite{Chacaltana:2012zy,Balasubramanian:2014jca}.

\item The quivers $T^{\vec \sigma}_{\vec \rho}(G)$ contains orthogonal gauge groups as nodes. The distinction between an $SO(N)$ and an $O(N)$ gauge group is important. Theories with $SO(N)$ gauge groups  have typically more BPS gauge invariant operators compared with the 
same theory with gauge group $O(N)$. 
We often have different interesting  quivers which we can write under the name of $T^{\vec \sigma}_{\vec \rho}(G)$ and that differ in the choice of $O/SO$ factors. Their Coulomb branch  is typically a covering of \eref{orbit}.  We discuss examples in section \eref{USp4}. 


\item There is a  Springer map  $T^*(G/P)\rightarrow \bar{O}_{\vec \sigma^\vee} $,  where the parabolic group $P$ is related to $\vec\sigma^\vee$  by yet another nontrivial map that we discuss below. In other types, the Springer map is not necessarily one-to-one. We can always write a generalisation of the Hall-Littewood function \eref{mainHS} for generic classical group $G$, which computes the Hilbert series of some covering of the moduli space \eref{orbit}.%
\footnote{ Further complications might arise for {\it non-normal} orbits, the first example of
which is the orbit $(3,2,2)$ of $SO(7)$; we discuss this case briefly in \tref{tab:paraC3}.} 
The Hall-Littlewood formula computes the Coulomb branch Hilbert series of the quiver $T^{\vec \sigma}_{\vec \rho}(G)$ for a specific choice of  $SO/O$ factors.

\end{itemize}

All these features are discussed in the explicit examples which are discussed below. We first discuss some general properties of partitions of a classical group $G$, we write a generalised Hall-Littlewood function and we present the  results for $USp(4)$ and $SO(5)$. Other groups of low rank are discussed in Appendix \ref{scan}.   We provide mirror pairs and  we test mirror symmetry by
evaluating the monopole formula in the Coulomb branch of $T^{\vec\sigma}(G)$ theories and the Molien-Weyl integral in the Higgs branch of $T_{\vec\sigma}(G)$. Whenever the $SO/O$ factors in the theory $T^{\vec \sigma}_{\vec \rho}(G)$ can be chosen in physically inequivalent ways, we put subscripts in order to differentiate the theories: we adopt the convention that the subscript  $({\rm I})$ refers to the theory which has  moduli space  \eref{orbit}; other subscripts correspond to various coverings of \eref{orbit}.

\subsection{Properties of partitions of a classical group $G$}\label{Jordan}

As discussed above, for any partition $\vec\sigma$ of $G$ we need to define two auxiliary objects. One is a partition $\vec\sigma^\vee$ of the dual group $G^\vee$.  The Coulomb branch of $T^{\vec \sigma}_{\vec \rho}(G)$ is expressed as an algebraic variety in terms of $\vec\sigma^\vee$ as in \eref{orbit}. The other is the Levi type $\vec\sigma^L$ of the parabolic group corresponding to  $\vec\sigma^\vee$, which is needed to write a resolution of the moduli space.

Recall that a partition $\vec\sigma$ of a classical group identifies the Jordan type of a nilpotent element of ${\rm Lie}(G)$ up to conjugacy.
The Jordan types of matrices in the Lie algebra of a classical group  are restricted  as follows \cite{collingwood1993nilpotent}. 
 A partition of type $A$ is just a non-increasing sequence of integers.
 A partition of type $B$ and $D$ is a non-increasing sequence of integers where all the even parts appear an even number of times. 
A partition of type $C$ is a non-increasing sequence of integers where all the odd parts appear an even number of times. For each  non-increasing sequence of integers  $\vec\sigma$  we can define a $B$-, $C$- and $D$- collapse as the maximal partition $\vec\tau\leq \vec\sigma$ of type  $B$, $C$ and $D$, respectively.

We then define the map \cite{spaltenstein,barbash,Chacaltana:2012zy,Balasubramanian:2014jca}
\bea\label{vee}
\vee \, :   \,\,  \vec\sigma \rightarrow \vec\sigma^\vee
\eea
as follows
\begin{itemize}
\item For $G=SU(N)$, $\vec\sigma^\vee$ is just the transpose of $\vec\sigma$
\item For $G=SO(2N+1)$,   $\vec\sigma^\vee$ is obtained by transposing $\vec\sigma$, subtracting $1$ from the last entry of the transpose partition and C-collapsing. 
\item For $G=USp(2N)$,   $\vec\sigma^\vee$ is obtained by transposing $\vec\sigma$, adding a new part equal to 1 to the transpose partition and B-collapsing. 
\item For $G=SO(2N)$,   $\vec\sigma^\vee$ is obtained by transposing $\vec\sigma$ and D-collapsing. 
\end{itemize}
$\vee$ is an inclusion-reversing map between the orbits of $G$ and $G^\vee$ which becomes one-to-one when restricted to the so-called  {\it special} orbits \cite{spaltenstein,barbash}. 

Consider now the nilpotent orbit associated with $\sigma^\vee$. We are interested in maps
\bea\label{resol}
T^*(G/P)\rightarrow \bar{O}_{\vec \sigma^\vee}
\eea
where $P$ is a parabolic subgroup of $G$ with ${\rm dim}  \,  \bar{O}_{\vec \sigma^\vee}= 2~ {\rm dim}\,  G/P$. Such $P$ is called a {\it polarization} of $\vec\sigma^\vee$.

For $G=SU(N)$, polarizations exist and are unique. As we already discussed in section \ref{loc}, we have $\vec\sigma^\vee=\vec\sigma^T$ and, up to conjugation, the associated parabolic group $P$   is the group of  upper triangular block matrices with blocks of size $\sigma_i$.
The algebra of  $P$  decomposes  as
\bea
{\rm Lie} P = {\rm Lie} L(\vec\sigma) \oplus n(P)
\eea
where $n(P)$ is the nil-radical of ${\rm Lie} P$ and $L(\vec\sigma)$ is the Levi subgroup. Here ${\rm Lie} L(\vec\sigma) =\prod_i U(\sigma_i)$  consists of block diagonal matrices of sizes $\sigma_i$. 
Notice that the original partition $\vec\sigma$ determines the structure of blocks in $P$ while the transpose partition $\vec\sigma^T$ determines the Jordan type of the nilpotent orbit. 

The conditions for the existence of a polarization of $\vec \sigma^\vee$ for other classical groups have been discussed in \cite{hesselink,fu,namikawa04}.  The structure of  parabolic groups is more complicated than for $SU(N)$, see for example section 2 of \cite{namikawa04}.  The Levi subgroup $L(\vec\sigma)$ is now of the form $L(\vec\sigma)=g \times\prod_{i=1}^n  U(l_i)^2$ where $g$ is a classical group of the same type ($B,C$ or $D$) as $G$ and each factor $U(l_i)$  appears an even number of times.
We denote with $\vec\sigma^L$ the set of numbers $(l_1,l_1,\cdots, l_n,l_n,p)$, where $p$ is the dimension of the block corresponding to $g$.  Notice that $\vec\sigma^L$ is not strictly a partition of $G$. There is yet another map from the  Levi type of a parabolic group $G$ to a partition of $G$ \cite{hesselink}
\bea\label{levimap}
S\, :  \vec\sigma^L \rightarrow \{ \, {\rm partitions \, of }  \, G\, \}
\eea
defined as follows. Let $\pi$ be the set  obtained by ordering the parts of $\vec\sigma^L$ in non-increasing order.  Define the set of indices
\bea\label{setI}
I ( \pi ) =  \{ j\in \mathbb{N} |~ j\neq n \, ({\rm mod} \,2),\,  \pi_j\,  {\rm even\, for\,  SO \, and\,  odd\,  for\,  USp}, \, \pi_j\geq \pi_{j+1}+2 \}
\eea 
where $n$ is the dimension of the matrix giving the classical representation of the group $G$. The map $S$ is defined by a series of moves. For all the indices $j$ belonging to  $I(\pi)$ we simultaneosly decrease by one unit the parts $\pi_j$ and increase by one unit the corresponding part $\pi_{j-1}$. $S(\vec\sigma^L)$ is then a partition
of $G$ (see for example, Theorem 2.7 in \cite{namikawa04}). Moreover, $\vec\sigma^L$ is the Levi type of a polarization of $\sigma^\vee$ if and only if 
$S(\vec\sigma^L ) =\vec\sigma^\vee$. All the polarizations of classical groups of small rank are explicitly tabulated in \cite{hesselink}, including all the cases considered in this paper.

In contrast with $G=SU(N)$, polarizations are not unique for other classical groups. Moreover the map \eref{resol} is not necessarily one-to-one and, therefore,  it is not necessarily a resolution. The degree of the map \eref{resol} can be explicitly computed  from \eref{setI}: it is $2^{|I(\pi)|}$ except for the special case of $\vec\sigma^L$ of $SO$ groups with no special part $p$ and all other parts odd where it is given by $2^{|I(\pi)|-1}$ (see Theorem 8 in \cite{namikawa04}). When the degree is one
the map \eref{resol} is a Springer resolution of $\bar{O}_{\vec \sigma^\vee}$.

\subsection{The generalised Hall-Littlewood formula for a classical group $G$}
 
The generalised Hall-Littlewood formula for a classical group  is expressed in terms of geometric data of  the dual group.
It is then convenient to write the generalised Hall-Littlewood formula for the dual group $G^\vee$. 

The Coulomb branch Hilbert series for $T^{\vec \sigma}_{\vec \rho}(G^\vee)$ is
\bea \label{mainHSG}
H[T^{\vec \sigma}_{\vec \rho}(G^\vee)] (t; \vec x; \vec n) =  t^{p_{\vec \sigma}(\vec n)} (1-t^2)^{r(G)} K_{\vec \rho}( \vec x;t) \widehat{Q}^{\vec n}_{\vec \sigma} ( \vec a_{\vec \rho} (t, \vec x); t)
\eea
where the notations are defined as follows.
\ben
\item ${\vec \sigma}$ is a partition of $G^\vee$ and $\vec \rho$ is a partition of $G$. 
\item Here $\widehat{Q}^{\vec n}_{\vec \sigma}$ is the modified dual Hall-Littlewood polynomial associated to a Lie group $G$, given by
\be\label{QG}
\begin{split}
&\widehat{Q}^{\vec n}_{\vec \sigma} (x_1, \ldots, x_r; t)  \\
&=\frac{1}{|W_{L(\vec \sigma)}|} \sum_{w \in W_G} {\vec x}^{w(\vec n)}
 {\purple \prod_{\vec \alpha \in \Delta_{\vec \sigma}} (1- {\vec x}^{-w(\vec \alpha)})(1-t^2 {\vec x}^{w(\vec \alpha)})} {\blue \prod_{\vec \gamma \in \Delta_+(G)} \frac{1-t^2 {\vec x}^{-w(\vec \gamma)}}{1-{\vec x}^{-w(\vec \gamma)}}}~,
\end{split}
\ee
where 
\bi
\item $\Delta_+(G)$ the set of positive roots of $G$.  
\item $L(\vec \sigma)$ denote the Levi subgroup associated with the partition $\vec \sigma^\vee$. $L(\vec\sigma)$ can be computed  as described above and is explicitly tabulated in \cite{hesselink} for all cases considered in this paper.
\item $\Delta_{\vec \sigma}$ is the set of positive roots in the diagonal blocks associated with $L( \vec \sigma)$.
\item $W_G$ denotes the Weyl group of $G$. 
\item $W_{L( \vec \sigma)}$ denotes the Weyl group of the Levi subgroup $L(\vec \sigma)$.
\item $\vec n= \sum_{i=1}^r n_i \vec{e}_i$, with $\{ \vec {\vec e}_1, \ldots, \vec {\vec e}_r\}$ the standard basis of the weight lattice and $r$ the rank of $G$. 
The Hall-Littlewood formula applies when $\vec n$ is a dominant weight of $G$ invariant under the action of $W_{L( \vec \sigma)}$.
\ei
\item The power $p_{\vec \sigma}(\vec n)$ is a linear function  of $\vec n$ that generalizes the expression \eref{powerG}. Examples are given in \tref{tab:paraC3}.
\item The argument $\vec a_{\vec \rho} (t, \vec x)$, which we shall henceforth abbreviate as $\vec a$, is determined by the following decomposition of the fundamental representation of $G$ to $G_{\vec \rho} \times {\vec \rho} (SU(2))$:
\bea \label{decompfundG}
\chi^G_{{\bf fund}} (\vec a_{\vec \rho} ) = \sum_{k}  \chi^{G_{\rho_k}}_{{\bf fund}}( \vec x_k) \chi^{SU(2)}_{[\rho_k -1]} (t)~,
\eea
where $G_{\rho_k}$ denotes a subgroup of $G_{\vec \rho}$ corresponding to the part $k$ of the partition $\vec \rho$.  Formula \eref{decompfund} determines $\vec a$ as a function of $t$ and $\{ \vec x_k \}$ as required.  Of course, there are many possible choices for $\vec a$; choices that are related to each other by outer automorphisms of $G$ are equivalent.
\item The prefactor $K_{\vec \rho}(\vec x; t)$ is independent of ${\vec n}$ and can be determined as follows.  The embedding specified by $\vec \rho$ induces the decomposition 
\begin{equation}
\chi^G_{\bf Adj} (\vec a) = \sum_{j \in \frac{1}{2}\bZ_{\ge 0}}  \chi^{G_{\vec \rho}}_{R_j}(\vec x_j)  \chi^{SU(2)}_{[2j]}(t)~, \label{decompadjG} 
\end{equation} 
where $\vec a$ on the left hand side is the same $\vec a$ as in \eref{decompfund}.  Each product in the previous formula gives rise to a term in the plethystic exponential 
\begin{equation}
K_{\vec \rho}(\vec x; t)=\PE \left[t^2 \sum_{j \in \frac{1}{2}\bZ_{\ge 0}}  t^{2j} \chi^{G_{\vec \rho}}_{R_j}({\vec x}_j )\right].  \label{KG}
\end{equation}
\een

\subsection{$T^{\vec \sigma}(SO(5))$ and $T^{\vec \sigma}(USp(4))$}\label{USp4}

We now consider the case of the theories $T^{\vec \sigma_1}(SO(5))$ and $T^{\vec \sigma_2}(USp(4))$ where many of the differences with the unitary case are
manifest. Here ${\vec \sigma_1}$ is a $B_2$-partition and ${\vec \sigma_2}$ is a $C_2$-partition. The possible partitions and the mirror pairs are summarized in \tref{tab:paraC2}.
Important data associated with each partition  and the Hilbert series for the Coulomb branch of the $T^{\vec\sigma}(G)$ theory  are  given in  \tref{tab:paraB2}.  

\begin{table}[H]
\hspace{-1cm}
{\footnotesize
\begin{tabular}{|c|c|c||c|c|c|}
\hline
$B_2$-part. ${\vec \sigma}$ & Quiver of $T^{\vec \sigma}(SO(5))$ &  $T_{\vec \sigma}(USp(4))$ & $C_2$-part. ${\vec \sigma}$ & Quiver of $T^{\vec \sigma}(USp(4))$ &  $T_{\vec \sigma}(SO(5))$ \\

\hline \hline
$(1^5)$ &$\node{}{O(1)}-\Node{}{2^*}-\node{}{O(3)}-\Node{\ver{}{5}}{4^*}$ &  $\node{}{2}-\Node{}{2}-\node{\Ver{}{4}}{4}$ & $(1^4)$ & $\node{}{2}-\Node{}{2}-\node{\Ver{}{4}}{4}$  & $\node{}{O(1)}-\Node{}{2^*}-\node{}{O(3)}-\Node{\ver{}{5}}{4^*}$ \\
\hline \hline
$(2^2,1)$ & $\node{}{O(1)}-\Node{}{2^*}-\node{\Ver{}{2}}{O(3)}-\Node{\ver{}{1}}{2^*}$  & $\node{}{4^*}-\sqNode{}{4}$ & $(2,1^2)$ & $\node{}{2}-\Node{\ver{}{1}}{2}-\node{\Ver{}{2}}{3}$ & $\node{}{3^*}-\Node{\ver{}{5}}{2}$ \\
\hline 
 $(3,1^2)$ & $\node{}{O(1)}-\Node{\ver{}{1}}{2^*}-\node{}{2}-\Node{\ver{}{2}}{2^*}\, \, $$_{({\rm II})}$ &  $\node{}{2}-\sqNode{}{4}\, \, $$_{({\rm II})}$ & $(2^2)$ & $\node{}{2}-\Node{\ver{}{2}}{2}-\node{}{2}\, \, $$_{({\rm II})}$ & \begin{tabular}[c]{@{}c@{}}  $\node{}{1}-\Node{\ver{}{5}}{2}\, \, $$_{({\rm II})}$ \\
 equiv. to~ $\Node{\ver{}{6}}{2}$ \end{tabular} \\
\hline \hline
\end{tabular}}
\caption{{\footnotesize Quiver diagrams for $T^{\vec \sigma}(SO(5))$ and their mirror theories. The asterisk $*$ indicates the gauge group that renders the quiver a `bad' theory. Each black node labeled by $M$ denotes a $USp(M)$ group, each gray node labeled by $N$ denotes an $SO(N)$ group and for an orthogonal group $O(N)$ is spelt out explicitly. Whenever a gauge group $O(N)$ is indicated by $*$, such a gauge group can be taken as $O(N)$ or $SO(N)$ without changing the Higgs branch; see Appendix \ref{sec:orthoSeiberg}. Subscripts near the quiver allow to distinguish theories with different  choices of $SO/O$ factors.  }}
\label{tab:paraC2}
\end{table}%

\begin{table}[H]
\hspace{-1cm}
{\scriptsize
\begin{tabular}{| l | l | l || l | l | l || l | c | l | l | }
\hline
$T^{\vec \sigma_1}(SO(5))$& ${\vec \sigma}_1^\vee$ & $ \sigma_1^L $: Levi subgroup &$T^{\vec \sigma_2}(USp(4))$ & ${\vec \sigma}_2^\vee$ & $ \sigma_2^L $: Levi subgroup    & Unrefined  Coulomb branch 
\\
  && $\Delta_{\vec \sigma_1} \subset \Delta_+(USp(4))$ & & &   $\Delta_{\vec \sigma_2} \subset \Delta_+(SO(5))$  & HS of for $T^{\vec\sigma} (G)$ from \eref{mainHSG}   
 \\
\hline
$(1^5)$ &$(4)$ & $(1^2, 1^2):U(1)^2$ &$(1^4)$& $(5)$  & $(1^2, 1^2,1): U(1)^2$  & $\frac{\left(1+t^2\right)^2 \left(1+t^4\right)}{\left(1-t^2\right)^8}$ 
\\
             &         & $\Delta_{(1^5)}=\emptyset$ &          &                           & $\Delta_{(1^4)}=\emptyset$   	&$=1 + 10 t^2 + 54 t^4 + 210 t^6 +\ldots$     
              \\
\hline \hline
 $(2,2,1)$& $(2,2)$ & $(2^2): U(2)$   &$(2,1,1)$ & $(3,1,1)$ & $(1^2,3): U(1) \times SO(3)$   	&	$\frac{(1 + t^2) (1 + 3 t^2 + t^4)}{(1 - t^2)^6}$ 
 \\
&  & $\{ {\vec e}_1-{\vec e}_2\}$ &  &   &$\{ \e_2 \}$ 	  & $=1+10 t^2+49 t^4+165 t^6+\ldots$ 
 \\
\hline
$(3,1,1)_{{\rm II}}$ &$(2,2)$ &$(1^2,2):U(1) \times USp(2)$ & $(2,2)_{{\rm II}}$  & $(3,1,1)$& $(2^2,1):U(2)$        & $\frac{(1 + t^2) (1 + 8 t^2 + t^4)}{(1 - t^2)^6}$ 
\\
                  &&$\{2 \e_2 \}$&       & &	$\{ {\vec e}_1-{\vec e}_2\}$			& $= 1+15 t^2+84 t^4+300 t^6+ \ldots$ 
                  \\
\hline \hline
\end{tabular}
}
\caption{For $SO(5)$, the positive roots are ${\vec e}_i - {\vec e}_j$, ${\vec e}_i + {\vec e}_j$ and ${\vec e}_i$, with $1\leq i < j \leq 2$.  For $USp(4)$, the positive roots are ${\vec e}_i - {\vec e}_j$, ${\vec e}_i + {\vec e}_j$ and $2{\vec e}_i$, with $1\leq i < j \leq 2$.   
The $C_2$-partition $\vec \sigma_1^\vee$ can be obtained from $\vec \sigma_1$ by transposing, deleting a box in the last tuple and C-collapsing. The $B_2$-partition $\vec \sigma_2^\vee$ can be obtained from $\vec \sigma_2$ by adding a box, transposing and B-collapsing. The subscript below  ${\vec \sigma}$ indicates that we are using the theory $T^{\vec\sigma}(G)_{{\rm II}}$; we are correspondingly using a polarization where the  the degree of map \eref{resol}, as computed using \eref{setI},  is  $2$. The HS given above are computed from the Hall-Littlewood formula. For comparison, the HS for the Higgs branch of $T_{(1,1,1,1)}(SO(5))$  and $T_{(1,1,1,1,1)}(USp(4))$ were explicitly computed in (4.34), (4.36) of \cite{Cremonesi:2014kwa} and the HS  for the Higgs branch of the quiver $[USp(4)]-SO(2)$ for the theory $T_{(3,1,1)}(USp(4))_{({\rm II})}$ was computed in  (D.4) of \cite{Cremonesi:2014kwa}.}
\label{tab:paraB2}
\end{table}%

We make some general observations about these tables.

\begin{itemize}
\item
We expect  that the $T^{\vec \sigma}_{\vec \rho}(G)$ theories for isomorphic groups should be equivalent with an appropriate mapping between the partitions, even if the quivers are different.  We  verify this explicitly at the level of Hilbert series for $SO(5)\sim USp(4)$. In the tables we report in parallel the results  for $SO(5)$ and $USp(4)$ and the correspondence between partitions.

\item 
In \tref{tab:paraC2}  we have taken some of orthogonal groups to be $SO$. The distinction between an $SO(N)$ and an $O(N)$ gauge group is important. Theories with $SO(N)$ gauge groups  have typically more BPS gauge invariant operators compared with the 
same theory with gauge group $O(N)$.  In the Higgs branch, an $SO(N)$ gauge symmetry allows for baryonic  operators and extra mesonic operators which are odd under parity.  In the Coulomb branch, the magnetic lattice of $SO(N)$ is different from that of $O(N)$.  As a general rule, we have only considered quivers without baryonic operators.   Even with this restriction, we have often different interesting  quivers which we can write under the name of $T^{\vec \sigma}_{\vec \rho}(G)$ and  differ in the choice of $O/SO$ factors. We use the notation  $T^{\vec \sigma}_{\vec\rho}(G)_{({\rm I})}$ and $T^{\vec \sigma}_{\vec\rho}(G)_{({\rm II})}$ to differentiate these theories. In our conventions, $T^{\vec \sigma}_{\vec\rho}(G)_{({\rm I})}$ is a quiver with Coulomb branch moduli space equal to \eref{orbit}. The Coulomb branch of $T^{\vec \sigma}_{\vec\rho}(G)_{({\rm II})}$ is instead  a double cover of \eref{orbit}. 

\item 
In \tref{tab:paraB2} we present the Hilbert series for the Coulomb branch of the $T^{\vec\sigma} (G)$ theory based on the Hall-Littlewood formula \eref{mainHSG}. 
The last two rows in \tref{tab:paraB2} contain the  two non-trivial partitions $\vec\sigma$ of $SO(5)$ or $USp(4)$. We see that they both correspond to the same ${\vec\sigma}^\vee$. The Hilbert series in the last two rows in \tref{tab:paraB2}  correspond to two different polarizations of ${\vec\sigma}^\vee$, one of degree one and one of degree two.  We have chosen the $O/SO$ factors in the quivers \tref{tab:paraC2} in order to match the two different Hilbert series. This involves choosing the theory $T^{\vec \sigma}(G)_{({\rm II})}$ in some cases.

\item
We have explicitly computed the Hilbert series for the Higgs branch (using the Molien-Weyl integral)   and the Coulomb branch (using the monopole formula) of all the quivers given in the tables whenever they are well defined.  The result obviously coincides with that given in \tref{tab:paraB2} based on the Hall-Littlewood formula. The monopole formula fails when the quiver is bad. In particular,  we can  only compute the monopole formula the Coulomb branch Hilbert series of $T^{\vec \sigma}(USp(4))$, since in general the $T^{\vec \sigma}(SO(5))$ theories are bad. Recall  that, in general, a linear quiver theory is {`bad'} if it contains one of the following items:
\bi
\item $SU(N_c)$ gauge group with $N_f < 2N_c-1$;
\item $SO(N_c)$ gauge group with $N_f < N_c-1$;
\item $USp(2N_c)$ gauge group with $N_f < 2N_c+1$.
\ei
In the case of a bad quiver, there is no complete Higgsing along the Higgs branch and the F-flat moduli space is not a complete intersection. As a result the Hilbert series needs to be computed using other techniques, for example using  {\tt Macaulay2} \cite{M2}.%
\footnote{For example, for $T_{(2,1,1)}(SO(5)): [SO(5)]-(USp(2))-(SO(3))$, the gauge group $SO(3)$ is not completely broken on the hypermultiplet moduli space; rather, at a generic point, $SO(3)$ is broken to $SO(2)$.  Indeed, {\tt Macaulay2} reveals that 
Higgs branch HS of [USp(2)]-(SO(3))  =  Higgs branch HS of ~[USp(2)]-(O(1)) = HS of~$\BC^2/\BZ_2$. Hence, upon gluing these quivers with $[SO(5)]-[USp(2)]$ via the $USp(2)$ group, we reach the conclusion that
Higgs branch HS of $T_{(2,1,1)}(SO(5)): [SO(5)]-(USp(2))-(SO(3))$   = Higgs branch HS of $[SO(5)]-(USp(2))-(O(1))$, 
with the unrefined Hilbert series presented in \tref{tab:paraB2}.}
\end{itemize}

To fully appreciate the differences between the quivers and the subtlelties about  $O/SO$ factors we need  a longer discussion which is given in the next subsections.




\subsubsection{Relations between quivers with $O$ and $SO$ gauge groups: the Higgs branch of $T_{\vec \sigma}(G)$}\label{SO/O1}

The Higgs branch of the theories $T_{\vec \sigma}(SO(N))$ when all orthogonal gauge groups are of $O$ type (and not $SO$) was explicitly shown to be the nilpotent orbit $O_{\vec\sigma^\vee}$ in \cite{Benini:2009mz}. The argument can be generalized to $T_{\vec \sigma}(USp(N))$ \cite{Chacaltana:2012zy}.   One can see with methods similar to those in Appendix \ref{sec:orthoSeiberg} that the
presence of  $SO$ groups in \tref{tab:paraC2}, \tref{tab:paraZ2} and  \tref{tab:paraZ22} does not introduce extra baryonic operators in the chiral ring and in most of the cases does not affect the Higgs branch. 

For some particular theories, ungauging the parity in a group $O$ might introduce extra mesonic operators.  
Consider for example the theory $T_{(3,1,1)}(USp(4))$.  We have two choices for the corresponding quiver,  $[USp(4)]-O(2)$, which we call $T_{(3,1,1)}(USp(4))_{({\rm I})}$ and  $[USp(4)]-SO(2)$, which we call $T_{(3,1,1)}(USp(4))_{({\rm II})}$. As discussed in details in Appendix \ref{sec:orthoSeiberg},  the Higgs branch of $[USp(4)]-O(2)$ is the nilpotent orbit  $(2,2)$ of $USp(4)$. The Higgs branch of $[USp(4)]-SO(2)$ is obviously a two-fold covering of the nilpotent orbit  $(2,2)$.  This is an example of the theories that we call 
$T_{\vec \sigma}(G)_{({\rm I})}$ and $T_{\vec \sigma}(G)_{({\rm II})}$, with $T_{\vec \sigma}(G)_{({\rm I})}$ giving the hyperK\"ahler quotient  description of a nilpotent orbit and $T_{\vec \sigma}(G)_{({\rm II})}$ a covering of it.  

Let us also notice that hyperK\"ahler quotient constructions
for all the nilpotent orbits of all classical groups have been given in the  mathematical literature a long time ago \cite{kobak}. The corresponding hyperK\"ahler quotient is sometimes different from ours, allowing for $USp$ groups with odd number of half-hypermultiplets. The quiver corresponding to the $T_{\vec \sigma}(G)$ theories have always
an even number of half-hypermultiplets for any $USp$ gauge group in order to cancel parity anomalies and provide a somehow non-minimal (in terms of groups in the quiver)  hyperK\"ahler quotient construction of the nilpotent orbits of $G$. One can use the result in Appendix \ref{sec:orthoSeiberg} to show that the various different formulations for the hyperK\"ahler quotient construction of the same nilpotent orbit are equivalent.

\subsubsection{Relations between quivers with $O$ and $SO$ gauge groups: the Coulomb branch of $T^{\vec \sigma}(USp(4))$}\label{SO/O2}


In this section we focus for simplicity on the theories $T^{\vec\sigma}(USp(4))$.
A parallel analysis can be done for $T^{\vec\sigma}(SO(5))$.

Part of the story about the theories that we have called $T^{\vec \sigma}(G)_{({\rm I})}$ and $T^{\vec \sigma}(G)_{({\rm II})}$ is related to the fact the map $\vee$ is not injective. The $C_2$-partitions $\vec \sigma= (2,1,1)$ and $\vec \sigma = (2,2)$ correspond both to the orbit  $\vec \sigma^\vee =(3,1,1)$. We should expect that the theories $T^{ (2,1,1)}(USp(4))$ and $T^{ (2,2)}(USp(4))$ describe the same physics although they have different quivers.
We now discuss in what sense this is true.  To understand the following discussion, it is important to notice that  $\vec \sigma^\vee$ has two different polarizations,
of Levi type $\vec\sigma^L=(1^2,3)$ and $(2^2,1)$ corresponding to maps \eref{resol} of degree one and two respectively.
\bi
\item The quivers for $T^{(2,1,1)}(USp(4))$ and for its mirror  $T_{(2,1,1)}(SO(5))$ are given in the second row of  \tref{tab:paraC2}. The Coulomb branch 
of $T^{(2,1,1)}(USp(4))$  is the nilpotent orbit $O_{(3,1,1)}$ of $SO(5)$ and its Hilbert series is given by  the Hall-Littlewood formula \eref{mainHSG} for the choice
of parabolic group associated to $\vec\sigma^L=(1^2,3)$ which gives rise to a smooth resolution $T^*(G/P)$ of the orbit $O_{\vec\sigma^\vee}$.
\item The quivers for $T^{(2,2)}(USp(4))_{({\rm II})}$ and  its mirror  $T_{(2,2)}(SO(5))_{({\rm II})}$ are given in the third row of  \tref{tab:paraC2}. 
The Coulomb branch  of $T^{(2,2)}(USp(4))_{({\rm II})}$  is a double cover of the nilpotent orbit $O_{(3,1,1)}$ of $SO(5)$ and its Hilbert series is given by  the Hall-Littlewood formula \eref{mainHSG} for the choice of parabolic group $P$ corresponding to $\vec\sigma^L=(2^2,1)$.
If we further gauge a $\BZ_2$ parity in both the quiver for $T^{(2,2)}(USp(4))_{({\rm II})}$ and $T_{(2,2)}(SO(5))_{({\rm II})}$ we obtain the pair of mirror quivers $T^{(2,2)}(USp(4))_{({\rm I})}$ and   $T_{(2,2)}(SO(5))_{({\rm I})}$ given in  \tref{tab:paraZ22}. The effect of the gauging on the moduli space is a $\BZ_2$ quotient  which now makes it equivalent to the  moduli space of  $T^{(2,1,1)}(USp(4))$ and  $T_{(2,1,1)}(SO(5))$ respectively.  
\ei

\begin{table}[H]
\begin{center}
{\footnotesize
\begin{tabular}{|c|c||c|c|}
\hline
  $T^{(2,1,1)}(USp(4))$ &$T^{(2,2)}(USp(4))_{({\rm I})}$   & $T_{(2,1,1)}(SO(5))$ & $T_{(2,2)}(SO(5))_{({\rm I})}$  \\
\hline
$\node{}{2}-\Node{\ver{}{1}}{2}-\node{\Ver{}{2}}{3}$  & $\node{}{2}-\Node{\ver{}{2}}{2}-\node{}{O(2)}\, \, $$_{({\rm I})}$  & $\node{}{3^*}-\Node{\ver{}{5}}{2}$ & $\node{}{O(1)}-\Node{\ver{}{5}}{2}\, \, $$_{({\rm I})}$\\
\hline
\end{tabular}}

\end{center}
\caption{$T^{\vec \sigma}(USp(4))/T_{\vec \sigma}(SO(5))$ theories with the same Coulomb/Higgs branches for vanishing background charges. 
}
\label{tab:paraZ22}
\end{table}

Let us check explicitly that  the Coulomb branch of $T^{(2,2)}(USp(4))_{({\rm II})}$ is a double cover of the  Coulomb branch of $T^{(2,1,1)}(USp(4))$ at the level of Hilbert series with vanishing background fluxes. This can be seen by gauging the parity in an $SO(2)$ factor in the quiver $T^{(2,2)}(USp(4))_{({\rm II})}$. The result is the quiver $T^{(2,2)}(USp(4))_{({\rm I})}$ given in  \tref{tab:paraZ22}, which is different from the quiver of $T^{(2,1,1)}(USp(4))$ but it is has the same Coulomb branch. The two quivers indeed differ by replacing $SO(3)$ gauge group with $O(2)$ together with shifts in flavor symmetries. We argue now that this move does not change the monopole formula. The reason is the following. As we discuss in appendix \ref{orth} the weight lattice and the classical $P$ factors for $O(2)$ are the same as those for $SO(3)$. Moreover, the dimension 
of the monopole operator, as a function of the dynamical magnetic charges (but for vanishing background magnetic charges), does not change because the shift in flavors compensates the contribution of the vector multiplet that has been changed. In this way, the monopole formula for the two theories is the same.   

To illustrate this, we give the formulae for the dimension of the monopole operators in the two quivers below:
\be \label{examplecancel1}
\begin{split}
\Delta  \left[ \node{}{2}-\Node{\ver{}{1}}{2}-\node{\Ver{}{2}}{3} ~\right] &= |u-a|+|u+a|+ \frac{1}{2} \sum_{s=0}^1 \sum_{\beta=-1}^1 \left |(-1)^{s} a- \beta b \right |+ \frac{1}{2}(|a| +|-a|) \\
& {\blue +\frac{1}{2} \sum_{s=0}^1 \sum_{\beta=-1}^1 |\beta b-(-1)^{s} n|} -2|2a| {\blue -2|b|}~,
\end{split}
\ee
where $u$, $a$, $b$ are topological charges for $SO(2)$, $USp(2)$ and $SO(3)$ gauge groups, and $n$ is the background monopole flux for the global symmetry $USp(2)$.
\bea  \label{examplecancel2}
\Delta \left[ \node{}{2}-\Node{\ver{}{2}}{2}-\node{}{O(2)} \right] &= |u-a|+|u+a| +|a-b|+|-a-b|+ |a-n|+|-a-n| -2|2a|~,
\eea
where $u$, $a$, $b$ are topological charges for $SO(2)$, $USp(2)$ and $O(2)$ gauge groups, and $n$ is the background monopole flux for the global symmetry $SO(2)$.  Observe that when $n=0$, the two blue terms in \eref{examplecancel1} cancel with each other and the equality between \eref{examplecancel1} and \eref{examplecancel2} can be established. It would be interesting to understand better the role of background fluxes in these theories.

The mirror statement is that the Higgs branch of $T_{(2,2)}(SO(5))_{({\rm II})}$ is a $\BZ_2$ covering of the  Higgs branch of $T_{(2,1,1)}(SO(5))$. This again can be seen by gauging the parity in $T_{(2,2)}(SO(5))_{({\rm II})}$. The result is the quiver $T_{(2,2)}(SO(5))_{({\rm I})}$ given in  \tref{tab:paraZ22}.
The fact that the Higgs branch of $T_{(2,2)}(SO(5))_{({\rm I})}$ coincides with the Higgs branch of $T_{(2,1,1)}(SO(5))$ is proven in Appendix \ref{sec:orthoSeiberg}. 

\begin{table}[H]
\begin{center}
{\footnotesize
\begin{tabular}{|c|c||c|c|}
\hline
 $T^{(2,2,1)}(SO(5))$ & $T^{(3,1,1)}(SO(5))_{({\rm I})}$ & $T_{(2,2,1)}(USp(4))$ & $T_{(3,1,1)}(USp(4))_{({\rm I})}$ \\
\hline
$\node{}{O(1)}-\Node{}{2^*}-\node{\Ver{}{2}}{O(3)}-\Node{\ver{}{1}}{2^*}$  & $\node{}{O(1)}-\Node{\ver{}{1}}{2^*}-\node{}{O(2)}-\Node{\ver{}{2}}{2^*}\, \, $$_{({\rm I})}$  & $\node{}{4^*}-\sqNode{}{4}$ & $\node{}{O(2)}-\sqNode{}{4}\, \, $$_{({\rm I})}$\\
\hline
\end{tabular}}

\end{center}
\caption{$T^{\vec \sigma}(SO(5))/T_{\vec \sigma}(USp(4))$ theories  with the same Coulomb/Higgs branches for vanishing background charges.
The Hilbert series for the Higgs branch of the quiver $[USp(4)]-O(2)$ was given in    (D.10) of \cite{Cremonesi:2014kwa} and coincides with the second row of  \tref{tab:paraB2}.
The equality between  the Higgs branches of the theories $T_{\vec\rho} (USp(4))$ is  proved in Appendix \ref{sec:orthoSeiberg}.}
\label{tab:paraZ2}
\end{table}

A similar analysis applies to $SO(5)$ and the relevant quivers are given in \tref{tab:paraZ2}. In all other examples considered in Appendix \ref{scan}, whenever two partitions $\vec\sigma_1$ and $\vec\sigma_2$ correspond to the same $\vec\sigma^\vee$, we have two different quivers.  One has moduli space   \eref{orbit}. The other comes  in two versions, related by ungauging the parity in one of the $O(N)$ gauge groups,  with moduli space  \eref{orbit} or a covering of it, respectively.  It is interesting to notice that, in the mathematical literature, the map $\vee$ comes equipped with a {\it local system}, typically a set of discrete symmetries, which is non-trivial precisely when $\vee$ is not injective \cite{MR1927953,MR1976456}. It would be interesting to see if there is a relation with our results.

\acknowledgments{ 
We thank Yuji Tachikawa and all the contributors to the MathOverflow discussion \cite{MathOverflow} that sparked this work, Ben Webster for useful discussions, and the following institutes and workshops for hospitality and partial support: KITP Program on New Methods in Nonperturbative Quantum Field Theory (AH and NM); University of Texas at Austin (NM); Perimeter Institute (NM); LPTHE, Universit\'e Pierre et Marie Curie (AZ); Workshop on Gauge theories: quivers, tilings and Calabi-Yaus at International Centre for Mathematical Sciences, Edinburgh (SC, AH, NM and AZ); Workshop on Localisation \& Gauge/Gravity, King's College London (SC, AH and NM); Imperial College London and Queen Mary University of London (NM); Carg\`ese Summer Institute 2014 (NM); Exact Quantum Fields and the Structure of M-theory, University of Crete (NM and AZ); the Simons Center for Geometry and Physics and the 2014 Workshop (AH and NM); the CERN-Korea Theory Collaboration funded by National Research Foundation (Korea) and the workshop Exact Results in SUSY Gauge Theories in Various Dimensions at CERN (SC and NM); the XLIV\`eme Institut d'\'Et\'e at ENS Paris (SC); Journ\'ees de Physique Math\'ematique Lyon 2014 (NM); University of Milano-Bicocca (NM); the IV Workshop on Geometric Correspondences of Gauge Theories at SISSA (SC and NM); the 6th Bethe Center Workshop on Topological Strings and Applications (NM).  AZ is  supported in part by INFN and  by the MIUR-FIRB grant RBFR10QS5J ``String Theory and Fundamental Interactions''. }

\appendix
\section{Monopole formula for orthogonal and special orthogonal gauge groups}  \label{orth}
We state the following general observation
\begin{quote}
{\bf The $P$-factor and the GNO lattice of magnetic charges (in the summation of the monopole formula) of an $O(2k)$ group are the equal to those of $SO(2k+1)$ group.}\footnote{On the other hand, the weights and the roots of $O(2k)$ gauge group appearing in the dimension formula are the same as those of $SO(2k)$ gauge group.}
\end{quote}
We demonstrate this with an example. Let us compare the following data for $SO(4)$, $O(4)$ and $SO(5)$ groups.   

\bi
\item For $SO(4)$ group, the magnetic fluxes is $(m_1, m_2)$ with $m_1 \geq |m_2| \geq 0$ and $-\infty < m_2 < \infty$.  The residual gauge symmetries in the presence of various magnetic charges are presented in \tref{tab:PSO4}. 
\begin{table}[H]
\begin{center}
\begin{tabular}{|c|c|c|}
\hline
Monopole fluxes &  Residual gauge symmetry & $P$-factor\\
\hline
$(0,0)$ & $SO(4)$ & $\frac{1}{(1-t^4)^2}$ \\
$(1, \pm1)$ & $U(2)$ & $\frac{1}{(1-t^2)(1-t^4)}$ \\
$(1,0)$ & $U(1) \times SO(2)$ & $\frac{1}{(1-t^2)^2}$ \\
$(2, 1)$& $U(1)^2$ & $\frac{1}{(1-t^2)^2}$ \\
\hline
\end{tabular}
\end{center}
\caption{Data for $SO(4)$ group}
\label{tab:PSO4}
\end{table}%
For $SO(4)$ group, there are two independent Casimir invariants, namely $\delta^{jk} \delta^{il} \phi_{ij} \phi_{kl}$ and $\epsilon^{ijkl} \phi_{ij} \phi_{kl}$, with $i,j,k,l =1,\ldots, 4$.
\item  Let us now go from $SO(4)$ to $O(4)$.  The parity symmetry $\BZ_2$ identifies the monopole fluxes $(n, m)$ and $(n, -m)$. Therefore, the monopole fluxes associated with $O(4)$ becomes $(m_1, m_2)$ with $m_1 \geq m_2 \geq 0$. We emphase that $m_2 \geq 0$ instead of $-\infty < m_2 < \infty$.
\begin{table}[H]
\begin{center}
\begin{tabular}{|c|c|c|}
\hline
Monopole fluxes &  Residual gauge symmetry & $P$-factor\\
\hline
$(0,0)$ & $O(4)$ & $\frac{1}{(1-t^4)(1-t^8)}$ \\
$(1,1)$ & $U(2)$ & $\frac{1}{(1-t^2)(1-t^4)}$ \\
$(1, 0)$ & $U(1) \times O(2)$ & $\frac{1}{(1-t^2)(1-t^4)}$ \\
$(2, 1)$& $U(1)^2$ & $\frac{1}{(1-t^2)^2}$ \\
\hline
\end{tabular}
\end{center}
\caption{Data for $O(4)$ group}
\label{tab:PO4}
\end{table}%
For $O(4)$, the invariants involving epsilon tensors are projected out by the parity and the Casimir invariant at order 4 becomes another independent one; hence we have 
\bea
P_{O(4)} (t; 0,0) = \frac{1}{(1-t^4)(1-t^8)}~.
\eea
For $U(1) \times O(2)$, the $P$-factor receives two contribution: one from $U(1)$, namely $1/(1-t^2)$, and the other from $O(2)$, namely $1/(1-t^4)$.  Thus,
\bea
P_{O(4)} (t; m,0) = \frac{1}{(1-t^2)(1-t^4)}~, \qquad m >0~.
\eea
\item Let us now compare $O(4)$ with $SO(5)$.  The monopole fluxes for $SO(5)$ gauge group take the form $(m_1, m_2)$ with $m_1 \geq m_2 \geq 0$.  The relevant data for $SO(5)$ are tabulated in \tref{tab:PSO5}.
\begin{table}[H]
\begin{center}
\begin{tabular}{|c|c|c|}
\hline
Monopole fluxes &  Residual gauge symmetry & $P$-factor\\
\hline
$(0,0)$ & $SO(5)$ & $\frac{1}{(1-t^4)(1-t^8)}$ \\
$(1, 1)$ & $U(2)$ & $\frac{1}{(1-t^2)(1-t^4)}$ \\
$(1, 0)$ & $U(1) \times SO(3)$ & $\frac{1}{(1-t^2)(1-t^4)}$ \\
$(2, 1)$& $U(1)^2$ & $\frac{1}{(1-t^2)^2}$ \\
\hline
\end{tabular}
\end{center}
\caption{Data for $O(5)$ group}
\label{tab:PSO5}
\end{table}%
\ei
This example generalizes to all orthogonal groups.

\section{Different theories with the same Higgs branch}\label{seiberg}

In this section we analyze the Higgs branch of various ${\cal N}=4$ theories with single gauge group in three dimensions, focussing on pairs of theories that have the same Higgs branch. 
The results for orthogonal groups are useful to understand the Higgs branch of the $T_{\vec \sigma}(G)$ theories, the choice of $SO/O$ factors and the equivalence of Higgs branches of different theories.

\subsection{Unitary gauge groups}

We analyze a $U(N_c)$ theory with $N_f$ flavors in terms of nilpotent orbits.  The $F$-terms relevant to the Higgs branch of this theory are
\bea
 \tQ^b_{~i} Q^i_{~a} = 0~, \label{FtermsU34flv}
\eea
where $a, b=1,\cdots , N_c$ are $U(N_c)$ gauge indices and $i,j=1,\cdots , N_f$ are $U(N_f)$ flavor indices.
The Higgs branch of this theory is generated by the mesons 
\bea
M^i_{~j} = Q^i_{~a} \tQ^a_{~j}~
\eea
satisfying the relations \cite{Argyres:1996eh}
\bea
M^i_{~i} =0~, \qquad {\rm rank}~ M \leq \min \left\{N_c,\left[\frac{N_f}{2}\right]\right\} ~, \qquad M^2=0~.
\eea
These conditions imply that the Higgs branch of this theory corresponds to a nilpotent orbit of $SU(N_f)$. The orbit is specified by the Jordan
type of $M$, which is determined by the previous equations. For a matrix $M$ with Jordan blocks of sizes $(n_1, n_2, \ldots, n_b)$ with $n_1 \geq n_2 \geq \dots \geq n_b$, the rank of $M^p$ is $\sum_{p : n_p \geq p} (n_p- p)$. Since $M^2=0$ the $n_i$ can only be $1$ or $2$.

We now analyse the following two theories:
\bi
\item {\bf $U(N_c)$ gauge theory with $N_f = 2N_c-k$ flavors.}
The meson $M$ in this theory satisfies $M^2=0$ and ${\rm rank}\; M \leq N_c$.  If $k$ is odd, then the maximal dimensional Higgs branch corresponds to the orbit $(2^{N_c-(k+1)/2},1)$, in which case $M$ has rank $N_c-(k+1)/2$.  On the other hand, if $k$ is even, the orbit is $(2^{N_c-k/2})$ and $M$ has rank $N_c-k/2$. 
\item {\bf $U(N_c-k)$ gauge theory with $2N_c-k$ flavors.}
The meson $M'$ in this theory satisfies $M'^2=0$ and ${\rm rank}\; M' \leq N_c-k$.  Hence the maximal dimensional Higgs branch corresponds to the orbit $(2^{N_c-k}, 1^{k})$, in which the meson $M'$ has rank $N_c-k$.
\ei
$(2^{N_c-k}, 1^{k})$ is a suborbit of $(2^{N_c-(k+1)/2},1)$ when $k$ is odd and of $(2^{N_c-k/2})$ when $k$ is even,%
\footnote{An orbit of Jordan type $\vec\rho$ is a subvariety of an orbit of Jordan type $\vec\rho^\prime$ if $\vec\rho\, <\, \vec\rho^\prime$ \cite{collingwood1993nilpotent}.} hence the Higgs branch of the latter theory is a subvariety of that of the former theory.  Note that the two varieties coincide for 
$k=1$ (\ie~when the $U(N_c)$ gauge theory is ugly).

In the general case they are different but the two varieties coincide when FI terms are turned on. Indeed, part of the moduli space of the first theory is lifted by the presence of FI terms and   it reduces to the suborbit  $(2^{N_c-k}, 1^{k})$.%
\footnote{This can be understood from the general analysis of how mesonic and baryonic branches intersect \cite{Argyres:1996eh},  or by comparing baryonic Hilbert series.} In fact, this is a special case of the general isomorphism between the Higgs branches of $U(N_c)$ gauge theory with $N_f$ flavors and $U(N_f-N_c)$ gauge theory with $N_f$ flavors at non-vanishing FI parameter \cite{Antoniadis:1996ra}, which follows from the Grassmannian duality ${\rm Gr} (N_c, N_f) \cong {\rm Gr} (N_f-N_c, N_f)$. 

Let us mention certain features of the Higgs branch corresponding to the orbit $(2^{N_c-k/2})$, where $k$ is even.  As discussed above, the rank of the meson is $r= N_c-k/2$. According to (2.14) of \cite{Argyres:1996eh}, this corresponds to a submanifold with enhanced gauge group $U(k/2)$. 

\subsection{Symplectic gauge groups}
We now consider the Higgs branches of $USp(2N_c)$ and $USp(2N_c-2k)$ gauge theories with $2N_c+1-k$ flavors.  
\bi
\item The Higgs branch of $USp(2 N_c)$ gauge theory with $SO(4N_c+2-2k)$ flavor symmetry has a meson $M$ as the generator.  It is a matrix of size $(4N_c+2-2k) \times (4N_c+2-2k)$ satisfying $M^2=0$.  The maximal rank of $M$ can be $2N_c+1-k$. Keeping the constraints on $D$-partitions into account, we conclude that the Higgs branch corresponds to the orbit $(2^{2N_c+1-k})$ for odd $k$ and $(2^{2N_c-k},1^{2})$ for even $k$. 
\item The Higgs branch of $USp(2 N_c-2k)$ gauge theory with $SO(4N_c+2-2k)$ flavor symmetry  has the meson of the maximal rank $2N_c-2k$.  Hence, this branch of the moduli space corresponds to the orbit $(2^{2N_c-2k},1^{2k+2})$.
\ei
%
These orbits correspond to the dual partitions $\vec \rho^\vee$ which can be obtained by transposing and $D$-collapsing of the following partitions $\vec \rho$:
\be
\begin{split}
\vec \rho=(2N_c+1-k, 2N_c+1-k)   \quad &\longrightarrow \qquad \vec \rho^\vee= \begin{cases} (2^{2N_c+1-k})~, & \text{$k$ odd} \\  (2^{2N_c-k},1^{2})~ & \text{$k$ even} \end{cases}
  \\
\vec \rho=(2N_c+1, 2N_c-2k+1)   \quad &\longrightarrow \qquad {\vec \rho}^\vee= (2^{2N_c-2k},1^{2k+2}) ~.
\end{split}
\ee
Using the rule given in section \ref{sec:quivers}, we conclude that
\bi
\item The Higgs branch of $USp(2 N_c)$ gauge theory with $SO(4N_c+2-2k)$ flavor symmetry is equal to that of $T_{(2N_c+1-k, 2N_c+1-k)} [SO(4N_c+2-2k)]$, even though the quiver of the latter is not the same as that of the former,  having gauge group $USp(2(N_c-[k/2]))$.
\item  The theory $USp(2 N_c-2k)$ with $SO(4N_c+2-2k)$ flavor symmetry is identical to $T_{(2N_c+1, 2N_c-2k+1)} (SO(4N_c+2-2k))$.
\ei

%
Since the orbit $(2^{2N_c-2k},1^{2k+2})$ is a suborbit of $(2^{2N_c+1-k})$ and of $(2^{2N_c-k},1^{2})$, we reach the conclusion that the Higgs branch of the $USp(2N_c-2k)$ gauge theory is a subvariety of the $USp(2N_c)$ gauge theory.  

\subsection{Orthogonal gauge groups} \label{sec:orthoSeiberg}

Now we turn to $O(N_c)$ and $O(2N_f-N_c+2)$ gauge theories with $N_f = N_c-k$ flavors. 
\bi
\item The Higgs branch of $O(N_c)$ gauge theory with $USp(2N_c-2k)$ flavor symmetry has a meson $M$ as the generator.  It is a matrix of size $(2N_c-2k) \times (2N_c-2k)$ satisfying $M^2=0$.  The maximal rank of $M$ can be $N_c-k$.  Hence the Higgs branch corresponds to the orbit $(2^{N_c-k})$. 
\item The Higgs branch of $O(2N_f-N_c+2)$ gauge theory with $USp(2N_c-2k)$ flavor symmetry  has the meson of the maximal rank $2N_f-N_c+2$.  Hence, this branch of the moduli space corresponds to the orbit $(2^{N_c-2k+2},1^{2k-4})$.
\ei
In some cases, these orbits correspond to the dual partitions $\vec \rho^\vee$ which can be obtained by transposing and $C$-collapsing of the following partitions $\vec \rho$:
\be
\begin{split}
\vec \rho=(N_c-k,N_c-k,1) \quad &\longrightarrow \qquad 
\vec \rho^\vee = (2^{N_c-k}) \\
\vec \rho=(N_c-1,N_c-2k+1,1) \quad &\longrightarrow \qquad  {\vec \rho}^\vee = (2^{N_c-2k+2},1^{2k-4}), \qquad   \, N_c\,  {\rm even}.
\end{split}
\ee
Using the rule given in section \ref{sec:quivers}, we conclude that
\bi
\item The Higgs branch of $O(N_c)$ gauge theory with $USp(2N_c-2k)$ flavor symmetry is equal to that of $T_{(N_c-k,N_c-k,1)} [USp(2N_c-2k)]$, even though the quiver of the latter is not the same as that of the former.
\item The theory $O(2N_f-N_c+2)$ with $USp(2N_c-2k)$ flavor symmetry with of $N_c$ even is identical to $T_{(N_c-1,N_c-2k+1,1)} (USp(2N_c-2k))$. 
\ei
Since $(2^{N_c-2k+2},1^{2k-4})$ is a suborbit of $(2^{N_c-k})$, the Higgs branch of the $O(2N_f-N_c+2)$ gauge theory is a subvariety of the $O(N_c)$ gauge theory. When $k=2$ they are the same.

\subsubsection*{$O(N)$ vs $SO(N)$ gauge group.}  
For $SO(N)$ gauge theory with $N_f$ flavor, as discussed below (2.7) of \cite{Argyres:1996hc}, the baryon can acquire a non-zero vacuum expectation value on the branch on which the meson has rank $r$ only if $r =N \leq N_f$.  If this condition does not hold, the gauge group $SO(N)$ can be taken as $SO(N)$ or $O(N)$ with no distinction on the Higgs branch.  

As a consequence, for the orbit $(2^{N_c-k})$ with $k\geq 1$, the corresponding gauge theory can be taken as either $O(N_c)$ or $SO(N_c)$ gauge group with $N_c-k$ flavors; they both have the same Higgs branch.
\paragraph{Example: $SO(4)$ and $O(2)$ gauge theories with $2$ flavors}
$O(4)$ with two flavors and $O(2)$ with two flavors are a pair of theories as discussed above with $N_c=4$ and $k=2$. Therefore they have the same hypermultiplet moduli space.  Moreover, Since $k\geq 1$, $SO(4)$ with two flavors is the same as $O(4)$ with two flavors by the previous remark. 
\paragraph{Example: $SO(3)$ and $O(3)$ gauge theories with $1$ flavor}
 Using {\tt Macaulay2}, we find that the Hilbert series of the $F$-flat moduli space is
\bea
H[\fflat] (t; z; x) = \left[ 1-t^2\chi^{SO(3)}_{[2]} (z) +t^3 \chi^{USp(2)}_{[1]}(x) \right] \PE \left[t \chi^{SO(3)}_{[2]} (z) \chi^{USp(2)}_{[1]}(x) \right]~.
\eea
Observe that the $F$-flat space is a 4 complex dimensional non-complete-intersection space.  Integrating over the $SO(3)$ gauge group, we obtain the Hilbert series of $\BC^2/\BZ_2$.
\bea
H[\text{$SO(3)$ w/ $1$ flv}](t; x) = \int {\rm d} \mu_{SO(3)}(z) H[\fflat] (t; z; x) = \sum_{p=0}^\infty \chi^{USp(2)}_{[2m]} t^{2m}~.
\eea
The generator of this space contains only the meson, which does not involve a contraction with the epsilon tensor (\ie~ no baryon).   This moduli space may as well be viewed as the Higgs branch of $O(3)$ gauge theory with one flavor, since the parity symmetry does not project out any gauge invariant quantity from the $SO(3)$ counterpart.

It is worth pointing out that the gauge symmetry is not completely broken at a generic point on the hypermultiplet moduli space; rather $SO(3)$ or $O(3)$ gauge symmetry is broken to $SO(2)$ so that the space is $3-(3-1)=1$ quaternionic dimensional.

As argued above, the $O(1)$ gauge theory with $1$ flavor has the same Higgs branch. The Higgs branch of $O(1)$ with $1$ flavor is the reduced moduli space of 1 $USp(2)$ instanton on $\BC^2$; this space is $\BC^2/\BZ_2$, in agreement with the above computation.

\section{Computing residues in the Higgs branch} \label{app:Higgsresidue}

In this appendix we derive the baryonic generating function of $T^{\vec \rho^\prime}_{\vec \sigma}(SU(N))$ from that of $T^{\vec \rho}_{\vec \sigma}(SU(N))$, where $\vec\rho^\prime$ is obtained from $\vec\rho$ by moving the last box to a previous column, by computing residues at certain poles of the latter. Since all partitions can be obtained by the partition $(1,1\dots,1)$ by repeatedly moving a single box, it suffices to consider 
\bea\label{movebox}
\vec \rho = (\rho_1, \ldots, \rho_{d-h}, H, 1^h)~, \qquad \vec \rho' = (\rho_1, \ldots, \rho_{d-h}, H+1, 1^{h-1})~,
\eea
where 
the lengths of partitions $\vec \rho$ and $\vec \rho'$ are $d+1$ and $d$ respectively, and
\bea
H= \rho_{d-h+1}~.
\eea

Let us suppose that $H>1$ and return to the special case of $H=1$ later.  The quiver diagram of $T^{\vec \rho}_{\vec \sigma}(SU(N))$ is as follows:
\bea
\begin{tikzpicture}[font=\footnotesize]
\begin{scope}[auto,%
  every node/.style={draw, minimum size=1.3cm}, node distance=0.6cm];
\node[circle] (UN1) at (0, 0) {$N_1$};
\node[circle, right=of UN1] (UN2) {$N_2$};
\node[draw=none, right=of UN2] (dots) {$\cdots$};
\node[circle, right=of dots] (UNlm1) {$N_{H}$};
\node[circle, right=of UNlm1] (UNl) {$N_{H+1}$};
\node[draw=none, right=of UNl] (dots2) {$\cdots$};
\node[rectangle, below=of UN1] (UM1) {$h$};
\node[rectangle, below=of UNlm1] (UMlm1) {$1$};
\node[rectangle, below=of UNl] (UMl) {$M_{H+1}$};
\end{scope}
\draw (UN1) -- (UN2)
(UN2)--(dots)
(dots)--(UNlm1)
(UNlm1)--(UNl)
(UN1)--(UM1)
(UNlm1)--(UMlm1)
(UNl)--(dots2)
(UNl)--(UMl);
\end{tikzpicture}
\eea
The quiver diagram of $T^{\vec \rho'}_{\vec \sigma}(SU(N))$ is
\bea
\begin{tikzpicture}[font=\footnotesize]
\begin{scope}[auto,%
  every node/.style={draw, minimum size=1.5cm}, node distance=0.6cm];
\node[circle] (UN1) at (0, 0) {$N_1-1$};
\node[circle, right=of UN1] (UN2) {$N_2-1$};
\node[draw=none, right=of UN2] (dots) {$\cdots$};
\node[circle, right=of dots] (UNlm1) {$N_{H}-1$};
\node[circle, right=of UNlm1] (UNl) {$N_{H+1}$};
\node[draw=none, right=of UNl] (dots2) {$\cdots$};
\node[rectangle, below=of UN1] (UM1) {$h-1$};
\node[rectangle, below=of UNl] (UMl) {$M_{H+1}+1$};
\end{scope}
\draw (UN1) -- (UN2)
(UN2)--(dots)
(dots)--(UNlm1)
(UNlm1)--(UNl)
(UN1)--(UM1)
(UNl)--(dots2)
(UNl)--(UMl);
\end{tikzpicture}
\eea
Let us start with the baryonic generating function of $T^{\vec \rho}_{\vec \sigma}(SU(N))$ given by \eref{Molien-Weyl}.
Taking fugacities $w_{1,h}$ for the last Cartan $U(1)$ in the flavor symmetry $U(h)$ and $w_{H,1}$ for the flavor symmetry $U(1)$ to be as follows
\bea
w_{1,h} = (tz)^{-H} \tilde{w}~, \qquad w_{H,1} = (tz) \tilde{w}~,
\eea 
we find that there is a pole at $z=1$ with the contributions from the residues
\bea
s_{i, N_i} = t^{i-H} \tilde{w} z^{-H} = t^i w_{1,h}~, \qquad \text{with $i=1,\ldots, H$}~.
\eea
Evaluating the residues, we obtain
\be\label{resgener}
\begin{split}
& \underset{z\rightarrow 1}{{\rm Res}}  \,  g[T^{\vec \rho}_{\vec \sigma}(SU(N))] (t; \vec w_1, \ldots, \vec w_{\hat{\ell}}; \vec B) \Bigg |_{\substack{w_{1,h}= (tz)^{-H} \tilde{w} \\ w_{H,1}= (tz) \tilde{w}}}  \\
&= \frac{1}{H+1} \tilde{w}^{-\sum_{i=1}^H B_i}  t^{\sum_{i=1}^H (H-i) B_i}{\rm PE}  \left[t^2+t \sum_{q=1}^{h-1} (t^{1-H} \tilde{w} w_{1,q}^{-1}+t^{H-1} \tilde{w}^{-1} w_{1,q})\right] \times \\
&\qquad g[T^{\vec \rho'}_{\vec \sigma}(SU(N))] (t;  \tilde{\vec w}_1, \ldots,  \tilde{\vec w}_{\hat{\ell}'}; \vec B) \Bigg |_{\substack{\tilde{w}_{H+1, M_{H+1}+1}= \tilde{w}}} ~,
\end{split}
\ee
Note that the prefactor $\tilde{w}^{-\sum_{i=1}^H B_i}$ becomes $1/{\vec x}^{\vec s(\vec n)}$ given by \eref{prefactxs} of the new quiver $T^{\vec \rho'}_{\vec \sigma}(SU(N))$ after substituting $B_i=n_i-n_{i+1}, \; \tilde{w} = \tilde{w}_{H+1, M_{H+1}+1}$. 

For the case of $H=1$, the quiver diagrams of $T^{\vec \rho}_{\vec \sigma}(SU(N))$ and $T^{\vec \rho'}_{\vec \sigma}(SU(N))$ are respectively as follows:
\bea
\begin{tikzpicture}[scale=0.8, transform shape]
\begin{scope}[auto,%
  every node/.style={draw, minimum size=1.3cm}, node distance=0.6cm];
\node[circle] (UN1) at (0, 0) {$N_1$};
\node[circle, right=of UN1] (UN2) {$N_2$};
\node[draw=none, right=of UN2] (dots) {$\cdots$};
\node[rectangle, below=of UN1] (UM1) {$h+1$};
\node[rectangle, below=of UN2] (UM2) {$M_{2}$};
\end{scope}
\draw (UN1) -- (UN2)
(UN1) -- (UM1)
(UN2)--(UM2)
(UN2)--(dots);
\draw[thick,->] (6,0) -- (7,0);
\end{tikzpicture}
\qquad \qquad
\begin{tikzpicture}[scale=0.8, transform shape]
\begin{scope}[auto,%
  every node/.style={draw, minimum size=1.3cm}, node distance=0.6cm];
\node[circle] (UN1) at (0, 0) {$N_1-1$};
\node[circle, right=of UN1] (UN2) {$N_2-1$};
\node[draw=none, right=of UN2] (dots) {$\cdots$};
\node[rectangle, below=of UN1] (UM1) {$h$};
\node[rectangle, below=of UN2] (UM2) {$M_{2}+1$};
\end{scope}
\draw (UN1) -- (UN2)
(UN1) -- (UM1)
(UN2)--(dots)
(UN2)--(UM2);
\end{tikzpicture}
\eea
Formula \eref{resgener} becomes
\bea
& \underset{z\rightarrow 1}{{\rm Res}}  \,  g[T^{\vec \rho}_{\vec \sigma}(SU(N))] (t; \vec w_1; \vec B) \Bigg |_{\substack{w_{1,1}= t z x_1 \\ w_{1,q}=x_q \,\, q=2,\cdots N-1\\ w_{1,N}=  (tz)^{-1} x_1} } =  \\
& \frac{1}{2} x_1^{-B_1} {\rm PE}  \left[t^2+t \sum_{q=1}^{h-1} (x_1 x_{q}^{-1}+ x_1^{-1} x_{q})\right] g[T^{\vec \rho'}_{\vec \sigma}(SU(N))] (t; \tilde{\vec w}_1, \ldots,  \tilde{\vec w}_{\hat{\ell}'}; \vec B)
 \nn~,
\eea
where the first line receives the contribution from the residue:
\bea
s_{1, N_1} = t w_{1,h} = x_1 z^{-1}~.
\eea
For $H=1$ and $h=N$, we reproduce \eref{H1hNm1} presented in the main text.

\section{More examples for 
orthogonal and symplectic groups}\label{scan} 

We present another set of examples for $USp(6)$ and $SO(8)$.  The relevant information are contained in the following series of tables.
 For partitions $\vec\sigma$ with the same $\vec\sigma^\vee$ we have two different quivers; one of the two comes in two versions,  (I) and (II),  corresponding to a moduli space that is a nilpotent orbit or its double covering. The same subscripts are used to distinguish an  orbit and its covering in \tref{tab:paraC3} and \tref{tab:paraD4}. We list in the tables only the quivers whose Hilbert series can be obtained in terms of the Hall-Littlewood formula. 

\subsection{$T^{\vec \sigma}(USp(6))$}
We present the quiver diagrams of $T^{\vec \sigma}(USp(6))$ and their mirror duals $T_{\vec \sigma}(SO(7))$ for various $C_3$-partitions $\vec \sigma$ in Table 9.  Information about the associated nilpotent orbits are provided in \tref{tab:paraC3}. All statements of equality of Coulomb/Higgs branches between different theories hold for vanishing background charges. 

\begin{table}[H]
\hspace{-2cm}
{\scriptsize
\begin{tabular}{|c|c|c|c|c|}
\hline
$C_3$-part. ${\vec \sigma}$ & Quiver of $T^{\vec \sigma}(USp(6))$ & Quiver with the same  & Quiver of $T_{\vec \sigma}(SO(7))$ & Quiver with the same\\
 & & Coulomb branch as $T^{\vec \sigma}(USp(6))$& &  Higgs branch of $T_{\vec \sigma}(SO(7))$ \\ 
\hline \hline
$(4,1^2), \; (4,2)$ &  $(4,1^2):\; \node{}{2}-\Node{\ver{}{1}}{2}-\node{}{3}-\Node{}{2}-\node{\Ver{}{2}}{3}$ &  $(4,2): \; \node{}{2}-\Node{\ver{}{1}}{2}-\node{}{3}-\Node{\ver{}{1}}{2}-\node{}{\substack{O(2)}}$~ (I) & $(4,1^2): \; \node{}{3^*}-\Node{}{2}-\sqnode{}{7}$ & $(4,2): \; \node{}{O(1)}-\Node{}{2}-\sqnode{}{7}$ ~ (I)   \\
\hline \hline
$(3^2)$ & $\node{}{2}-\Node{}{2}-\node{\Ver{}{2}}{4}-\Node{}{2}-\node{}{2}$ && $\node{}{1}-\Node{}{4^*}-\sqnode{}{7}$~ equivalent to~ $\Node{}{4^*}-\sqnode{}{8}$ & \\
\hline 
 $(2^3)$ &$\node{}{2}-\Node{}{2}-\node{}{4}-\Node{\ver{}{3}}{4}-\node{}{3}$ && $\node{}{3}-\Node{}{4}-\sqnode{}{7}$ &\\
\hline \hline
 $(2^2,1^2)$  & $\node{}{2}-\Node{}{2}-\node{}{4}-\Node{\ver{}{2}}{4}-\node{\Ver{}{2}}{4}$~(II)& &  $\node{}{O(1)}-\Node{}{2}-\node{}{3}-\Node{}{4}-\sqnode{}{7}$~ (II)&\\
 \hline  
 $(2,1^4)$ & $\node{}{2}-\Node{}{2}-\node{}{4}-\Node{\ver{}{1}}{4}-\node{\Ver{}{4}}{5}$ &  $\node{}{2}-\Node{}{2}-\node{}{4}-\Node{\ver{}{2}}{4}-\node{\Ver{}{2}}{O(4)}$~(I)&  $\node{}{3}-\Node{}{2}-\node{}{5*}-\Node{}{4}-\sqnode{}{7}$ & $\node{}{O(1)}-\Node{}{2}-\node{}{O(3)}-\Node{}{4}-\sqnode{}{7}$~(I) \\
\hline \hline
$(1^6)$ &$\node{}{2}-\Node{}{2}-\node{}{4}-\Node{}{4}-\node{\Ver{}{6}}{6}$ && $\node{}{O(1)}-\Node{}{2}-\node{}{O(3)}-\Node{}{4}-\node{}{O(5)}-\Node{}{6}-\sqnode{}{7}$ &\\
\hline
\end{tabular}
\label{tab:quivcc3}
\caption{Quiver diagrams for $T^{\vec \sigma}(USp(6))$ and their mirror theories. The asterisk $*$ indicates the gauge group that renders the quiver a `bad' theory. Each black node labeled by $M$ denotes a $USp(M)$ group, each gray node labeled by $N$ denotes an $SO(N)$ group and for an orthogonal group $O(N)$ is spelt out explicitly. Whenever a gauge group $O(N)$ is indicated by the asterisk, such a gauge group can be taken as $O(N)$ or $SO(N)$ without changing the Higgs branch; see Appendix \ref{sec:orthoSeiberg}.}}
\end{table}%

\begin{table}[H]
\hspace{-1.2cm}
{\scriptsize
\begin{tabular}{|l|l|l|l|l|c|}
\hline
$T^{\vec \sigma}(USp(6))$ & $B_3$-part. ${\vec \sigma}^\vee$ & ${\vec \sigma}^L$: Levi subgroup & $\Delta_{\vec \sigma} \subset \Delta_+(SO(7)) $ & Unrefined Coulomb HS   \\ 
              & &                                            &  & of $T^{\vec \sigma}(USp(6))$ from \eref{mainHSG} \\
\hline
$(4,2)$  & $(3,1^4)$ &  &$\{\e_2-\e_3,\e_2+\e_3,\e_2, \e_3 \}$   & $\frac{(1+t^2) (1+10 t^2+20 t^4+10 t^6+t^8)}{(1-t^2)^{10}}$     \\
$(4,1^2)$  & &{$(1^2,5):U(1) \times SO(5)$} &   &   $=1 + 21 t^2 + 195 t^4 + 1155 t^6 +\ldots $  \\
\hline \hline
$(3^2)$ & $(3,2^2)^\dagger$&$(3^2,1):U(3)$  &  $\{\e_1-\e_2,\e_1-\e_3,\e_2-\e_3 \}$       & $\frac{(1 + t^2)^2 (1 + 14 t^2 + 36 t^4 + 14 t^6 + t^8)}{(1 - t^2)^{12}}$   \\ 
              &&                                        &                 &$=1 + 28 t^2 + 335 t^4 + 2492 t^6 +\ldots$  \\ 
\hline
 $(2^3)$ & $(3^2,1)$ &$(2^2,3):U(2) \times SO(3)$  &   $\{\e_1-\e_2, \e_3 \}$                 &$\frac{(1+t^2) (1+6 t^2+21 t^4+28 t^6+21 t^8+6 t^{10}+t^{12})}{(1 - t^2)^{14}}$  \\
                    &        & &       &  $= 1 + 21 t^2 + 230 t^4 + 1722 t^6 +\ldots$  \\
\hline \hline
 $(2^2,1^2)_{{\rm II}}$ & {${(5,1^2)}$}  &$(2^2,1^2,1): U(2) \times U(1)$  &   $\{ \e_1-\e_2 \}$  or $\{ \e_2-\e_3 \}$        &  $\frac{\left(1+t^2\right)^2 \left(1 + 3 t^2 + 14 t^4 + 20 t^6 + 14 t^8 + 3 t^{10} + t^{12}\right)}{(1-t^2)^{16}}$  \\
      & &        &       &  $=1 + 21 t^2 + 237 t^4 + 1883 t^6+ \ldots$ \\
 \hline  
 $(2,1^4)$ & {$(5,1^2)$} &$(1^2,1^2,3): U(1)^2 \times SO(3)$   &  $\{ \e_3 \}$      &  $\frac{(1 + t^2)^2 (1 + t^4) (1 + 3 t^2 + 6 t^4 + 3 t^6 + t^8)}{(1-t^2)^{16}}$  \\
   & &        &       &   $=1 + 21 t^2 + 230 t^4 + 1743 t^6+ \ldots$  \\
\hline \hline
$(1^6)$ & $(7)$ & $(1^2,1^2,1^2,1): U(1)^3$ &	$\emptyset$	 & $\frac{(1 + t^2)^3 (1 - t + t^2) (1 + t + t^2) (1 + t^4) (1 - t^2 + t^4)}{(1 - t^2)^{18}}$  \\
    & &        &       &$ =1 + 21 t^2 + 230 t^4 + 1750 t^6 + \ldots$   \\
\hline
\end{tabular}
}
\caption{For $SO(7)$, the positive roots are ${\vec e}_i - {\vec e}_j$, ${\vec e}_i + {\vec e}_j$ and ${\vec e}_i$, with $1\leq i < j \leq 3$.   The $B_3$-partition $\vec \sigma^\vee$ can be obtained from $\vec \sigma$ by adding a box, transposing and B-collapsing. The subscript below  ${\vec \sigma}$ indicates that we are using the theory $T^{\vec\sigma}(G)_{{\rm II}}$; we are correspondingly using a polarization where the  the degree of map \eref{resol}, as computed using \eref{setI},  is  $2$. The orbit marked by $\dagger$ is non-normal, see Page 543 of \cite{kraftprocesi}; this should probably explain an anomalous point in the table: although the degree of the map \eref{resol} is one, the Hall-Littlewood formula compute the Hilbert series for a covering of the nilpotent orbit. The resolution \eref{resol} has the same holomorphic functions of the normalization of $O_{\vec\sigma^\vee}$, which are more than for $O_{\vec\sigma^\vee}$ itself.
 For $\vec \sigma=(4,2), (4,1^2), (2^3), (1^6)$, we have checked that the Coulomb branch HS given above is equal to the Higgs branch HS of $T_{\vec \sigma}(SO(7))$.}
\label{tab:paraC3}
\end{table}%

\begin{table}[H]
\begin{center}
\begin{tabular}{|c|c|c|}
\hline
$C_3$-partition ${\vec \sigma}$ & Possible background fluxes $\vec n^{\vec\sigma}$ & $p_{\vec \sigma}(\vec n^{\vec\sigma})$ \\
\hline
$(4,2)$  &  $(0,0,0)$  & $0$\\
$(4,1^2)$ &$(n,0,0), \; n\geq 0$ & $3n$     \\
\hline
$(3^2)$ & $(n,n,n), \; n\geq 0$& $4n$ \\
\hline 
 $(2^3)$ & $(n, n, 0), \; n\geq 0$ & $4n$ \\
\hline \hline
 $(2^2,1^2)$  & $(n_1, n_1, n_2), \; n_1, n_2 \geq 0$& $4n_1+2n_2$ \\
 \hline  
 $(2,1^4)$ & $(n_1, n_2,0), \; n_1 \geq n_2 \geq 0$ & $4n_1+2n_2$ \\
\hline \hline
$(1^6)$ &$(n_1,n_2, n_3), \; n_1 \geq n_2 \geq n_3 \geq 0$ & $5 n_1 + 3 n_2 + n_3$\\
\hline
\end{tabular}
\caption{The possible background fluxes $\vec n^{\vec\sigma}$ that can be turned on in \eref{mainHS} and the corresponding power of $t$, $p_{\vec \sigma}(\vec n^{\vec\sigma})$, appearing in  \eref{mainHS}.}
\label{tab:paraC3a}
\end{center}
\end{table}%

\subsection{$T^{\vec \sigma}(SO(8))$}
We present the quiver diagrams of $T^{\vec \sigma}(SO(8))$ and their mirror duals $T_{\vec \sigma}(SO(8))$ for various $D_4$-partitions $\vec \sigma$ in \tref{tab:quiverD4}.  Information about the associated nilpotent orbits is provided in \tref{tab:paraD4}.

\begin{table}[H]
\hspace{-1.6cm}
{\scriptsize
\begin{tabular}{|c||c|c||c|c|}
\hline
$D_4$-partition ${\vec \sigma}$ & Quiver of $T^{\vec \sigma}(SO(8))$ &Quiver with the same  & Quiver of $T_{\vec \sigma}(SO(8))$ &Quiver with the same  \\
 & & Coulomb branch as $T^{\vec \sigma}(SO(8))$& & Higgs branch as $T_{\vec \sigma}(SO(8))$ \\
\hline \hline
$(3^2,1^2)$ &$\node{}{2}-\Node{}{2}-\node{}{4}-\Node{\ver{}{2}}{4}-\node{}{4}-\Node{\ver{}{2}}{2}$ \, (II)  && $\node{}{2}-\Node{\ver{}{8}}{4}$\, (II) &   \\
\hline
$(3,2^2,1)$ & $\node{}{2}-\Node{}{2}-\node{}{4}-\Node{\ver{}{1}}{4}-\node{\Ver{}{2}}{5}-\Node{\ver{}{1}}{2}$ & $\node{}{2}-\Node{}{2}-\node{}{4}-\Node{\ver{}{2}}{4}-\node{}{O(4)}-\Node{\ver{}{2}}{2}$ \, (I)  & $\node{}{4^*}-\Node{\ver{}{8}}{4}$ & $\node{}{O(2)}-\Node{\ver{}{8}}{4}$ \, (I)  \\
\hline
\hline
$(4,4)$ & $\node{}{2}-\Node{}{2}-\node{\Ver{}{2}}{4}-\Node{}{2}-\node{}{2}$ && $\Node{\ver{}{8}}{4}$ &\\
\hline
\end{tabular}}
\caption{Quiver diagrams for $T^{\vec \sigma}(SO(8))$ and their mirror theories. The asterisk $*$ indicates the gauge group that renders the quiver a `bad' theory. Each black node labeled by $M$ denotes a $USp(M)$ group, each gray node labeled by $N$ denotes an $SO(N)$ group and for an orthogonal group $O(N)$ is spelt out explicitly. Whenever a gauge group $O(N)$ is indicated by the asterisk, such a gauge group can be taken as $O(N)$ or $SO(N)$ without changing the Higgs branch; see Appendix \ref{sec:orthoSeiberg}.}
\label{tab:quiverD4}
\end{table}%

\begin{table}[H]
\hspace{-2cm}
{\scriptsize
\begin{tabular}{|l|l|l|c|l|l|l|l|}
\hline
$D_4$-part. ${\vec \sigma}$ & $D_4$-part. ${\vec \sigma}^\vee$ & ${\vec \sigma}^L$: Levi subgroup&$\Delta_{\vec \sigma} \subset \Delta_+(SO(8)) $ &  Coulomb branch HS of $T^{\vec \sigma}(SO(8))$ from \eref{mainHSG}  \\
\hline
 $(3^2,1^2)_{{\rm II}}$ & {${(3^2,1^2)}$} &$(3^2,1^2): U(3) \times U(1)$  &   $\{ \e_1-\e_2,  \e_1-\e_3, \e_2-\e_3 \}$      &  $\frac{\left(1+t^2\right)^2 \left(1 + 9 t^2 + 73 t^4 + 227 t^6 + 340 t^8 + 227 t^{10 }+ 73 t^{12} + 
 9 t^{14} + t^{16}\right)}{(1-t^2)^{18}}$ \\
   & & see (2.21) of \cite{Chacaltana:2012zy}   &         &  $=1 + 28 t^2 + 433 t^4 + 4626 t^6 + 37374 t^8 + \ldots$\\
 \hline  
 $(3,2^2,1)$ & {${(3^2,1^2)}$} &$(2^2,4): U(2) \times SO(4)$  &  $\{ \e_1-\e_2,  \e_3-\e_4, \e_3+\e_4 \}$       &  $\frac{(1 + t^2)^2 (1 + 9 t^2 + 45 t^4 + 109 t^6 + 152 t^8 + 109 t^{10} + 45 t^{12} + 
  9 t^{14} + t^{16})}{(1-t^2)^{18}}$   \\
               &  & see (2.20) of \cite{Chacaltana:2012zy}                                      &                 &$=1 + 28 t^2 + 405 t^4 + 3976 t^6 +29652 t^8 + \ldots$ \\ 
\hline \hline
 $(4,4)$ & {${(2^4)}$} &$(4^2): U(4)$   &  $\{ \e_i-\e_j\}_{1\leq i <j \leq 4}$      & $\frac{(1 + t^2)^2 (1 + 14 t^2 + 36 t^4 + 14 t^6 + t^8)}{(1-t^2)^{12}}$ \\
   & &        &       &   $=1 + 28 t^2 + 335 t^4 + 2492 t^6 + 13524 t^8 + \ldots$ \\
\hline   
\end{tabular}
}
\caption{For $SO(8)$, the positive roots are ${\vec e}_i - {\vec e}_j$, ${\vec e}_i + {\vec e}_j$, with $1\leq i < j \leq 4$.   The $D_4$-partition $\vec \sigma^\vee$ can be obtained from $\vec \sigma$ by adding a box, transposing and D-collapsing.}
\label{tab:paraD4}
\end{table}%

\section{More on $T^{\vec \rho}_{\vec \sigma}(USp'(2N))$ theories} \label{sec:TsigrhoUSpP_Higgs}

In this appendix we provide more details on $T^{\vec \rho}_{\vec \sigma}(USp'(2N))$ theories, which are realised on the worldvolume of $N$ D3 branes parallel to an $\widetilde{O3}^+$ plane and ending on systems of half D5 branes and of half NS5 branes. ${\vec \sigma}$ and ${\vec \rho}$, which determine how the D3 branes end on the D5 and on the NS5 branes, are both C-partitions of $USp(2N)$.  

Some examples of the $T^{\vec \sigma}_{\vec \rho}(USp'(2N))$ theory were given in sections 7 and 9 of \cite{Feng:2000eq}. In section \eqref{sec:TsigrhoUSpP} we provided a prescription to write down the  quiver diagram for general C-partitions $\vec \sigma$ and $\vec \rho$. Let us present some examples here: 
\bi
\item If $\vec \sigma = \vec \rho =(1^{2N})$, we refer to the theory as $T(USp'(2N))$.  The quiver diagram is 
\bea
&T(USp'(2N)): \nn \\
& [USp(2N)]-(O(2N+1))-\cdots -(USp(4))- (O(5))-(USp(2))-(O(3))  ~.
\eea
\item If $\vec \sigma = (1^{2N})$ and $\rho=(2,1^{2N-2})$, the quiver diagram is
\bea
& T^{(1^{2N})}_{(2,1^{2N-2})}(USp'(2N)): \nn \\
& [USp(2N)]-(O(2N-1))- \cdots -(USp(4))-(O(3))-(USp(2))-(O(1)) ~.
\eea
\ei

\subsection{The Higgs branch}

The Coulomb branch of the quiver gauge theory associated to $T^{\vec \sigma}_{\vec \rho}(USp'(2N))$ cannot be studied using the monopole formula because the theory is bad. We can however study the Higgs branch of the mirror theory $T^{\vec \rho}_{\vec \sigma}(USp'(2N))$, which is protected against quantum corrections. In analogy to \eqref{orbit}, it is natural to expect that the Higgs branch of $T^{\vec \rho}_{\vec \sigma}(USp'(2N))$ is given by the closure of a nilpotent orbit intersected with a Slodowy slice:
\bea\label{orbit_SpPrime}
\bar{O}_{\vec \sigma^\vee} \cap S_{\vec \rho}\, .
\eea
$S_{\vec \rho}$ is the Slodowy slice associated to the homomorphism
$\vec\rho : {\rm Lie}(SU(2))\rightarrow {\rm Lie}(USp(2N))$, where $\vec\rho$ is a C-partition. The type $\vec\sigma^\vee$ of the nilpotent orbit is given by a C-partition that is determined by a map $\vee : \sigma \rightarrow \sigma^\vee$ from C-partitions to C-partitions, because $USp'$ is self-dual. 
We propose that this map is defined as follows. We first map the C-partition $\vec\sigma$ into a $B$-partition $\tilde{\vec \sigma}$ that is defined as the B-collapse of the partition $(\vec \sigma, 1)$. Then we apply to $\tilde{\vec\sigma}$ the previously defined map \eqref{vee} from B-partitions to C-partitions \cite{spaltenstein,barbash} to obtain the desired $\sigma^\vee$. 

According to our proposal, the Higgs branch of $T^{\vec\rho}_{\vec \sigma}(USp'(2N))$, with $\vec \sigma$ and $\vec \rho$ two C-partitions of $2N$, is equal to that of $T^{\vec\rho}_{\tilde{\vec \sigma}}(USp(2N))$, with $\tilde{\vec \sigma}$ the B-partition of $2N+1$ defined above. We have checked explicitly the equalities of the Higgs branch Hilbert series of the following sets of theories 
\bea
&\{T_{(1^5)}(USp(4)), T_{(1^4)}(USp'(4)), T_{(2,1^2)}(USp'(4))\}, \nn \\
& \{T^{(2,1^2)}_{(1^5)}(USp(4)), T^{(2,1^2)}_{(1^4)}(USp'(4)), T^{(2,1^2)}_{(2,1^2)}(USp'(4))\} \nn\\ 
&\{T_{(2^2,1)}(USp(4)), T_{(2^2)}(USp'(4))\} \nn \\
&\{ T_{(1^7)}(USp(6)), T_{(1^6)}(USp'(6)), T_{(2,1^4)}(USp'(6))\} \nn \\
&\{T_{(2^2,1^3)}(USp(6)), T_{(2^2,1^2)}(USp'(6)), T_{(2^3)}(USp'(6)) \} \nn \\
& \{T^{(2,1^4)}_{(2^2,1^3)}(USp(6 )), T^{(2,1^4)}_{(2^2,1^2)}(USp'(6)), T^{(2,1^4)}_{(2^3)}(USp'(6 ))\} \nn\\
&\{ T_{(3^2,1)_{\rm II}} (USp(6)), T_{(3^2)}(USp'(6)), T_{(4,2)}(USp'(6))\} ~, \nn
\eea
where indeed the B-partition $\tilde{\vec \sigma}$ in $T^{\vec\rho}_{\tilde{\vec \sigma}}(USp(2N))$ is obtained as the B-collapse of the partition $(\vec \sigma, 1)$, with $\vec\sigma$ the C-partition appearing in $T^{\vec\rho}_{{\vec \sigma}}(USp'(2N))$. Note that the $\sim$ map from C-partitions to B-partitions is not injective, therefore $T^{\vec\rho}_{{\vec \sigma}}(USp'(2N))$ theories with different $\vec \sigma$ can have the same Higgs branch.



\bibliographystyle{ytphys}
\bibliography{ref}

\end{document}